\documentclass[10pt,preprint,sort&compress]{elsarticle}
\pdfoutput=1
\usepackage[latin1]{inputenc}
\usepackage{graphicx}
\usepackage{latexsym}
\usepackage[T1]{fontenc}
\usepackage{amsmath}
\usepackage{amssymb}
\usepackage{endnotes}
\usepackage{lmodern}
\usepackage{float}
\usepackage[rightcaption]{sidecap}

\journal{Comptes Rendus Physique}

\begin{document}

\begin{frontmatter}

\title{Electronic nematic susceptibility of iron-based superconductors} 

\author{Anna E. Böhmer}
\fntext[myfootnote]{Now at Ames Laboratory/ Iowa State University}
\ead{anna.boehmer@kit.edu}
\author{Christoph Meingast}
\ead{christoph.meingast@kit.edu}
\address{Institut für Festkörperphysik, Karlsruhe Institute of Technology, 76021 Karlsruhe, Germany}

\begin{abstract}
We review our recent experimental results on the electronic nematic phase in electron- and hole-doped BaFe$_2$As$_2$ and FeSe. The nematic susceptibility is extracted from shear-modulus data (obtained using a three-point-bending method in a capacitance dilatometer) using Landau theory and is compared to the nematic susceptibility obtained from elastoresistivity and Raman data. FeSe is particularly interesting in this context, because of a large nematic,
i.e., a structurally distorted but paramagnetic, region in its phase diagram. Scaling of the nematic susceptibility with the spin lattice relaxation rate from NMR, as predicted by the spin-nematic theory, is found in both electron- and hole-doped BaFe$_2$As$_2$, but not in FeSe. 
The intricate relationship of the nematic susceptibility to spin and orbital degrees of freedom is discussed.\\

\end{abstract}


\end{frontmatter}

\section{Introduction}

Electronic nematic phases have recently attracted considerable attention, especially in connection to high-temperature superconductivity \cite{Fradkin2010,Fernandes2014}. For these phases, the term ``nematic'', which originally refers to a liquid-crystal phase, in which rotational symmetry is broken while translational symmetry is preserved, has been borrowed from its original context to describe a symmetry breaking due to electronic effects\cite{Fernandes2012}. In the iron-based systems, the stripe-type antiferromagnetic phase\cite{Cruz2008,Rotter2008III,Nomura2008,Lumsden2010}, which occurs in close proximity to the superconducting phase, reduces the $C_4$ rotational symmetry of the high-temperature tetragonal state to $C_2$ and is therefore referred to as Ising-nematic. It is necessarily accompanied by an orthorhombic lattice distortion\cite{Cano2010}. Simultaneous magnetic and structural phase transitions occur, for example, in underdoped Ba$_{1-x}$K$_x$Fe$_2$As$_2$ \cite{Avci2011}, suggesting that the small lattice distortion ($\lesssim 4\times10^{-3}$) is simply induced by magnetoelastic coupling. However, it was observed early on that the (nearly \cite{Kim2011}) concomitant magneto-structural phase transition of the parent compound BaFe$_2$As$_2$ splits into two well-defined transitions upon Co substitution \cite{Ni2008,Chu2009}. Neutron diffraction studies demonstrated that the structural transition temperature $T_s$ is several K higher than the magnetic transition temperature $T_N$ \cite{Lester2009,Kreyssig2010}. Hence, there is a small region of an orthorhombic---i.e., ``nematic''---but paramagnetic phase in Ba(Fe$_{1-x}$Co$_x$)$_2$As$_2$, which is, e.g., also observed in other transition-metal doped 122-systems \cite{Ni2010}, in F-doped 1111-compounds \cite{Luetkens2009} and in Co-doped NaFeAs \cite{Parker2010}. Intuitively, $T_s>T_N$ suggests that the structural instability is the primary one, even though a scenario in which the structural transition is nevertheless a consequence of magnetic interactions has been proposed soon after\cite{Nandi2010,Fernandes2010}. The observation certainly puts the simple picture of the orthorhombic distortion as a mere consequence of stripe-type antiferromagnetism into question and has sparked great interest in the nematic phase of the iron-based materials, including an intensive debate about its microscopic origin \cite{Fernandes2014}. Of particular interest in this debate is FeSe, which undergoes a similar structural distortion as the other compounds, but does not order magnetically\cite{McQueen2009}. 

Here we review and discuss our recent measurements of the elastic shear modulus, which is shown to be a particularly sensitive probe of the incipient structural distortion, using a novel three-point bending setup in a capacitance dilatometer\cite{Boehmer2014,Boehmerthesis}. We use the notation of the two-iron (or the tetragonal) unit cell, in which the structural distortion occurs in the $B_{2g}$ channel and is related to the elastic shear modulus $C_{66}$. The contribution is organized as follows. In the remaining part of this Introduction, we review previous experimental and theoretical work on nematicity, in particular as related to the softening of the shear modulus above the structural transition. In the following, we detail how $C_{66}$ can be linked to the nematic susceptibility using Landau theory (Section \ref{sec:Landau}). In Section \ref{sec:technique} we present briefly the three-point-bending technique and in Section \ref{sec:DMA} we show how the elastic data is influenced by domain formation in the orthorhombic state using dynamical three-point bending measurements on BaFe$_2$As$_2$. In Section \ref{sec:results} we present comprehensive shear-modulus data of Ba(Fe$_{1-x}$Co$_x$)$_2$As$_2$ and (Ba$_{1-x}$K$_x$)Fe$_2$As$_2$\cite{Boehmer2014} and of FeSe\cite{Boehmer2015}. Analysis of these data shows that, while the Ba(Fe$_{1-x}$Co$_x$)$_2$As$_2$ system can be described in a quantum critical scenario, (Ba$_{1-x}$K$_x$)Fe$_2$As$_2$ is characterized by a first-order transition between different ground states on increasing K content. The inferred nematic susceptibility of FeSe is shown to be remarkably similar to the one of underdoped BaFe$_2$As$_2$. In Section \ref{sec:comparison} the nematic susceptibility, as determined by elastoresistivity measurements and electronic Raman scattering, is compared to our elastic data. In Section \ref{sec:scaling} we test the scaling of the nematic susceptibility with magnetic fluctuations, derived from NMR data, within the spin-nematic scenario, in which the structural transition is a direct consequence of strong magnetic fluctuations. This scaling is found to be well satisfied in both Ba(Fe$_{1-x}$Co$_x$)$_2$As$_2$\cite{Fernandes2013} and (Ba$_{1-x}$K$_x$)Fe$_2$As$_2$, but fails for FeSe. Finally, we present a summary and outlook in Section \ref{sec:summary}.

\subsection{Electronic in-plane anisotropy}

The anisotropic properties of the magnetic/nematic phase have been studied by various different experimental techniques. An early observation of electronic in-plane anisotropy in the orthorhombic phase was a large $a-b$ anisotropy of the resistivity in Ba(Fe$_{1-x}$Co$_x$)$_2$As$_2$ (nearly a factor of 2) \cite{Chu2010}. However, it was subsequently shown experimentally that this resistivity anisotropy strongly depends on the degree of disorder of the samples and the type of substitution \cite{Ying2011,Nakajima2012,Ishida2013,Ishida2013II,Blomberg2013,Liu2015}. Various theoretical works also place emphasis on the role of disorder in explaining these observations\cite{Fernandes2011,Blomberg2013,Gastiasoro2014}. 
Strongly anisotropic defects in the orthorhombic state, which are a candidate to produce this resistivity anisotropy, were, indeed, observed using scanning tunneling microscopy \cite{Song2012,Allen2013,Rosenthal2014} and recent measurements of the anisotropy of the Hall effect also point to a dominating role of the carrier mobility anisotropy in creating the resitivity anisotropy\cite{Deng2015}. 
However, optical conductivity studies show the importance of orbital anisotropy in addition to the anisotropy of scattering rates \cite{Dusza2011,Nakajima2012, Mirri2014, Mirri2014II}. Recent work stresses, in particular, the importance of the anisotropy of the Drude weight, rather than scattering rate, for the resistivity anisotropy\cite{Mirri2015}, as was also indicated by earlier disorder-dependent elastoresistivity measurements\cite{Kuo2014}. Similarly, early angular resolved photo emission spectroscopy (ARPES) measurements found a significant shift of the $d_{xz}$ ($d_{yz}$) orbitals to lower (higher) energies below $T_s$ \cite{Yi2011}. Further, the thermopower was shown to have an even larger anisotropy than the resistivity in the orthorhombic state, arising from an interplay of anisotropic scattering and orbital polarization\cite{Jiang2013}. Nuclear magnetic resonance (NMR)\cite{Fu2012} and inelastic neutron scattering studies \cite{Harriger2011,Lu2014} show significant anisotropy of the spin dynamics even for $T_s>T>T_N$. Inelastic neutron scattering as a probe of nematicity is described in another contribution to this issue\cite{Inosov2015}.

Altogether, the observed large electronic anisotropy in the presence of a rather small lattice distortion of $\delta=(a-b)/(a+b)<0.4\%$ supports the idea that an electronic order parameter, which has to be ``nematic'' by symmetry, drives the structural transition. 
To show this, the resistivity anisotropy induced by an externally imposed lattice distortion (applied to a sample via a piezo stack) was measured in Ref. \cite{Chu2012}. It was found that, indeed, the susceptibility of an electronic nematic order parameter diverges on approaching $T_s$ from above, in agreement with the assumption that it drives the structural transition. Similarly, a study of the electronic relaxation dynamics using femtosecond-resolved polarimetry suggests that nematicity is an independent electronic degree of freedom\cite{Patz2014}.

In order to measure the in-plane anisotropy of various physical quantities, the crystals need to be detwinned, which can be accomplished by the application of uniaxial stress along the tetragonal $[110]$ direction, $\sigma_{[110]}$ \cite{Chu2010,Fisher2011} or, in some cases, by a high magnetic field \cite{Ruff2012,Zapf2014}. An earlier review on detwinning and electronic in-plane anisotropy in the iron-based superconductors is given in Ref. \cite{Fisher2011}. 
However, the application of $\sigma_{[110]}$ also significantly smears out the structural transition and can induce a marked anisotropy of electronic properties even above $T_s$ \cite{Chu2010,Blomberg2012,Dhital2012,Hu2012}. This strong sensitivity of the system to $\sigma_{[110]}$ already demonstrates a large nematic susceptibility and can be used to determine it quantitatively. The nematic susceptibility has been evaluated by various techniques, including the above described strain-dependent resistivity\cite{Chu2012}, the stress-dependent optical reflectivity\cite{Mirri2014}, the elastic shear modulus\cite{Boehmer2014} and the Raman response \cite{Gallais2014}. 

\subsection{Two theoretical scenarios -- spins vs orbitals}
As to which electronic degrees of freedom, spin or orbital, underlie this electronic nematic order parameter, two main scenarios have been discussed. The first one\cite{Kontani2011,Yamase2013,Yoshizawa2012II} places emphasis on orbital degrees of freedom, in particular the iron $d_{xz}$ and $d_{yz}$ orbitals, which are degenerate in the high-symmetry tetragonal phase. The structural transition, in this scenario, occurs when these orbitals order and become inequivalent in energy. The nematic order parameter $\varphi$ is then given by the difference in orbital occupation. Orbital order may trigger a, secondary, magnetic transition \cite{Kontani2011}. This model naturally explains why the structural transition occurs at a higher temperature than the magnetic transition, yet a magnetic transition does not necessarily follow the structural one within this orbital scenario. 

On the other hand, magnetism is essential in the second model, where spin fluctuations are considered as the driving force for the structural transition. In this spin-nematic scenario\cite{Fernandes2010,Nandi2010,Fernandes2012}, the primary instability is that of the stripe-type magnetic phase, with the ordering vector either$Q_1=(\pi,0)$ or $Q_2=(0,\pi)$, in orthorhombic notation. This state has an additional degree of freedom with respect to, e.g., checkerboard-type antiferromagnetism, namely the orientation of the ferromagnetic stripes (i.e., whether the ordering vector is $Q_1$ or $Q_2$), and this is determined by the ``spin-nematic'' order parameter $\varphi$ \cite{Fernandes2010}. The calculation of Ref. \cite{Fernandes2010} has shown that spin fluctuations can induce a finite $\varphi$, and thus a finite orthorhombic distortion, at a higher temperature than the magnetic ordering temperature, explaining the observed sequence of phase transitions. The debate about whether orbital or magnetic degrees of freedom are ``in the driver's seat'' of the structural transition has been reviewed in Ref. \cite{Fernandes2014}. 

\subsection{Extreme cases: orthorhombic distortion without magnetic order in FeSe and tetragonal magnetic phase in hole-doped BaFe$_2$As$_2$}
In most iron-based materials, the onsets of magnetic ordering and the structural distortion occur very close to each other in the phase diagram, which indicates that they are strongly coupled and hampers the determination of the ``driver''\cite{Fernandes2014}. There are, however, two exceptions to this which have recently attracted a lot of attention. 
The first exception is the 11-type iron-based superconductor FeSe, which undergoes a tetragonal-to-orthorhombic structural phase transition at $T_s\sim90$ K, similar to that found in the underdoped 122 materials and with similar magnitude of the orthorhombic distortion \cite{Margadonna2009,McQueen2009,Boehmer2015}. Yet, whereas this transition always occurs in proximity to stripe-type antiferromagnetic order in the 122 systems, no static magnetism was found in FeSe at ambient pressure \cite{McQueen2009,Bendele2010}. However, an enhanced spin-lattice relaxation rate in NMR at low temperatures indicates the presence of strong spin fluctuations \cite{Imai2009}. In contrast to the 122-type systems, phase-pure superconducting FeSe cannot be obtained by self-flux growth techniques. Recently however, high-quality single crystals were obtained using low-temperature vapor transport\cite{Boehmer2013} or KCl/AlCl$_3$ flux\cite{Lin2011,Chareev2013}, enabling a multitude of experimental studies including those of Young's modulus, NMR, elastoresistivity, ARPES and inelastic neutron scattering \cite{Boehmer2015,Baek2015,Watson2015,Shimojima2014,Maletz2014,Nakayama2014,Rahn2015,Wang2015}.

FeSe has a rich temperature-pressure phase diagram \cite{Margadonna2009,Garbarino2009,Bendele2012,Miyoshi2014}, which has recently been obtained in new comprehensiveness using resistivity measurements\cite{Terashima2015,Knoener2015}. The structural transition is found to be suppressed by hydrostatic pressure at $\lesssim 2$ GPa. On the other hand, the low-temperature spin fluctuations are enhanced by pressure \cite{Imai2009} and probably static magnetic order sets in above $\sim 1$ GPa with a subsequent increase of $T_N$ \cite{Bendele2010,Bendele2012,Terashima2015}. $T_c$ has a non-monotonic pressure dependence, increasing initially and reaching a local maximum at $\sim 0.8$ GPa  followed by a slight decrease \cite{Bendele2012,Miyoshi2014,Terashima2015}. At higher pressures, $T_c$ increases again and the onset of $T_c$ reaches surprising 37 K at $\sim7$ GPa \cite{Mizuguchi2008,Medvedev2009,Margadonna2009,Garbarino2009}. Recent inelastic neutron powder \cite{Rahn2015} and single crystal \cite{Wang2015} diffraction results suggest that magnetic fluctuations at ambient pressure occur around the same $(\pi,0)$ stripe-type wave vector as in the other iron-based compounds. These puzzling results have resently attracted considerable attention and have resulted in various theoretical scenarios\cite{Yu2015,Glasbrenner2015,Wang2015II,Mukherjee2015,Chubukov2015}.

The second exception where magnetic order and orthorhombic distortion do not closely follow each other is the $C_4$-symmetric reentrant magnetic phase in Na-doped BaFe$_2$As$_2$\cite{Avci2014}. Within a certain range of Na content, the usual orthorhombic distortion first develops with the stripe-type magnetic order below $T_N$, but then suddenly disappears within experimental resolution upon entering this phase. This observation was taken as evidence that magnetic degrees of freedom drive the structural phase transition (the spin-nematic scenario) because the existence of such a phase can hardly be reconciled with orbital order being a prerequisite for magnetism\cite{Avci2014}. Yet, the detailed magnetic structure is still under intense study. Polarized neutron scattering indicates a spin-reorientation from in-plane to $c$-axis oriented and suggests that the magnetic structure might still be orthorhombic, while a truly tetragonal ``$2-Q$'' magnetic structure could not be ruled out \cite{Wasser2015}. Group theoretical analysis suggests that the question of who is in the driver's seat may be solved by determining the pattern of magnetic and orbital order within this phase \cite{Khalyavin2014}. Further theoretical works \cite{Kang2014,Gastiasoro2015} show that interplay between the two stripe-type magnetic ordering vectors $Q_1=(0,\pi)$ and $Q_2=(\pi,0)$ can lead to the observed phase diagram. This putatively tetragonal magnetic phase was also reported in Na-doped SrFe$_2$As$_2$\cite{Taddei2015} and in single crystals of K-doped BaFe$_2$As$_2$\cite{Boehmer2015II}, where the interplay with superconductivity was also studied. Measurements of the orthorhombic order parameter using capacitance dilatometry \cite{Boehmer2015II} of these samples yield an upper limit for the orthorhombic distortion of $\sim0.01\times10^{-3}$ in the new magnetic phase of Ba$_{1-x}$K$_x$Fe$_2$As$_2$, i.e., less than $1\%$ of the value of $\delta$ in the stripe-type magnetic phase.

\subsection{The elastic shear modulus -- soft mode of the structural transition}
It is clear that once a material undergoes the structural transition, all properties (e.g., lattice constants, orbital occupation or spin fluctuations) aquire in-plane anisotropy, which makes it difficult to distinguish between the two scenarios. As an alternative, the study of the susceptibility of the various quantities above the structural transition has been suggested as a possible viable approach to determine the driving force of the structural transition\cite{Chu2012,Fernandes2013,Fernandes2014}. 

In proximity to the structural phase transition of the iron-based materials (which can be classified as pseudoproper ferroelastic \cite{Salje1990,Chu2012}) the elastic shear modulus $C_{66}$ is expected to grow soft. Note that the inverse of an elastic modulus can be considered a structural susceptibility.
The shear modulus $C_{66}$ was studied extensively in the Ba(Fe$_{1-x}$Co$_x$)$_2$As$_2$ system using ultrasound \cite{Fernandes2010,Goto2011,Yoshizawa2012,Yoshizawa2012II,Simayi2013}. These measurements confirm $C_{66}$ as the soft mode of the structural transition, as expected from the symmetry of the lattice distortion. An ultrasound study on single-crystalline FeSe \cite{Zvyagina2013} shows the same soft mode. Further, signatures of a behavior which was termed ``structural quantum criticality'' were found around optimal doping in the Ba(Fe$_{1-x}$Co$_x$)$_2$As$_2$ system. Namely, the temperature and doping dependence of the structural susceptibility $S_{66}=C_{66}^{-1}$ was found to resemble closely the magnetic susceptibility in proximity to a magnetic quantum critical point \cite{Yoshizawa2012}. The elastic softening persists over a large part of the superconducting dome on the overdoped side which makes the associated fluctuations a possible candidate for the pairing glue in the iron-based systems \cite{Boehmer2014,Yamase2013}. A pronounced hardening of $C_{66}$ below the superconducting transition temperature $T_c$ in optimally doped Ba(Fe$_{1-x}$Co$_x$)$_2$As$_2$ was shown to reflect the competition between magnetism and superconductivity because the structural distortion and magnetism are strongly coupled in this system\cite{Fernandes2010}.

\section{Landau theory with bilinear strain-order parameter coupling \label{sec:Landau}}

In the Landau theory of a second-order phase transition, a structural distortion can be induced by a bilinear coupling in the free energy between elastic strain $\varepsilon$ and the primary (possibly electronic) order parameter $\varphi$. In the case of the iron-based materials, the structural distortion is given by $\delta=(a-b)/(a+b)$, or $\varepsilon_6=\frac{\partial u_x}{\partial y}+\frac{\partial u_y}{\partial x}$, the relevant component of the elastic strain tensor, and $\varphi$ is the electronic nematic order parameter. Note that the bilinear coupling between $\varepsilon_6$ and $\varphi$ is, indeed, only allowed if $\varphi$ is ``nematic'', i.e., it breaks the four-fold rotational symmetry of the high-temperature phase. However, this phenomenological approach is independent of the microscopic origin of $\varphi$ (spin or orbital). Due to the bilinear coupling, $\varepsilon_6$ is a measure of $\varphi$, i.e., the two quantities are proportional to each other \cite{Fernandes2010}.  
The stress-strain relation $\sigma_6=C_{66}\varepsilon_6$ in the tetragonal or orthorhombic system shows that the soft elastic mode is $C_{66}$, i.e., a second-order structural phase transition occurs when $C_{66}\rightarrow0$. 

The relation between the nematic susceptibility and $C_{66}$ can be obtained by considering the free energy density \cite{Fernandes2010,Cano2010,Chu2012}
\begin{equation}
F=F_0+\frac{1}{2}C_{66,0}\varepsilon_6^2-\lambda\varepsilon_6\varphi+\frac{1}{2}\left(\chi_\varphi\right)^{-1}\varphi^2+\frac{B}{4} \varphi^4. \label{eq:Felnem}
\end{equation}
Here, $-\lambda\varepsilon_6\varphi$ is the bilinear coupling term (with the coupling constant $\lambda$), $\frac{1}{2}C_{66,0}\varepsilon_6^2$ is the bare elastic energy (\textit{i.e.} without the coupling $\lambda$)  and $C_{66,0}$ is the bare elastic constant, which, by assumption, has no strong temperature dependence. The last two terms represent a Landau expansion in the nematic order parameter $\varphi$, with $\chi_\varphi$ the bare nematic susceptibility and $B>0$ the usual quartic coefficient of the Landau expansion. A second bilinearly coupled (nematic) order parameter can be included analogously to describe, e.g., the interplay between elastic, spin-nematic and orbital degrees of freedom, as in Refs. \cite{Liang2013,Liang2014}. Here, however, we restrict ourselves to only one nematic order parameter. 
In general, the effective, renormalized elastic constant $C_{66}$ is given by \cite{Rehwald1973,Benoit1986}
\begin{equation}
C_{66}=\frac{d^2F}{d\varepsilon_6^2}
=\frac{\partial^2F}{\partial\varepsilon_6^2}-\left(\frac{\partial^2 F}{\partial\varepsilon_6\partial \varphi}\right)^2\left(\frac{\partial^2F}{\partial \varphi^2}\right)^{-1}
\end{equation}
because the condition that $F$ is minimal with respect to both $\varepsilon_6$ and $\varphi$ couples the two order parameters. 
This results in \cite{Cano2010,Chu2012}
\begin{equation} 
C_{66}=C_{66,0}-\frac{\lambda^2}{\left(\chi_\varphi\right)^{-1}+3B\varphi^2}\label{eq:Ceff}
\end{equation}
which reduces to $C_{66}=C_{66,0}-\lambda^2\chi_\varphi$ above the ordering temperature when $\varphi=0$. Note that $\chi_\varphi$ is also renormalized by the coupling $\lambda$ and the effective nematic susceptibility $\tilde{\chi}_\varphi$ is given by $\left(\tilde{\chi}_\varphi\right)^{-1}=\left(\chi_\varphi\right)^{-1}-\lambda^2/C_{66,0}$. Rewriting $C_{66}$ in terms of $\tilde{\chi}_\varphi$, yields the expression of Ref. \cite{Fernandes2010}
\begin{equation}
\frac{1}{C_{66}}=\frac{1}{C_{66,0}}+\frac{\lambda^2}{C_{66,0}^2}\tilde{\chi}_\varphi.\label{eq:CeffFernandes}
\end{equation}
Note that Eq. \ref{eq:CeffFernandes} is not limited to a Landau expansion in $\varphi$ and is actually valid more generally, as long as bilinear coupling to a harmonic lattice is considered\cite{Fernandes2010}. Equations \ref{eq:Ceff} and \ref{eq:CeffFernandes} show that we can access the nematic susceptibility $\chi_\varphi$ and also $\tilde{\chi}_\varphi$ by measuring $C_{66}$.

\begin{figure}
\includegraphics[width=\textwidth]{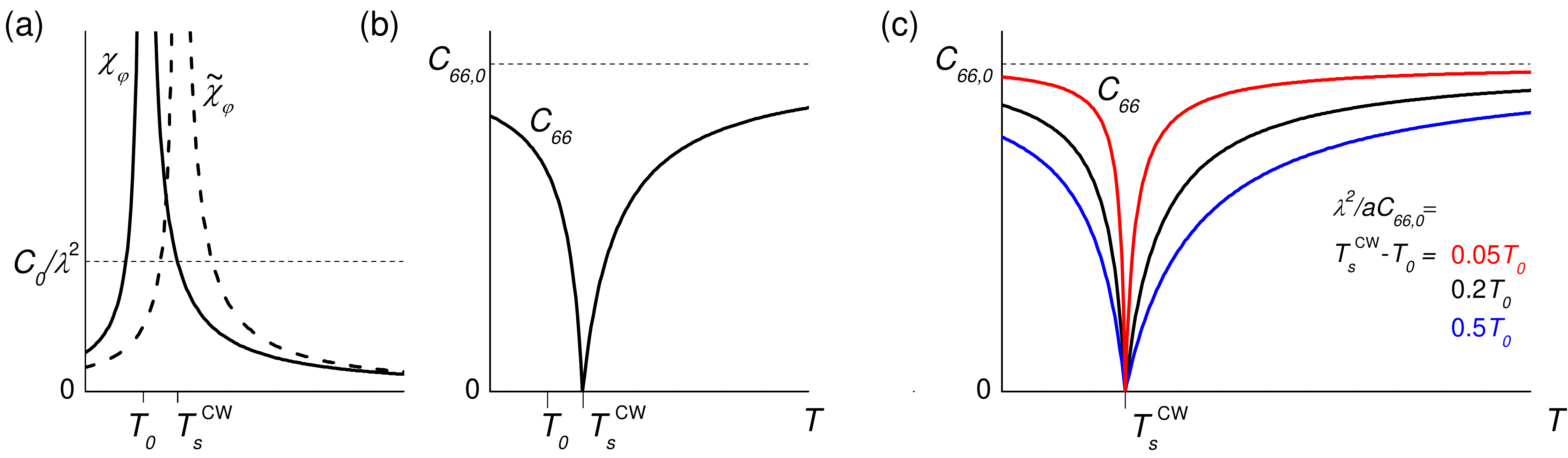}
\caption{(a) Temperature dependence of the nematic susceptibility $\chi_\varphi=1/a(T-T_0)$ (solid line) and the nematic susceptibility renormalized by bilinear coupling to the lattice $\tilde{\chi}_\varphi=1/a(T-T_s^{\mathrm CW})$ (dashed line) in a mean-field model. (b) Temperature dependence of the soft elastic mode $C_{66}=(T-T_s^{\mathrm{CW}})/(T-T_0)$ of the structural transition induced by this bilinear coupling between the strain component $\varepsilon_6$ and the nematic order parameter $\varphi$, eq. \ref{eq:Felnem}. The effect of the bilinear coupling is to increase the transition temperature from $T_0$ to $T_s^\mathrm{CW}$. (c) shows the temperature dependence of $C_{66}$ for a range of parameters $\lambda^2/aC_{66,0}=T_s^{\mathrm CW}-T_0$. The slope of $C_{66}$ just above $T_s$ is determined by the bilinear coupling strength $\lambda$.}
\label{fig:2}
\end{figure}

Before analysing real data, we illustrate the expected behavior of the shear modulus in the above Landau theory. We assume a mean-field Curie-Weiss-type divergence of $\chi_\varphi$ on approaching $T_0$, which would be the transition temperature in the absence of coupling to the elastic strain,
\begin{equation}
\chi_\varphi=\frac{1}{a(T-T_0)}.
\end{equation}
This leads to the temperature dependence of the soft elastic mode in a mean-field case \cite{Benoit1986}, 
\begin{align}
C_{66}=C_{66,0}\left(\frac{T-T_s^{\mathrm CW}}{T-T_0}\right)&\textnormal{ for } T>T_s^{\mathrm CW}\label{eq:CTdep}\\
C_{66}=C_{66,0}\left(\frac{2(T_s^{\mathrm CW}-T)}{3T_s^{\mathrm CW}-T_0-2T}\right)&\textnormal{ for } T<T_s^{\mathrm CW},
\end{align}
shown in Fig. \ref{fig:2}, with $T_s^{\mathrm CW}=T_0+\frac{\lambda^2}{a C_{66,0}}$ the new transition temperature, increased with respect to $T_0$ by the coupling to the strain. An energy scale characteristic of the coupling is given by $T_s^{\mathrm CW}-T_0$, and determines the curvature of $C_{66}(T)$ in Fig. \ref{fig:2} b, c. Note that the renormalized $\tilde{\chi}_\varphi=1/a(T-T_s^{\mathrm CW})$ naturally diverges at the new transition temperature. 

This above model, however, is limited to a description of a second-order nematic/structural transition and neglects the subsequent (or concomitant) magnetic transition occurring in most iron-based systems. To illustrate the expected temperature dependence of $C_{66}$ for BaFe$_2$As$_2$, the Landau model can be expanded to describe this split magneto-structural transition phenomenologically. This is achieved by the free energy density
\begin{equation}
F=F_0+\frac{a}{2}(T-T_0)\varphi^2+\frac{B}{4}\varphi^4
+\frac{C_{66,0}}{2}\varepsilon_6^2-\lambda\varepsilon_6\varphi-\varepsilon_6\sigma
+\frac{u}{2}(T-T_{N,0})M^2+\frac{v}{4}M^4-\mu\varphi M^2, \label{eq:Felnemmag}
\end{equation}
which is equivalent to equation \ref{eq:Felnem} concerning the elastic and nematic contributions and adds a magnetic order parameter $M$. Additionally, the contribution of conjugated uniaxial stress $-\varepsilon_6\sigma$ is also included to model the behavior under finite stress. For parameters values $a=1$, $T_0=1.025$, $B=1$, $C_{66,0}=1$, $\lambda=0.2$, $u=1$, $T_{N,0}=1$, $v=1$ and $\mu=0.2$ the model reproduces the second order structural and a first-order magnetic transition slightly below the structural one. Note that the coupling between $M$ and $\varphi$ causes the magnetic transition to be first order despite a positive fourth-order coefficient $v$. A solution which minimizes $F$ is calculated numerically for varying values of $\sigma$ and is shown together with the resulting elastic modulus, $C_{66}$ (still given by equation \ref{eq:Ceff}) in Fig. \ref{fig:3}b. Note that the temperature dependence of $C_{66}$ is unchanged with respect to the solution of eq. \ref{eq:Felnem} in the high-temperature region where $M=0$. 

In principle, the structural transition may arise either from a divergence of $\chi_\varphi$, as in an electronically-driven transition, or from a vanishing of $C_{66,0}$, as in a bare lattice instability. It has been proposed that the temperature dependence of $C_{66}$ close to the phase transition can distinguish the two cases \cite{Cano2010}, namely $C_{66}$ should vanish linearly in the case of a bare elastic instability or with a curvature as in Fig. \ref{fig:2} in the case of an electronically driven transition, which is assumed here.

We note that the above Landau formalism is a mean-field treatment, in which fluctuations of the order-parameter are ignored. In fact, this often turns out to be a good approximation for second-order structural phase transitions, because the long-range elastic interactions strongly suppress fluctuations \cite{Cowley1976,Folk1976,AlsNielsen1977}.

\section{Technique: Three-point bending in a capacitance dilatometer\label{sec:technique}}

Three-point bending is a long-standing and widely used mechanical test to study elastic properties of diverse materials. The technique is particularly appealing by its simplicity. A platelet- or beam-shaped sample is supported along two lines, while a force is applied at a third, middle, line and the induced deflection is measured. Small forces result in a sizable deflection so that the elastic modulus can be measured comparatively easily. In  the limit of thin samples, the measured elastic modulus is proportional the material's Young modulus, which also determines the sample stiffness in a uniaxial tension/compression experiment \cite{Kityk1996}, even though samples are subject to non-uniform stress in three-point bending\cite{Rossiter1991}. We have developed a three-point bending setup in a capacitance dilatometer\cite{Meingast1990} to measure the Young modulus of iron-based materials with very high resolution \cite{Boehmer2014}. The advantage of this technique over the ultrasound measurements used more traditionally to determine elastic constants, is that the sample requirements are not as stringent. In particular, ultrasound velocity measurements require samples of considerable size and quality. On the other hand, the typically quite thin platelet-like single crystals available for the iron-based based materials are perfectly suited for the three-point bending technique. This has allowed us to investigate a larger variety of materials than would have been possible using ultrasound.

\begin{figure}
\includegraphics[width=\textwidth]{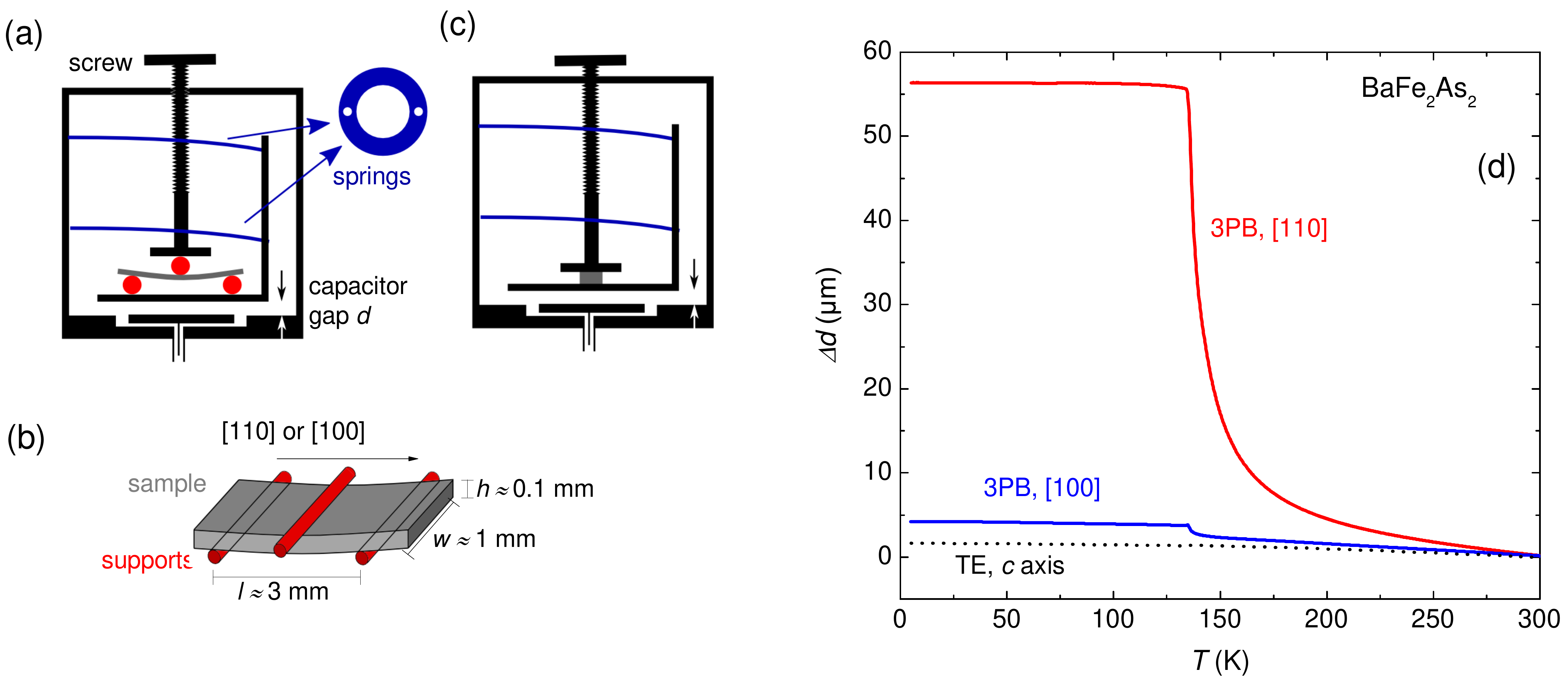}
\caption{(a) Schematic representation of the capacitance dilatometer with a sample inserted for three-point bending. The sample (see panel (b), of dimensions $l\times w\times h\approx 3\times1\times0.1$ mm$^3$) is supported by three wires and pressed against one movable plate of a plate-type capacitor using a screw with a force of $\approx 0.2$ N. The movable capacitor plate is suspended via a set of circular parallel springs and the change of the capacitor gap $d$ (indicated by arrows) on changing temperature is measured. (c) shows the regular setup with a sample inserted for thermal expansion measurement. (d) Temperature evolution of the  capacitor gap $d(T)$, expected in a $c$-axis thermal-expansion (TE) experiment of a $0.1$ mm thick BaFe$_2$As$_2$ sample\cite{Meingast2012} (black dashed line), and measured in three-point bending (3PB) of such a sample oriented along [100] (blue line) and [110] (red line). An exceedingly large effect is observed in the latter case, a result of the strong temperature dependence of the shear modulus $C_{66}$ of BaFe$_2$As$_2$.}
\label{fig:1}
\end{figure}

In a typical capacitance dilatometer, the sample is held in place by a small uniaxial force from the two leaf springs of the parallelogram arrangement holding the moveable capacitor plate (see Fig. \ref{fig:1}c). Hence, sample-length changes (due to, e.g., thermal expansion) lead to a change of the capacitor gap $d$. A very high length resolution of $0.1-0.01$ \AA\ can be achieved by measuring the resulting change of capacitance \cite{Meingast1990}. By placing a sample in three-point bending configuration (Fig. \ref{fig:1}b) into the dilatometer, as shown in Fig. \ref{fig:1}a, one no longer measures the thermal expansion of the sample, but rather its elastic bending modulus. This is because the small force from the dilatometer causes the sample to bend and the effective height of the arrrangement is determined by the sample's flection, hence, its elastic modulus. Notably, when the elastic modulus of such a sample becomes soft (i.e., decreases), the sample bends more strongly so that its effective height along the axis of the dilatometer decreases. Fig \ref{fig:1} (d) shows examples of such measurements. The red and the blue curves show three-point bending experiments with samples oriented with their tetragonal [110] and [100] directions, respectively, perpendicular to the supports. An exceedingly large effect is observed in three-point bending along [110]. This is because the Young's modulus along [110],
\begin{equation}
Y_{[110]}=4\left(\frac{1}{C_{66}}+\frac{1}{\gamma}\right)^{-1}\textnormal{ with } \gamma=\frac{C_{11}}{2}+\frac{C_{12}}{2}-\frac{C_{13}^2}{C_{33}}\label{eq:Y110}\\
\end{equation}
 is dominated by the elastic shear modulus $C_{66}$ as long as $C_{66}$ is smaller than the other $C_{ij}$, and $C_{66}$ decreases strongly on cooling towards $T_s$\cite{Yoshizawa2012}.  In contrast, the black dashed line shows the much smaller temperature dependent change of the capacitance gap $\Delta d$ expected from just the thermal expansion of a 100 $\mu$m thick sample of BaFe$_2$As$_2$\cite{Meingast2012}. A detailed description of the quantitative analysis of the data is given in \cite{Boehmerthesis}. The technique is particularly well suited to access large changes of the Young modulus of a sample occurring over a broad temperature range, as well as any sharp anomalies. As we will show, this technique even has the resolution to detect the often very small anomalies at the onset of superconductivity. 

In addition to the static three-point bending measurements in the capacitance dilatometer, ultralow-frequency dynamic three-point bending measurements of BaFe$_2$As$_2$ were performed using a Diamond DMA (dynamical mechanical analyzer) from PerkinElmer \cite{Salje2010}. Here, samples are subjected to a force $F=F_S+F_D\exp(i\omega t)$ with a dynamic component at the very low frequency of 1 Hz. From the measured displacement $u=u_S+u_D\exp(i(\omega t-\delta))$, the dynamic stiffness $k_s=F_D/u_D\exp(i\delta)\propto Y_{[110]}$ is calculated. $k_s$ may be a complex number in the presence of dissipation, and its real part corresponds to the results of the static measurements.

\section{Dynamical three-point bending measurements of B\MakeLowercase{a}F\MakeLowercase{e}$_2$A\MakeLowercase{s}$_2$\label{sec:DMA}}

\begin{figure}
\includegraphics[width=\textwidth]{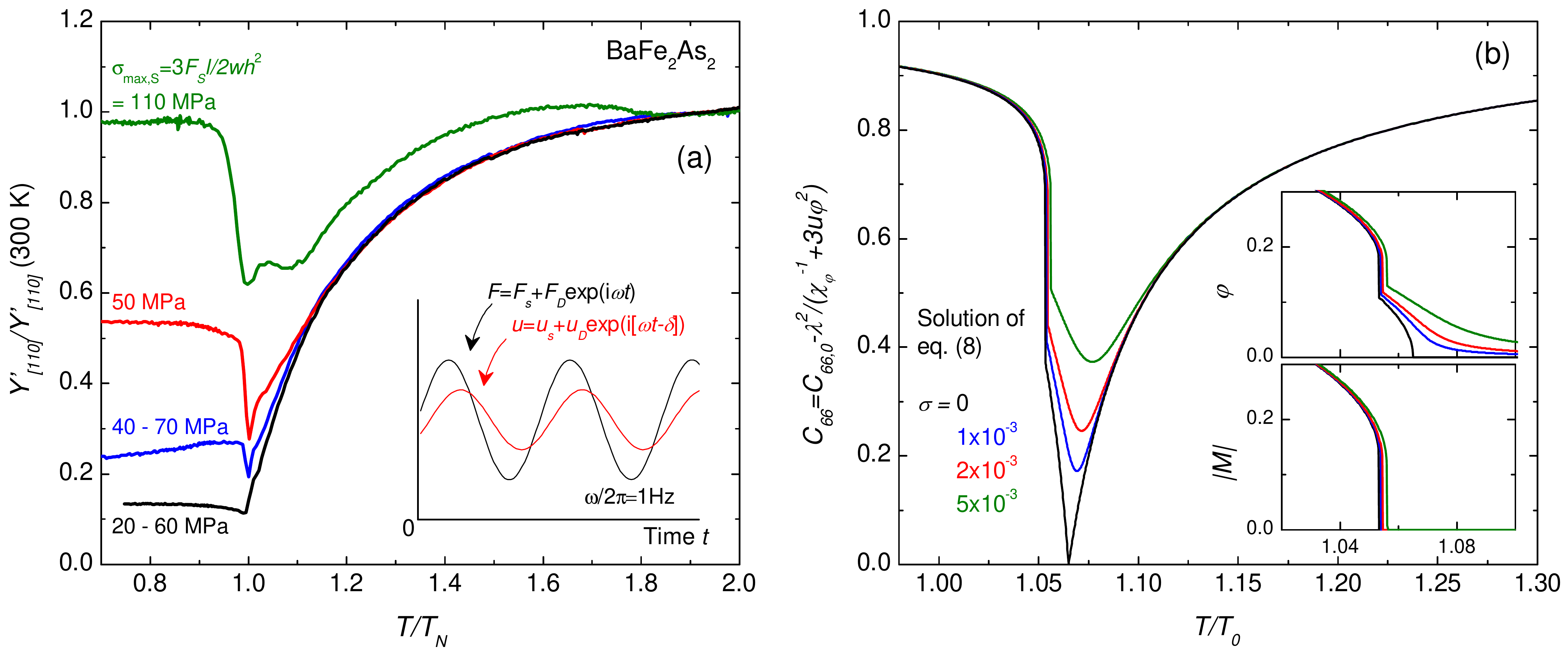}
\caption{(a) Real part of the Young modulus $Y_{[110]}'$ of BaFe$_2$As$_2$ obtained in dynamic three-point bending experiments in which the applied force $F=F_S+F_D\exp(i\omega t)$ varies at a frequency of $\omega/2\pi=1$ Hz (see inset). $Y_{[110]}\propto F_D/u_D\exp(i\delta)$ is calculated from the measured displacement $u=u_S+u_D\exp(i(\omega t-\delta))$ under the applied force (see inset). Data are taken at different $F_S$ and constant ratio $F_D/F_S$. The maximal uniaxial stress at the sample surface arising from $F_S$, $\sigma_{max,S}=\frac{3F_Sl}{2wh^2}$ \cite{Rossiter1991} is given as a characteristic parameter. The two measurements with the lowest force were conducted in ``constant amplitude'' mode, in which $F_S$ and $F_D$ are smaller when the sample is soft and increase when the sample becomes harder, thus increasing the resolution. (b) Soft elastic mode in a mean-field picture for closely spaced structural and magnetic phase transitions, obtained via numerically minimizing eq. \ref{eq:Felnemmag} (see text for parameters) and then calculating $C_{66}$. Good agreement with the experimental data is obtained for the high-stress measurement, while the measurements at lower stress and below $T_s$ are dominated by the effect of structural twins. Insets show the order parameters $\varphi$ and $|\mathbf{M}|$ of the numerical solution.}
\label{fig:3}
\end{figure}

Before addressing the detailed static measurements of the Young's modulus in a capacitance dilatometer, we show in this section results of dynamic three-point bending experiments on BaFe$_2$As$_2$, which elucidate the role of structural twins in the ordered phase\cite{noteSchranz}. For these measurements the sample is subjected to the force $F=F_S+F_D\exp(i\omega t)$ with a dynamic component $F_D\sim (0.7-0.8)F_s$ at a frequency of $\omega/2\pi=1$ Hz. Fig. \ref{fig:3}a shows the real part of the Young modulus $Y_{[110]}'$ (equivalent to the result of static measurements) for different $F_S$ but constant ratio $F_D/F_S$.

A smooth softening of $Y_{[110]}$ upon cooling towards the structural transition is observed (Fig. \ref{fig:3}a), in agreement with the soft-mode behavior shown in Fig. \ref{fig:2}. However, at the lowest stress, a kink and a small, sharp, minimum may be ascribed to the structural and magnetic transition, respectively, and $Y_{[110]}$ is essentially temperature independent below $T_N$, in strong contrast to the behavior shown in Fig. \ref{fig:2}. With increasing stress, the data deviate from the low-stress curve below some $T>T_s$ and $Y_{[110]}$ does not soften as much in total. The structural transition is smeared out, while the magnetic transition is affected much less. Notably, a discontinuous hardening of the Young modulus on cooling through $T_N$ emerges with increasing stress. It is reminiscent of the phonon modes measured using inelastic neutron scattering \cite{Parshall2014}. At the highest stress, $Y_{[110]}$ indeed recovers its initial high-temperature value at $\sim0.9 T_N$. Note that the deviation of this curve from the others at higher temperatures is an experimental artefact.

The experimental data at high stress are well reproduced by the model in eq. \ref{eq:Felnemmag} of a split magneto-structural transition (Fig. \ref{fig:3}b). However, the experimental data obtained at lower stress are much too small in comparison to this model for temperatures $T<T_s$. These anomalously low values of the Young's modulus below $T_s$ likely arise from the 'superelastic' behaviour of the twinned samples, which is due to domain wall motion and typically found in ferroelastic materials\cite{Kityk1996,Schranz2012,Schranz2012II}. The effect can be understood by considering two kinds of structural domains, $'+'$-type domains which are elongated along [110] and $'-'$-type type domains which are shortened along this direction. Application of a small compressive stress $\sigma_{[110]}$ increases the fraction of $'-'$-type domains at the expense of the $'+'$-type domains, so that the total sample length decreases through domain wall motion. A relatively small applied force can thus induce significant length changes, which means that the elastic modulus appears very small at low frequencies. In the three-point bending experiment, samples are essentially subjected to stress along the [110] direction, which suppresses domain formation and leads to a strong variation of the measured $Y_{[110]}$ with applied stress. The true monodomain elastic properties of the system can only be obtained in these dynamic three-point bending experiments if the applied stress is strong enough to force the system into a single domain state.

\section{Young's modulus of B\MakeLowercase{a}(F\MakeLowercase{e},C\MakeLowercase{o})$_2$A\MakeLowercase{s}$_2$, (B\MakeLowercase{a},K)F\MakeLowercase{e}$_2$A\MakeLowercase{s}$_2$ and F\MakeLowercase{e}S\MakeLowercase{e}\label{sec:results}}

\begin{figure}
\includegraphics[width=\textwidth]{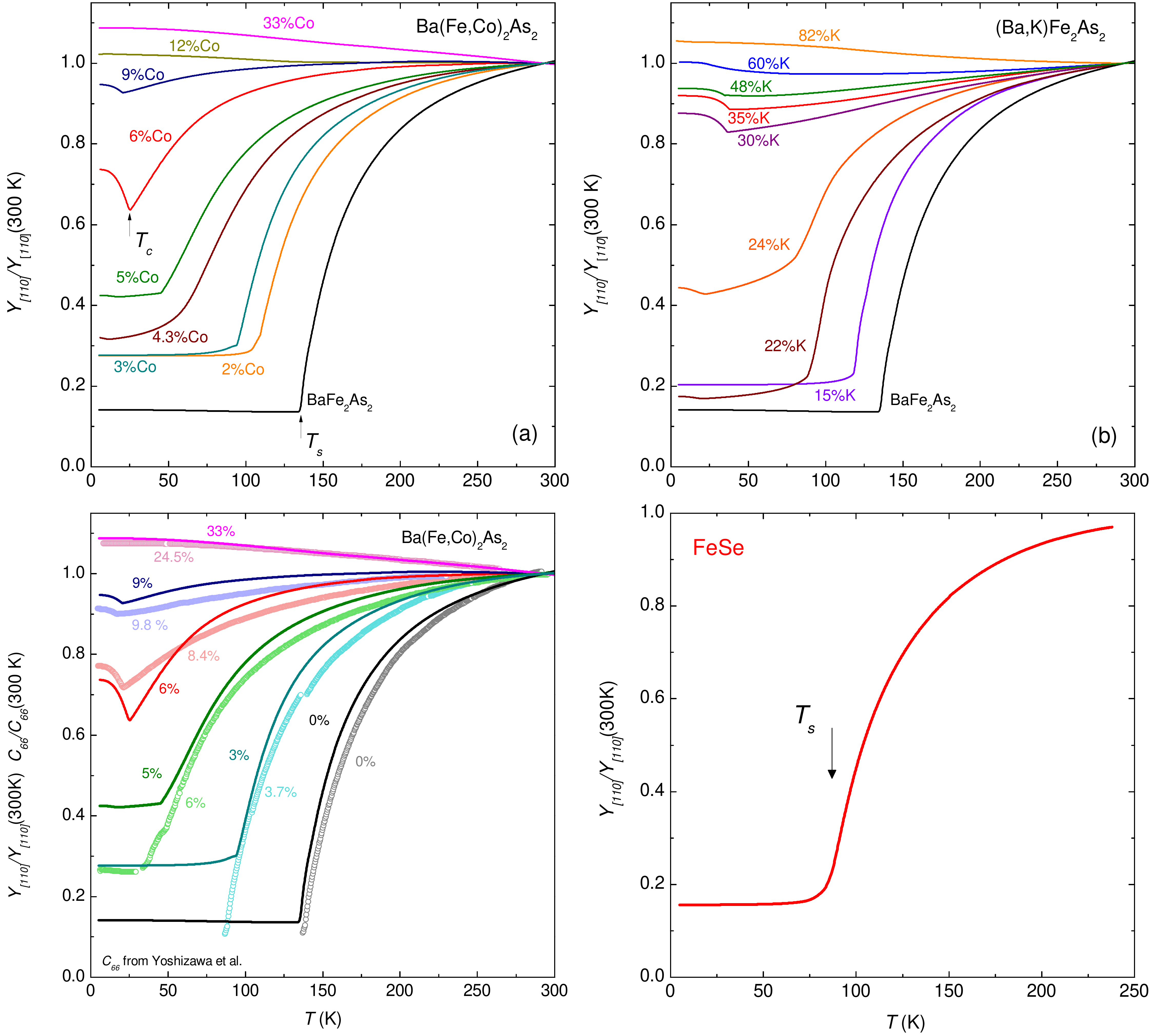}
\caption{Young's modulus along the tetragonal [110] direction $Y_{[110]}$ of (a) Ba(Fe$_{1-x}$Co$_{x}$)$_2$As$_2$ and (b) Ba$_{1-x}$K$_{x}$Fe$_2$As$_2$ for a wide doping range, measured using the static three-point bending in a capacitance dilatometer described in section \ref{sec:technique}. The data for $T<T_s$ are dominated by the effect of structural twins, similar to the low-stress measurements of Fig. \ref{fig:3}. (c) Young's modulus $Y_{[110]}$ (lines) and shear modulus $C_{66}$ (symbols, from Ref. \cite{Yoshizawa2012}) for Co-doped BaFe$_2$As$_2$. Both quantities are normalized at room temperature. The indicated Co content is taken from Ref. \cite{Yoshizawa2012} and samples with similar transition temperature are compared. All essential features, namely the strong softening of $Y_{[110]}$ on cooling towards $T_s$, and the marked hardening below $T_c$ for moderately overdoped samples, are reproduced in the Young's modulus data. Those curves show, however, a slightly stronger curvature than $C_{66}$, presumably because of the contributions from the other elastic constants (see eq. \ref{eq:Y110}). (d) Young's modulus $Y_{[110]}$ of single crystalline FeSe, which shows very similar softening on cooling towards $T_s$.}
\label{fig:4}
\end{figure}

Figure \ref{fig:4} shows the Young's modulus $Y_{[110]}$ of Co- and K-doped BaFe$_2$As$_2$ and of FeSe, as measured by the three-point bending technique in a capacitance dilatometer described in section \ref{sec:technique} \cite{Boehmer2014,Boehmer2015}. Fig. \ref{fig:4}c shows that the temperature dependence of $Y_{[110]}$ obtained by three-point bending is very similar to $C_{66}$ (see eq. \ref{eq:Y110}) as determined by ultrasound measurements for Ba(Fe$_{1-x}$Co$_x$)$_2$As$_2$ \cite{Yoshizawa2012}, the system for which such data is available. Notably, all essential features of $C_{66}$ are also observed in the Young's modulus data, confirming the capacity of our experiment to determine the temperature dependence of the shear modulus. In BaFe$_2$As$_2$, strong softening of $Y_{[110]}$ upon approaching the transition at $T_s$ from above is observed. The softening gradually disappears with both Co- and K-doping. Interestingly, the Young modulus $Y_{[110]}$ of FeSe, which undergoes a similar structural transition as the iron-arsenides but no magnetic phase transition, shows the same pronounced softening on cooling towards $T_s\approx 90$ K as underdoped BaFe$_2$As$_2$ (Fig. \ref{fig:4}(d)). For moderately overdoped samples, a marked softening on cooling is still observed, while $Y_{[110]}$ hardens considerably below $T_c$. The Young's modulus of a non-superconducting, strongly overdoped Ba(Fe$_{0.67}$Co$_{0.33}$)$_2$As$_2$ sample is found to harden by a few $\%$ on cooling from room temperature, which is the typical behavior induced by phonon hardening\cite{Varshni1970}. The flat temperature dependence in the orthorhombic state below $T_s$ is ascribed to the multidomain effect outlined in the previous section and, in the following, we will therefore focus on data at $T>T_s$. We note that we find no evidence of a higher-temperature ($T>T_s$) nematic transition as proposed by Kasahara et al., \cite{Kasahara2012} in any of our high-resolution shear modulus data. This is in agreement with previous high-resolution thermal-expansion measurements on both Co- and P-substituted BaFe$_2$As$_2$\cite{Meingast2012,Boehmer2012} and heat-capacity measurements in BaFe$_2$(As$_{1-x}$P$_x$)$_2$\cite{Luo2015}, which also do not find evidence of an additional phase transition.

\subsection{Doping and temperature dependence of the nematic susceptibility}

\begin{figure}
\includegraphics[width=\textwidth]{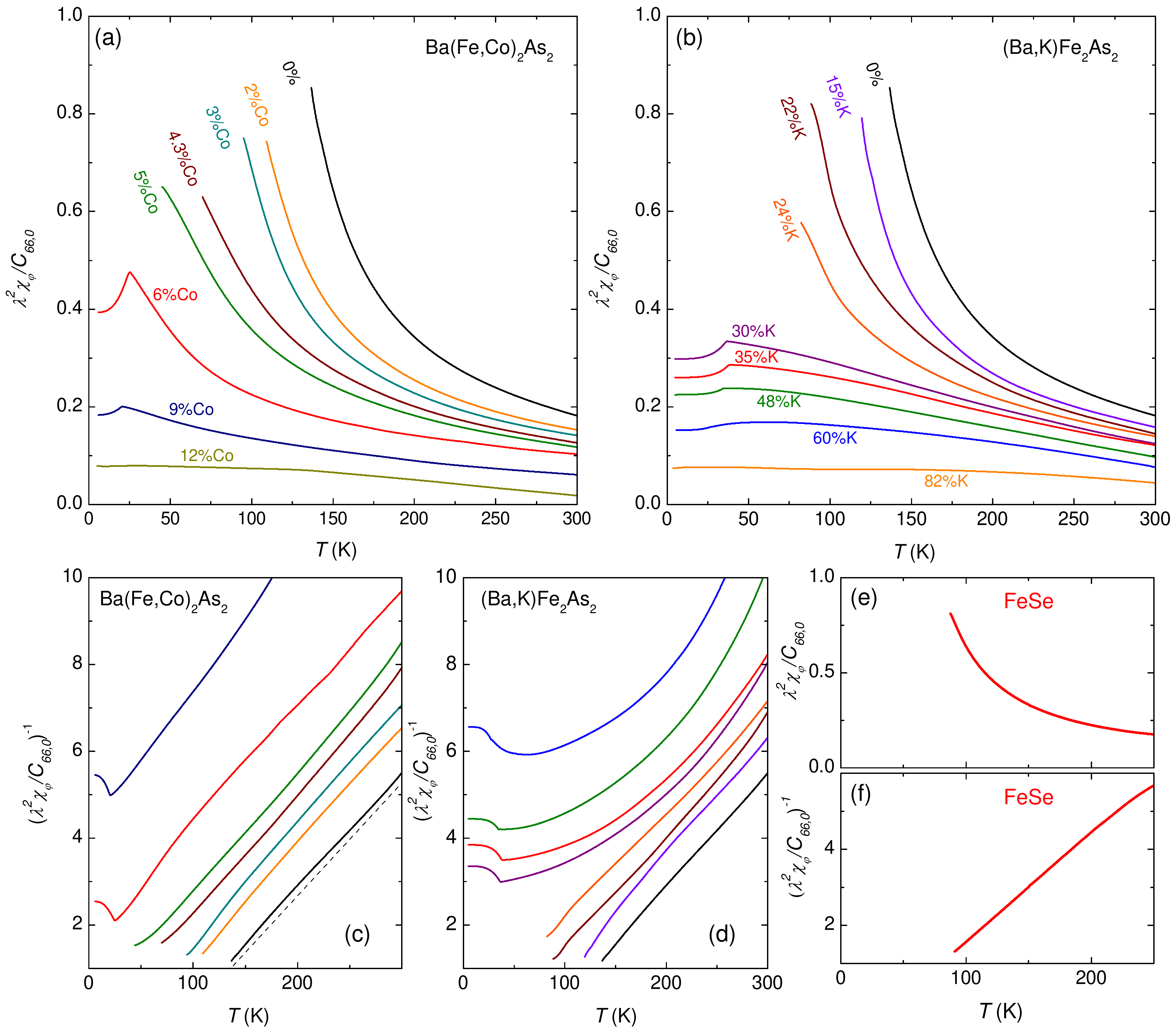}
\caption{Nematic susceptibility $\chi_\varphi$ in units of $C_{66,0}/\lambda^2$ of (a) Ba(Fe$_{1-x}$Co$_{x}$)$_2$As$_2$ and (b) Ba$_{1-x}$K$_{x}$Fe$_2$As$_2$ obtained from the data in Fig. \ref{fig:4}. For a second order structural phase transition $C_{66,0}/\lambda^2$ should reach 1 at $T_s$ and it is unclear why the experimental data do not reach this value. (c), (d) show the inverse $\left(\lambda^2\chi_\varphi/C_{66,0}\right)^{-1}$ revealing a Curie-Weiss-like temperature dependence of the nematic susceptibility for all samples except Ba$_{1-x}$K$_{x}$Fe$_2$As$_2$ with $\geq 30$\% K content. The dashed straight line in (c) is a guide to the eye. (e) and (f) show similar data for FeSe, which resemble closely underdoped BaFe$_2$As$_2$.}
\label{fig:6}
\end{figure}

\begin{SCfigure}
\includegraphics[width=8.6cm]{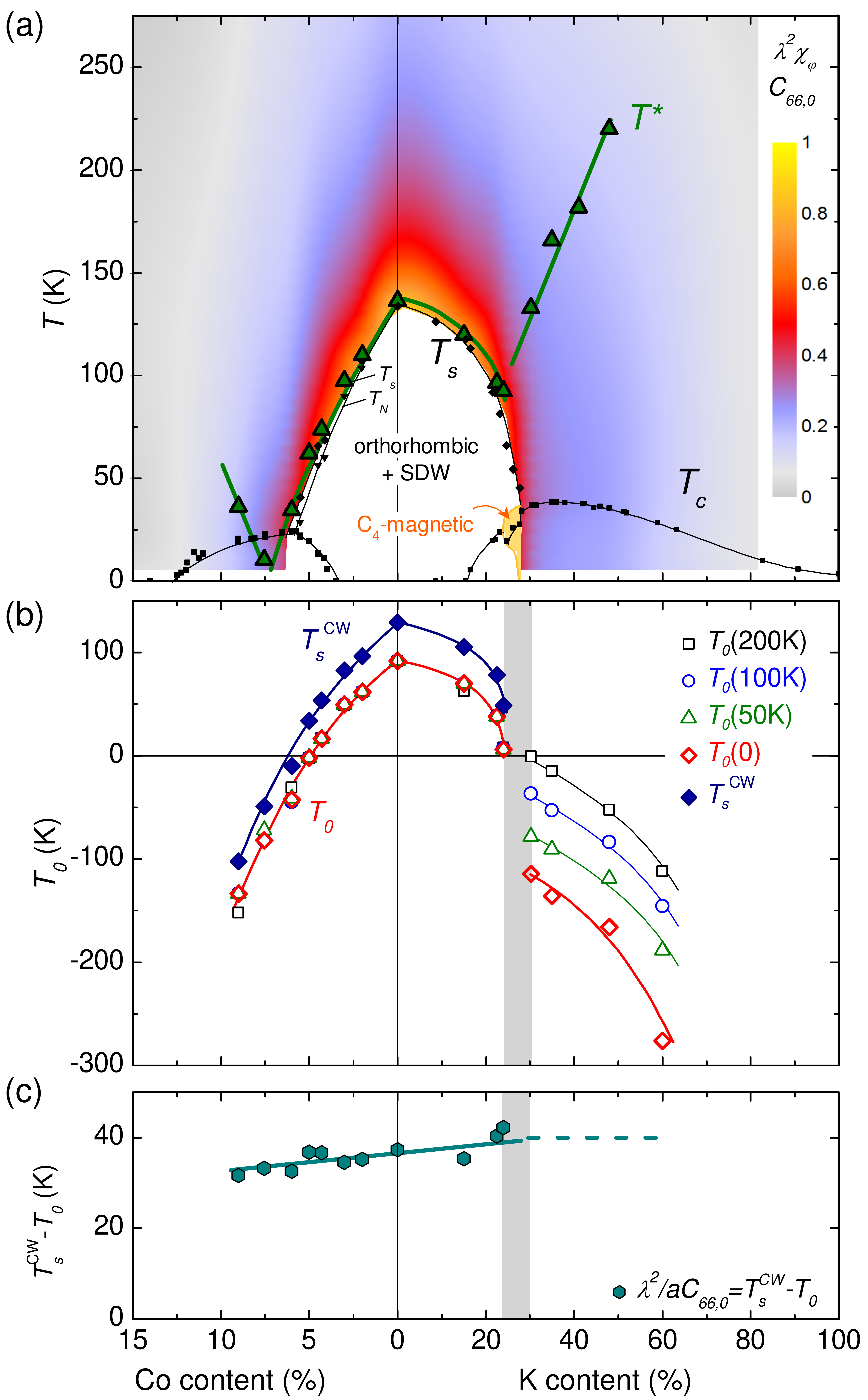}
\caption{(a) Nematic susceptibility as a color-coded map in the composition-temperature phase diagram of Ba(Fe$_{1-x}$Co$_{x}$)$_2$As$_2$ and Ba$_{1-x}$K$_{x}$Fe$_2$As$_2$. Green triangles show the inflection point of $\chi_\varphi(T)$ as a lower limit for the validity of Curie-Weiss law $\lambda^2\chi_\varphi/C_{66,0}=\frac{\lambda^2/aC_{66,0}}{T-T_0}$ (see text). (b) Weiss temperature $T_0$ and temperature at which $C_{66}$ extrapolates to zero, $T_s^{\mathrm CW}$. When the data deviate from the simple Curie-Weiss law, as for samples with $\geq30$ \% K content, this deviation is ascribed to a temperature dependent $T_0$ and $T_0$ (extrapolated) at various temperatures is reported. The dashed area indicates the doping range, in which an abrupt change suggests a first order transition between different ground states. (c) Curie constant $\lambda^2/aC_{66,0}=T_s^{\mathrm CW}-T_0$. The dashed line shows the extrapolation for the region where the data do not follow the Curie-Weiss law, used to obtain $T_0(T)$ in (b). }
\label{fig:7}
\end{SCfigure}

Eq. \ref{eq:Ceff} directly links $C_{66}$ to the nematic susceptibility $\chi_\varphi$ under the assumption that the structural transition is induced by bilinear coupling to a nematic order parameter, eq. \ref{eq:Felnem}. In real systems the bare elastic constant $C_{66,0}$ is slightly temperature dependent because of phonon anharmonicity and typically hardens by a few \% upon cooling between room temperature and zero temperature \cite{Varshni1970}. In order to remove this background contribution and to connect with the Young modulus, we make the approximation
\begin{equation}
\frac{Y_{[110]}}{Y_0}\approx\frac{C_{66}}{C_{66,0}}. \label{eq:Yel}
\end{equation}
Here, $Y_0$ is the non-critical contribution to $Y_{[110]}$ for which we use the data from a 33\% Co substituted sample and further assume that it is independent of doping\cite{Meingast2012}. Additional details can be found in \cite{Boehmer2014,Boehmerthesis}. 
Assuming that the structural transition of FeSe is also induced by bilinear coupling between a nematic order parameter and the lattice, $\varphi$ is inferred from the Young's modulus data of FeSe in the same way, assuming similar $Y_0$\cite{Boehmer2015}. 
The obtained bare nematic susceptibility $\chi_\varphi$ is plotted in Fig. \ref{fig:6} (a), (b), (e) in units of $\lambda^2/C_{66,0}$. Note that the nematic susceptibility renormalized by the coupling to the lattice can be inferred easily using $\left(\lambda^2\tilde{\chi}_\varphi/C_{66,0}\right)^{-1}=\left(\lambda^2\chi_\varphi/C_{66,0}\right)^{-1}-1$. $\chi_\varphi$ increases strongly with decreasing temperature for all but the most strongly overdoped BaFe$_2$As$_2$ samples. Interestingly, FeSe is found to resemble closely underdoped BaFe$_2$As$_2$. 

The inverse $\left(\chi_\varphi\right)^{-1}$, plotted in Fig. \ref{fig:6} (c), (d), (f), evidences a mean-field-type temperature dependence $\chi_\varphi=1/(a(T-T_0))$ for Ba(Fe$_{1-x}$Co$_2$)$_2$As$_2$ up to $x=0.09$, in agreement with Ref. \cite{Yoshizawa2012}, for Ba$_{1-x}$K$_{x}$Fe$_2$As$_2$ up to $x=0.24$ and for FeSe. However, and somewhat surprisingly, $Y_{[110]}$ does not follow a Curie-Weiss law in the higher doped Ba$_{1-x}$K$_{x}$Fe$_2$As$_2$ samples in which the structural transition is suppressed. Instead, the temperature dependence is found to be less ``critical'' and $Y_{[110]}$ displays a clear inflection point.  
The detailed doping and temperature dependence of $\chi_\varphi$ is presented in Fig. \ref{fig:7}. The color map of $\chi_\varphi$ in Fig. \ref{fig:7}a shows that the nematic susceptibility is enhanced fairly symmetrically over most of the superconducting domes of both Ba(Fe$_{1-x}$Co$_{x}$)$_2$As$_2$ and Ba$_{1-x}$K$_{x}$Fe$_2$As$_2$, suggesting its possible role in promoting superconductivity\cite{Boehmer2014}. To describe the temperature dependence of $\chi_\varphi$ for different doping levels, we consider two parameters. First, we define $T^*$ as either the inflection point of $\chi_\varphi(T)$ or its maximum value, whichever is greater. $T^*$ is thus a lower limit for the validity of a Curie-Weiss law. Second, deviations from the Curie-Weiss law are ascribed to a temperature dependence of the parameter $T_0$, whose value at fixed temperatures is plotted in Fig. \ref{fig:7}b. $T^*$ reaches near zero temperature around optimal doping, consistent with a quantum critical scenario \cite{Yoshizawa2012} only in the Ba(Fe$_{1-x}$Co$_{x}$)$_2$As$_2$ system. In contrast, $T^*$ does not go below $\sim75$ K in Ba$_{1-x}$K$_{x}$Fe$_2$As$_2$.
The findings suggest a first-order transition between different ground states preempting a quantum critical point in Ba$_{1-x}$K$_{x}$Fe$_2$As$_2$, in particular the step-like change of $T_0$ between 24\% and 30\% K content (Fig. \ref{fig:7}b). Interestingly, a new, $C_4$-symmetric, magnetic phase was subsequently found to emerge in between these two doping levels \cite{Boehmer2015II}. 

In Ref. \cite{Kontani2014}, $C_{66}(T)$ was calculated including vertex corrections and was found not to follow the mean-field-type Curie-Weiss law. The data in Figs.\ref{fig:4} and \ref{fig:6} can be well described in this approach, for which the change of behaviour of Ba$_{1-x}$K$_x$Fe$_2$As$_2$ between 24\% and 30\% K content would be a consequence of a simple change of parameters as might be due to, e.g., a Lifshitz transition\cite{Kontani2014}. Finally, it is notable that the Curie constant $\lambda^2/aC_{66,0}=T_s^{\mathrm CW}-T_0\approx 30-40$ K, which is the characteristic energy of the electron-lattice coupling, is found to be nearly doping independent up to 9\% Co content and 24\% K content, the compositions for which it can be evaluated. Note that a value of $\sim50-60$ K is found for same parameter by evaluating $C_{66}$ ultrasound data in the Ba(Fe$_{1-x}$Co$_{x}$)$_2$As$_2$ system \cite{Yoshizawa2012}. This value is comparable but slightly larger than our result and reflects the lower curvature of the $C_{66}$ with respect to the $Y_{[110]}$ data in Fig. \ref{fig:4}c.

\subsection{Young's modulus around $T_c$}

\begin{figure}
\begin{center}
\includegraphics[width=\textwidth]{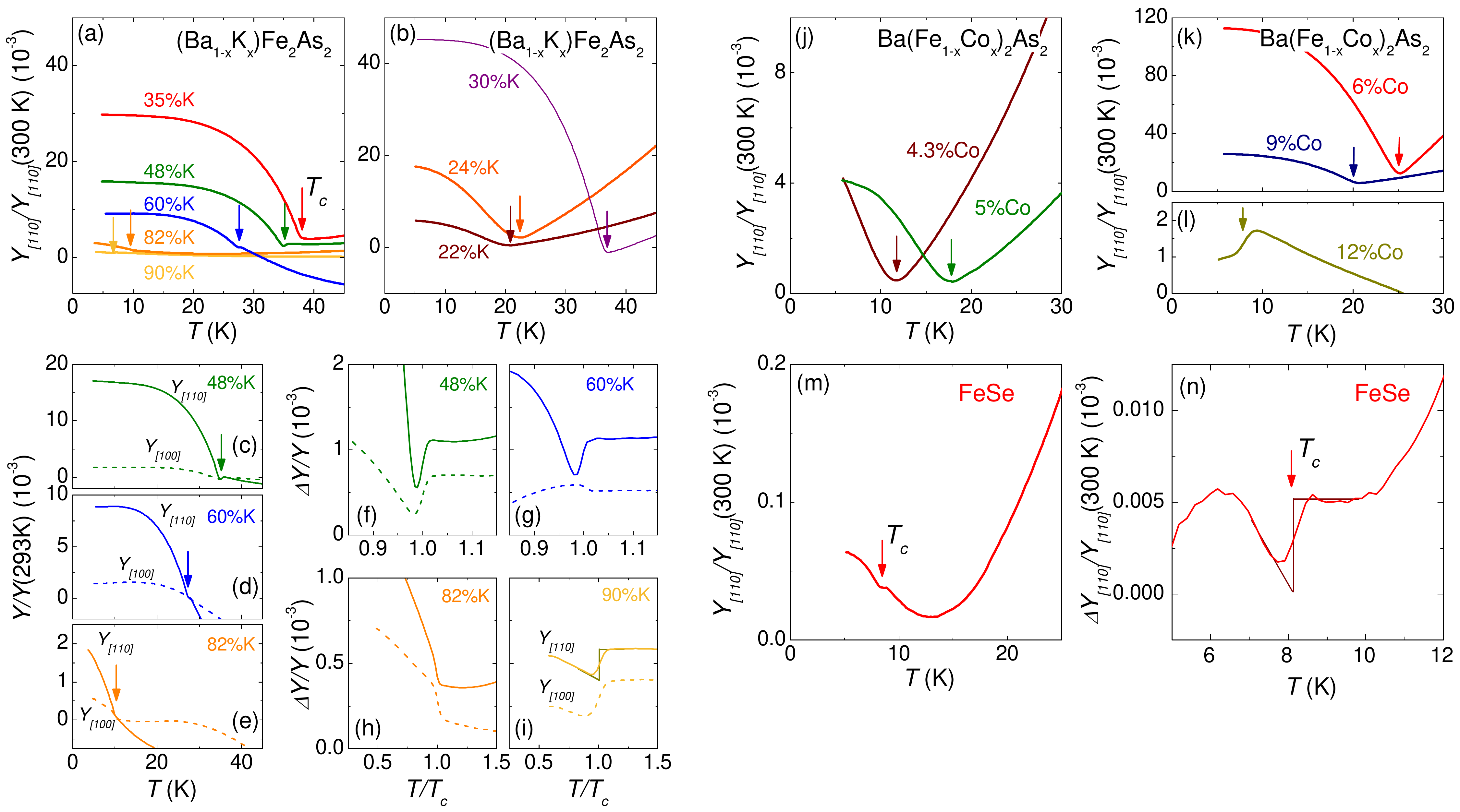}
\caption{Enlarged view of the low-temperature Young modulus $Y_{[110]}$ of (a) overdoped and (b) underdoped Ba$_{1-x}$K$_{x}$Fe$_2$As$_2$, showing a pronounced hardening below $T_c$ up to 82\% K content. (c)-(e) Low-temperature Young's modulus along [110], $Y_{[110]}$, (solid line) and along [100], $Y_{[100]}$, (dashed line) of overdoped Ba$_{1-x}$K$_{x}$Fe$_2$As$_2$, showing that the hardening occurs only in the [110] direction. (f)-(i) Enlarged view of the data close to $T_c$ showing an additional small step-like anomaly $\Delta Y_i$ at $T_c$. A linear background has been subtracted from the data. (j)-(l) Same as in (a), (b) for Ba(Fe$_{1-x}$Co$_{x}$)$_2$As$_2$. (m) Enlarged view of the low-temperature Young modulus of FeSe. (f) Even further enlarged view of the Young modulus of FeSe around $T_c$, with a linear background subtracted. Vertical arrows mark $T_c$ in all panels. Data in (a)-(e) and (j)-(m) have been shifted vertically for better comparison. Lines in (i), (n) indicate an ideal second-order phase transition.}
\label{fig:8}
\end{center}
\end{figure}

Besides the softening due to the nematic transition, the high resolution of our three-point bending experiment can also be used to study the behavior at and below the superconducting transition. In Fig. \ref{fig:8}, we show our Young's modulus data for Co- and K-doped BaFe$_2$As$_2$ as well as for FeSe around $T_c$ in detail. For most samples, a pronounced hardening of $Y_{[110]}$ below $T_c$ is observed, while for some compositions of strongly overdoped BaFe$_2$As$_2$ and FeSe a small step-like softening of $Y_{[110]}$ at $T_c$ is also visible. A hardening of the elastic shear modulus $C_{66}$ below $T_c$ was reported previously in overdoped Ba(Fe$_{1-x}$Co$_{x}$)$_2$As$_2$ \cite{Fernandes2010,Yoshizawa2012}. As explained below, this effect is unusual in its sign, shape and magnitude for the effect of superconductivity on an elastic modulus.

The usual thermodynamic signature of a second-order phase transition in the Young's modulus or any other elastic modulus is a small sudden softening upon cooling, which is related to the strain/stress dependence of $T_c$. Such a discontinuity is expected because the Young's modulus $Y_i$ is the inverse of a component of the elastic compliance $S_{ii}$, which, in turn, is a second derivative of the free energy. 
An Ehrenfest-type relation relates the size of this discontinuity at $T_c$, $\Delta Y_{i}$, to the uniaxial-pressure derivative $dT_c/dp_i$,
\begin{equation}
\Delta Y_i=-Y_i^2\left(\frac{dT_c}{dp_i}\right)^2\Delta C_p/T_c\label{eq:EhrenfestY}.
\end{equation}
Here, $\Delta C_p>0$ is the specific-heat discontinuity at $T_c$ and $i$ stands for the direction. Note that necessarily $\Delta Y_i<0$, i.e., $Y_i$ shows a step-like decrease on cooling through the transition. Also, the shear modulus $C_{66}$ alone is not expected to have such discontinuity, because the first derivative of $T_c$ with respect to a shear deformation necessarily vanishes \cite{Pippard1955}. Hence, any discontinuity in the Young's modulus arises from the contribution of the longitudial elastic constants (the '$\gamma$' in Eq. \ref{eq:Y110}) and $\Delta Y_{[110]}/Y_{[110]}^2=\Delta Y_{[100]}/Y_{[100]}^2$. It has been pointed out in Ref. \cite{Fernandes2013III} that for an exotic superconducting state which mixes nematic fluctuations and $s$- and $d$-wave superconductivity and breaks tetragonal symmetry, even $C_{66}$ can have a discontinuity at $T_c$. Such a state might occur in overdoped Ba$_{1-x}$K$_{x}$Fe$_2$As$_2$ \cite{Fernandes2013III}, however, the Young's modulus is not an ideal probe to search for this effect, since it also contains the contribution from longitudinal elastic constants, as mentioned above. 

Fig. \ref{fig:8} presents our Young's modulus data close to $T_c$ in detail. The hardening of $Y_{[110]}$ below $T_c$ is resolved for Ba$_{1-x}$K$_{x}$Fe$_2$As$_2$ up to 82\% K content and Ba(Fe$_{1-x}$Co$_{x}$)$_2$As$_2$ up to 9\% Co-content (panels (a), (b), (j), (k)). Note that the effect is even observed for underdoped samples, when $T_c$ lies within the orthorhombic state and the Young's modulus is actually dominated by the effect of structural twins (see section \ref{sec:DMA}). As detailed in Ref. \cite{Fernandes2010}, a hardening of the shear modulus $C_{66}$ shows that nematic fluctuations, which decrease $C_{66}$, are weakened in the superconducting state. Its observation up to 82\% K content indicates the presence of such nematic fluctuations over a large part of the superconducting dome in the phase diagram of Ba$_{1-x}$K$_{x}$Fe$_2$As$_2$, although the degree of hardening strongly decreases with doping. Since this kind of response of $Y_{[110]}$ to superconductivity derives from the contribution of the shear modulus $C_{66}$, it should not be present in the Young's modulus along [100], $Y_{[100]}$, which does not contain any contribution from $C_{66}$. Consistenly, panels (c), (d), (e) show that such a hardening is not present in $Y_{[100]}$ of overdoped Ba$_{1-x}$K$_{x}$Fe$_2$As$_2$ and very small effects may be due to sample misalignment. 

In FeSe, such a hardening of $Y_{[110]}$ setting in abruptly at $T_c$ is not observed, which can be taken as a sign that orthorhombic distortion and superconductivity do not compete in the same way as in doped BaFe$_2$As$_2$ in this material\cite{Boehmer2013,Boehmer2015}. However, there is a very slight hardening of $Y_{[110]}$ (two orders of magnitude smaller than for similar Ba(Fe$_{1-x}$Co$_{x}$)$_2$As$_2$ samples) with an onset well above $T_c$, which correlates with an anomalous uniaxial thermal expansion seen previously\cite{Boehmer2013}. The origin of this effect is unclear at this point and it may possibly be a consequence of the presence of very small Fermi energies in the system\cite{Maletz2014,Shimojima2014,Terashima2014,Kasahara2014}. 
For some compositions of strongly overdoped Ba$_{1-x}$K$_{x}$Fe$_2$As$_2$ (Fig. \ref{fig:8} (f), (g), (i)), for Ba(Fe$_{0.88}$Co$_{0.12}$)$_2$As$_2$ (Fig. \ref{fig:8} (l)), and for FeSe (Fig. \ref{fig:8} (n)), we can also resolve the small step-like softening of $Y_{[110]}$ and $Y_{[100]}$ expected from thermodynamics (Eq. \ref{eq:EhrenfestY}). Using additional specific-heat data \cite{Hardyunpublished,Hardy2010,Boehmer2015}, the pressure-derivative of $T_c$ is calculated via eq. \ref{eq:EhrenfestY} and shown in table \ref{tab:dTcdp}. The uniaxial pressure derivative $dT_c/dp_i$ can also be calculated by using uniaxial thermal expansion and specific heat via a similar Ehrenfest relation $dT_c/dp_i=V_m\Delta\alpha_i/\Delta(C_p/T)$. As shown in table \ref{tab:dTcdp}, the pressure derivatives derived from evaluating either the Young's modulus or the thermal-expansion\cite{Boehmer2013,Boehmer2015II,Meingastunpublished} data agree quite well. It is unclear why we do not resolve this step-like softening of the Young's modulus in the Ba$_{0.18}$K$_{0.82}$Fe$_2$As$_2$ sample nor in $Y_{[100]}$ of the Ba$_{0.4}$K$_{0.6}$Fe$_2$As$_2$ sample. These, instead, seem to show a step-like hardening that cannot be explained with eq. \ref{eq:EhrenfestY}. 

\begin{table}
\caption{Uniaxial in-plane pressure derivative $\left|dT_c/dp_a\right|$ (sixth column), inferred using the Ehrenfest relation, eq. \ref{eq:EhrenfestY}, from the discontinuity of the Young modulus at $T_c$ (third column, Fig. \ref{fig:8}) and the specific heat (fifth column). A high-temperature Young modulus $Y_i\textnormal{(300 K)}=80$ GPa has been assumed in all cases. The obtained value should be the same for the two in-plane inplane directions $i=[110]$ and $i=[100]$. For comparison, the last column shows $dT_c/dp_a$ inferred from uniaxial thermal-expansion and specific heat data. The sign of $dT_c/dp_a$ can only be obtained from thermal expansion.}
\begin{tabular}{llccccc}
\hline\hline
&$i$&$\frac{\Delta Y_i}{Y_i\textnormal{(300 K)}}$&$\frac{Y_i(T_c)}{Y_i\textnormal{(300 K)}}$&$\Delta C_p/T_c$&$\left|\frac{dT_c}{dp_a}\right|$&$\frac{dT_c}{dp_a}$\\
&&10$^{-3}$&&mJ mol$^{-1}$K$^{-2}$&K/GPa&K/GPa\\
&&&&Refs. \cite{Hardyunpublished,Boehmer2013,Hardy2010II}&eq. \ref{eq:EhrenfestY}&Refs. \cite{Meingast2012,Boehmer2013,Boehmer2015}\\
\hline
Ba$_{0.52}$K$_{0.48}$Fe$_2$As$_2$&${[100]}$&-0.62&1.03&126&1.9&-2.3\\
Ba$_{0.52}$K$_{0.48}$Fe$_2$As$_2$&${[110]}$&-1&0.92&126&2.7&-2.3\\  
Ba$_{0.40}$K$_{0.60}$Fe$_2$As$_2$&${[110]}$&-0.73&0.99&96&2.5&-2.5\\ 
Ba$_{0.10}$K$_{0.90}$Fe$_2$As$_2$&${[100]}$&-0.25&1.07&50&1.8&-2.2\\ 
Ba$_{0.10}$K$_{0.90}$Fe$_2$As$_2$&${[110]}$&-0.18&1.05&126&1.6&-2.2\\ 
Ba(Fe$_{0.88}$Co$_{0.12}$)$_2$As$_2$&${[110]}$&-0.8&1.02&9&8&$\approx12$\\
FeSe&${[110]}$&-0.0051&0.16&5.6&3.2&3\\ 
\hline \hline 
\end{tabular}
\label{tab:dTcdp}
\end{table}

\section{Different probes of the nematic susceptibility\label{sec:comparison}}

Quite generally, the nematic susceptibility can be probed by measuring the sensitivity of the electronic anisotropy of various quantities to uniaxial stress or strain\cite{Chu2012}. The most detailed data have been obtained using strain-dependent resistivity\cite{Chu2012,Kuo2013,Kuo2014,Kuo2015}. The Raman response in different symmetry channels provides another probe\cite{Gallais2014} without actually having to apply any stress or strain to the crystal. In this section, we briefly review these studies and compare them to the results of our shear-modulus measurements. We note that there are also other probes, which were used to obtain the nematic susceptibility, e.g., stress-dependent measurements of the optical reflectivity\cite{Mirri2014,Mirri2014II}. 

\subsection{Elastoresistivity}

The elastoresistivity is defined as the resistance change induced by sample deformation (strain) and is closely related to the piezoresistivity which is the resistance change due to stresses acting on the sample\cite{Kuo2013}. Since the in-plane resistance anisotropy can be taken as a proxy for the nematic order parameter in the Fe-based systems, there is an elastoresistivity coefficient, namely $m_{66}$, which is directly linked to the nematic susceptibility. In Refs. \cite{Chu2012,Kuo2013,Kuo2015}, the in-plane resistivity anisotropy $N$ of iron-based materials was measured as a function of strain $\varepsilon_6$, externally applied to the sample via a piezo stack. Making use of the bilinear coupling $\lambda$ between nematic order parameter and shear deformation (eq. \ref{eq:Felnem}), one obtains $2m_{66}=dN/d\varepsilon_6\propto d\varphi/d\varepsilon_6=\lambda\chi_\varphi$ \cite{Chu2012,Kuo2015}. The proportionality constant between $N$ and $\varphi$ is related to the details of the Fermi surface and electronic scattering and can, in principle, be temperature, as well as doping dependent. It was shown that $m_{66}$ follows a Curie-Weiss law $\sim1/(T-T_0)$ in BaFe$_2$(As$_{0.7}$P$_{0.3}$)$_2$ \cite{Kuo2015}, which advocates that this proportionality constant is only very weakly temperature dependent. Further, Ni- and Co-doped samples with the same $T_s$ were compared\cite{Kuo2014}. Ni doping induces larger scattering (lower RRR), however, it was found that the elastoresistivity $m_{66}$ is independent of this kind of disorder for $T>T_s$\cite{Kuo2014}. 

Elastoresistivity measurements of the optimally doped compounds BaFe$_2$(As$_{0.7}$P$_{0.3}$)$_2$, Ba$_{0.6}$K$_{0.4}$Fe$_2$As$_2$, Ba(Fe$_{0.93}$Ni$_{0.07}$)$_2$As$_2$, Ba(Fe$_{0.955}$Ni$_{0.045}$)$_2$As$_2$, Fe(Te$_{0.6}$Se$_{0.4}$) \cite{Kuo2015} as well as in FeSe\cite{Watson2015} show that the elastoresistivity coefficient $m_{66}$ diverges following an approximate Curie-Weiss law in all of these systems, though deviations below $\sim 100$ K are sometimes observed.  
We note that the quantity $\lambda^2\chi_\varphi/C_{66,0}$ obtained from the shear modulus is normalized such that it reaches (ideally) the value $1$ at the phase transition when $C_{66}$ is expected to vanish. In contrast, the elastoresistivity is not normalized and the value of $m_{66}$ is indeed found to be doping dependent. It seems, e.g., to peak around optimal doping in Ba(Fe$_{1-x}$Co$_x$)$_2$As$_2$ which was suggested to be due to enhanced fluctuations near a quantum critical point\cite{Kuo2015}. However, $m_{66}$ still as a similar magnitude in all of the studied iron-based systems, although its sign depends on the particular system.

As mentioned previously, the structural transition may, in principle, arise either from a divergence of $\chi_\varphi$, as in an electronically-driven transition (as assumed here), or from vanishing $C_{66,0}$, as in a bare lattice instability. The quantity obtained by the shear modulus measurements, $\lambda^2\chi_\varphi/C_{66,0}$ diverges in both cases. Notably, the elastoresistivity $m_{66}\propto\lambda\chi_\varphi$ is independent of the bare shear modulus $C_{66,0}$. If the lattice caused the transition, then $\lambda^2\chi_\varphi/C_{66,0}$ would diverge, but $\lambda\chi_\varphi$ would show no strong temperature dependence. The experiment by Chu et al.\cite{Chu2012} showed that, however, $\lambda\chi_\varphi$ also diverges, which was taken as a proof that $\varphi$ drives the transition and that the lattice distortion is just a consequence of the bilinear coupling to $\varphi$\cite{Chu2012}. 

\subsection{Electronic Raman scattering}

Electronic Raman scattering of Ba(Fe$_{1-x}$Co$_x$)$_2$As$_2$\cite{Yang2013,Gallais2014} finds an enhancement of the Raman response in the $B_{2g}$ symmetry channel (which is the symmetry that corresponds to the orthorhombic distortion) with respect to the $B_{1g}$ symmetry channel. Using Kramer's Kronig relations, the static nematic charge susceptibility $\chi_0^{x^2-y^2}$ was extracted from the data. $\chi_0^{x^2-y^2}$ of Ba(Fe$_{1-x}$Co$_{x}$)$_2$As$_2$ was found to increase on approaching the structural transition, though it does not diverge at $T_s$. It, indeed, follows a Curie-Weiss law for a wide doping range \cite{Gallais2014}. Recently, Raman scattering data and $\chi_0^{x^2-y^2}$ was also reported for FeSe and Ba$_{1-x}$K$_x$As$_2$As$_2$\cite{Gallais2015II}. While $\chi_0^{x^2-y^2}$ of FeSe and underdoped Ba$_{1-x}$K$_x$As$_2$As$_2$ shows the same Curie-Weiss-like divergence, deviations from the Curie-Weiss law were observed for close to optimally doped Ba$_{1-x}$K$_x$As$_2$As$_2$ samples. 

In Ref. \cite{Kontani2014}, both the shear modulus and the Raman susceptibility have been calculated in the five orbital model including vertex corrections. It was noted that the Raman susceptibility is less singular than the shear modulus because the photons cannot couple to the acoustic lattice vibrations due to the mismatch of their wavelength for the same frequency\cite{Kontani2014}. In this sense, the charge nematic susceptibility in Raman scattering should be similar to the bare nematic susceptibility $\chi_\varphi$ and not $\tilde{\chi}_\varphi$, which is renormalized by coupling to the lattice\cite{Gallaisprivatecommunication}. Electronic Raman scattering as a probe of nematicity in iron-based systems is discussed in another contribution to this issue \cite{Gallais2015}.

\begin{figure}
\begin{center}
\includegraphics[width=0.6\textwidth]{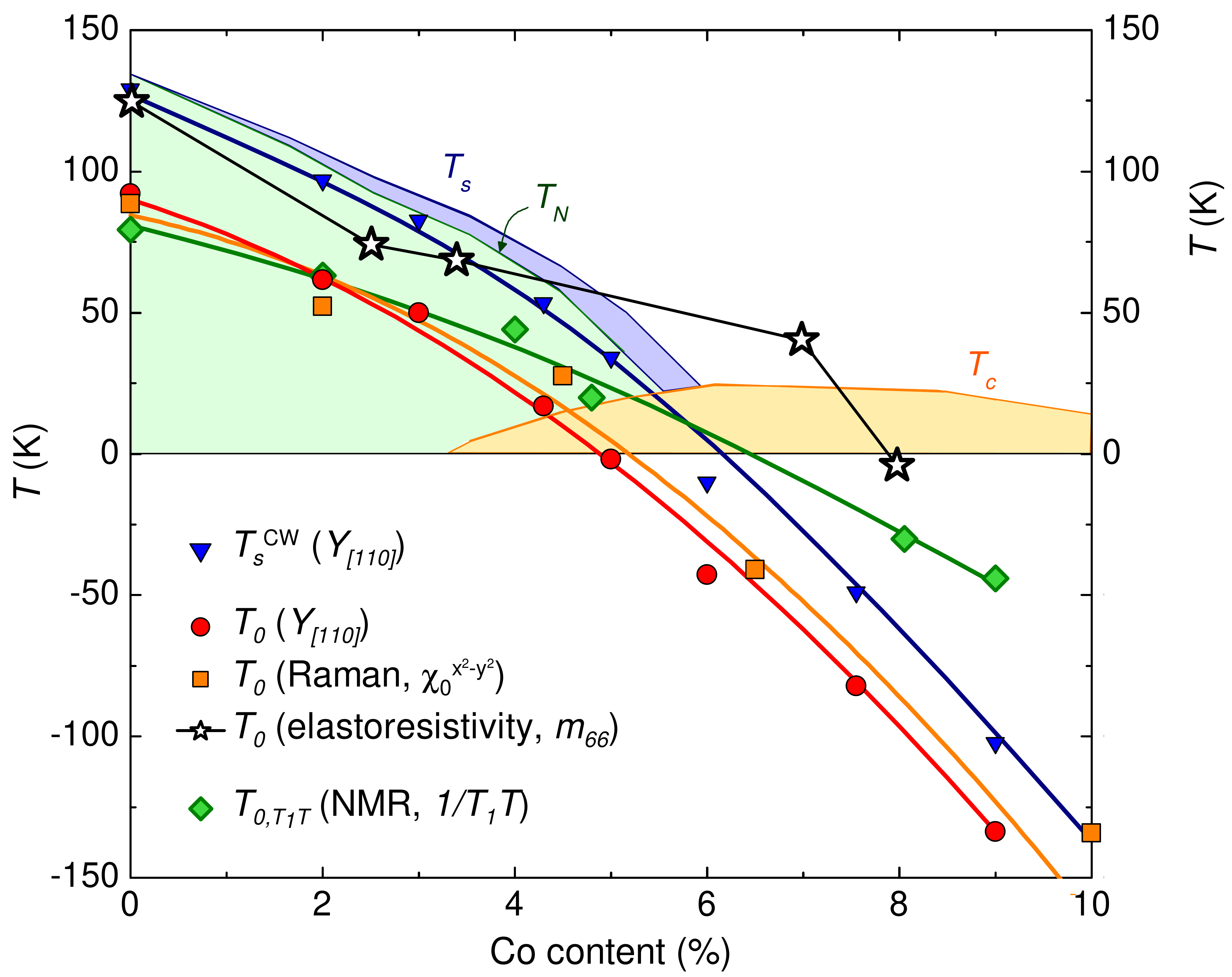}
\end{center}
\caption{Weiss-temperature $T_0$ obtained from fitting the nematic susceptibility as determined by Young's modulus $Y_{[110]}$\cite{Boehmer2014} (red circles),  electronic Raman scattering\cite{Gallais2014} (orange squares) and elastoresistivity $m_{66}$\cite{Kuo2015} (black open stars) measurements in Ba(Fe$_{1-x}$Co$_{x}$)$_2$As$_2$. Also given is the Weiss temperature $T_{0,T_1T}$ of the spin-lattice relaxation rate divided by temperature $1/T_1T$ in nuclear magnetic resonance\cite{Ning2010,Ning2014} (green diamonds). Lines are a guide to the eye. The phase transition temperatures $T_s$, $T_N$ and $T_c$\cite{Chu2009} are indicated by thin colored lines and areas. $T_s^{CW}$, the temperature at which $C_{66}$ extrapolates to zero, is indicated by blue triangles.}
\label{fig:13}
\end{figure}

There is very good general agreement between the nematic susceptibility from shear-modulus measurements $\chi_\varphi$ and the charge nematic susceptibility $\chi_0^{x^2-y^2}$ inferred from Raman scattering. First, $\chi_\varphi$ and $\chi_0^{x^2-y^2}$ both follow the familiar Curie-Weiss temperature dependence $\sim1/(T-T_0)$ \cite{Yang2013,Gallais2014} in Ba(Fe$_{1-x}$Co$_{x}$)$_2$As$_2$, as well as in FeSe\cite{Gallais2015II,Gallaisprivatecommunication}. In particular, the Weiss temperature $T_0$ obtained by the two probes agrees very well in the Ba(Fe$_{1-x}$Co$_{x}$)$_2$As$_2$ system (Fig. \ref{fig:13}, see below). For optimally doped Ba$_{1-x}$K$_{x}$Fe$_2$As$_2$, the shear-modulus data (see Fig. \ref{fig:6}) show an inflection point and strong deviations from a Curie-Weiss dependence of the nematic susceptibility.  Interestingly, a very similar temperature dependence of the Raman susceptibility $\chi_0^{x^2-y^2}$ has also been observed in close to optimally doped Ba$_{1-x}$K$_{x}$Fe$_2$As$_2$\cite{Gallais2015II}.
On the other hand, the elastoresistivity coefficient $m_{66}$ of all studied  materials, including optimally doped Ba$_{1-x}$K$_{x}$Fe$_2$As$_2$, shows an approximate Curie-Weiss-like divergence\cite{Kuo2015}. The origin of this difference is unclear to us. 

Among the above materials, the Ba(Fe$_{1-x}$Co$_{x}$)$_2$As$_2$ system has been studied most intensively. In this system, all three $\chi_\varphi$, $\chi_0^{x^2-y^2}$ and $m_{66}$ show a Curie-Weiss like divergence. To make a quantitative comparison, we show in Fig. \ref{fig:13} the values of the Weiss temperature $T_0$ obtained from fitting these data 
to a Curie-Weiss law $\sim 1/(T-T_0)$. Note that in case of the $m_{66}$ data, the low-temperature region, where $m_{66}$ deviates somewhat from a Curie-Weiss law, has been excluded from the fit\cite{Kuo2015}. We have also included $T_{0,T_1T}$ obtained from fitting the spin-lattice relaxation rate divided by temperature $1/T_1T$\cite{Ning2010,Ning2014}, since it also follows a Curie-Weiss temperature dependence \cite{Ning2010}. $1/T_1T$ is related to the strength of magnetic fluctuations, which the spin-nematic scenario links to the nematic susceptibility (these data are discussed in the following section \ref{sec:scaling}). 
Also shown in $T_s^{CW}$, the temperature at which $C_{66}$ extrapolates to zero, i.e., at which the structural transition would be expected (see section \ref{sec:Landau}). Why $C_{66}$ does not quite reach zero and $T_s^{\mathrm{CW}}$ is lower than $T_s$ is still an open question.

Intriguingly, $T_0$ as determined by electronic Raman scattering, agrees quite well with the elastic data over the whole doping range, which supports that the charge nematic susceptibility $\chi_0^{x^2-y^2}$ of the Raman experiment is very closely related to the bare nematic susceptibility $\chi_\varphi$ obtained using the shear-modulus data. In contrast, the values of $T_0$ derived from the elastoresistivity data are considerably higher at all doping levels than those from the elastic and Raman data, which is unexpected within the Landau analysis of this quantity \cite{Chu2012}. For this probe, $T_0$ appears even to cross the $T_s$ line at around 5\% Co content, which is also not expected in the simple Landau theory. This curiously high value of $T_0$ for the overdoped samples may partly be due to the exclusion of the low-temperature region from the fit of $m_{66}$. In this region, the divergence of $m_{66}$ appears to be suppressed at lower temperature, which was attributed to disorder effects\cite{Kuo2015}. Such an effect appears to be either absent or much less pronounced in the Raman and Young's-modulus data. 
The origin of the differences of the nematic susceptibility derived from the different experiments in both the $T_0$ values for Ba(Fe$_{1-x}$Co$_{x}$)$_2$As$_2$ and detailed temperature dependence for optimally doped Ba$_{1-x}$K$_{x}$Fe$_2$As$_2$ is unclear to us and deserves further attention.

\section{Magnetic correlations as the origin of nematicity?\label{sec:scaling}}

In the spin-nematic scenario of Refs. \cite{Nandi2010,Fernandes2010,Fernandes2012}, magnetic correlations are at the origin of nematicity and, ultimately, the structural phase transition. Hence, it should be possible to derive the elastic properties of iron-based materials from their magnetic properties. Indeed, a scaling relation between $1/T_1T$, as a measure of the strength of spin fluctuations, and the shear modulus $C_{66}$ has been derived in this theory, which provides a useful test of the scenario\cite{Fernandes2013}. Here, we briefly review the derivation of this scaling between spin-lattice relaxation rate $1/T_1$ and $C_{66}$ and then we check the scaling using experimental data in three different iron-based systems. 

\subsection{Scaling relation between $T_1T$ and $C_{66}$}

An expression for the nematic susceptibility in terms of the dynamic spin susceptibility $\chi$ is calculated in Ref. \cite{Fernandes2010} as, 
\begin{equation}
\left(\chi_\varphi\right)^{-1}=\frac{1}{\sum_q{\chi^2(q)}}-g_0,\label{eq:chispinnem}
\end{equation}
(see also Ref. \cite{Fernandes2012}). 
$\chi_\varphi$ is renormalized by bilinear coupling to the elastic system (as in equation \ref{eq:Felnem}) to
\begin{equation}
\left(\tilde{\chi}_\varphi\right)^{-1}=\frac{1}{\sum_q{\chi^2(q)}}-(g_0+\lambda^2/C_{66,0}),
\end{equation}
where $g=g_0+\lambda^2/C_{66,0}$ is the nematic coupling and the crucial parameter of the theory. $q=(\mathbf{q},\omega)$ stands for the momentum and frequency dependence. 
The magnetic transition occurs when $\sum_q{\chi^2(q)}$ diverges, but if $g>0$ it is sufficient for $\sum_q^2{\chi(q)}$ to reach a finite threshold value (i.e., $1/g$) to cause a divergence of $\tilde{\chi}_\varphi$ and induce the nematic/structural transition. This is the explanation why $T_s$ can be higher than $T_N$ in this scenario even though both transitions are driven by magnetic fluctuations.
The spin susceptibility $\chi$ can be accessed by the spin-lattice relaxation rate divided by temperature, as measured in NMR experiments,
\begin{equation}
\frac{1}{T_{1}T}=\gamma _{g}^{2}\lim_{\omega \rightarrow \omega_0}\sum_{\mathbf{q}%
}F^{2}\left( \mathbf{q}\right) \frac{\mathrm{Im}\,\chi \left( \mathbf{q}%
,\omega \right) }{\omega }  \label{T1T}
\end{equation}
Here, $\omega_0$ is the NMR frequency, which is considered to be very small and $F(\mathbf{q})$ is a momentum dependent form factor which peaks at the ordering wave vectors $Q_1$ and $Q_2$ when the magnetic field is applied parallel to the $ab$ plane \cite{Smerald2011}. As shown in detail in Ref. \cite{Fernandes2013}, $1/T_1T$ measured with the magnetic field in the $ab$ plance is proportional to $\sum_q{\chi^2(q)}$ under certain approximations (assuming overdamped dynamics $\left(\chi(\mathbf{q},\omega)\right)^{-1}=\left(\chi(\mathbf{q})\right)^{-1}-i\omega\Gamma$, vicinity to a finite temperature critical point so that $\sum_q\chi^2(q)$ can be replaced by $T_0\sum_\mathbf{q}\chi^2(\mathbf{q})$, and the replacement of the form factor $F(\mathbf{q})\rightarrow F(Q)$ because of the direction of the applied field). Using this proportionality, one can express $\chi_\varphi$ and, with the help of equation \ref{eq:Ceff}, $C_{66}$ in terms of $T_1$,
\begin{equation}
\frac{C_{66}}{C_{66,0}}=\frac{1}{1+\left(aT_1T-b\right)^{-1}} \label{eq:scaling}
\end{equation}
with the two parameters $a$ and $b$\cite{Fernandes2013}. The parameter $b$ is particularly interesting, since it provides a measure of the nematic coupling strength $b=\frac{C_{66,0}}{\lambda^2}g$.

\subsection{Test of the $1/T_1T$-$C_{66}$ scaling relation in Co- and K-doped BaFe$_2$As$_2$}

\begin{figure}
\includegraphics[width=\textwidth]{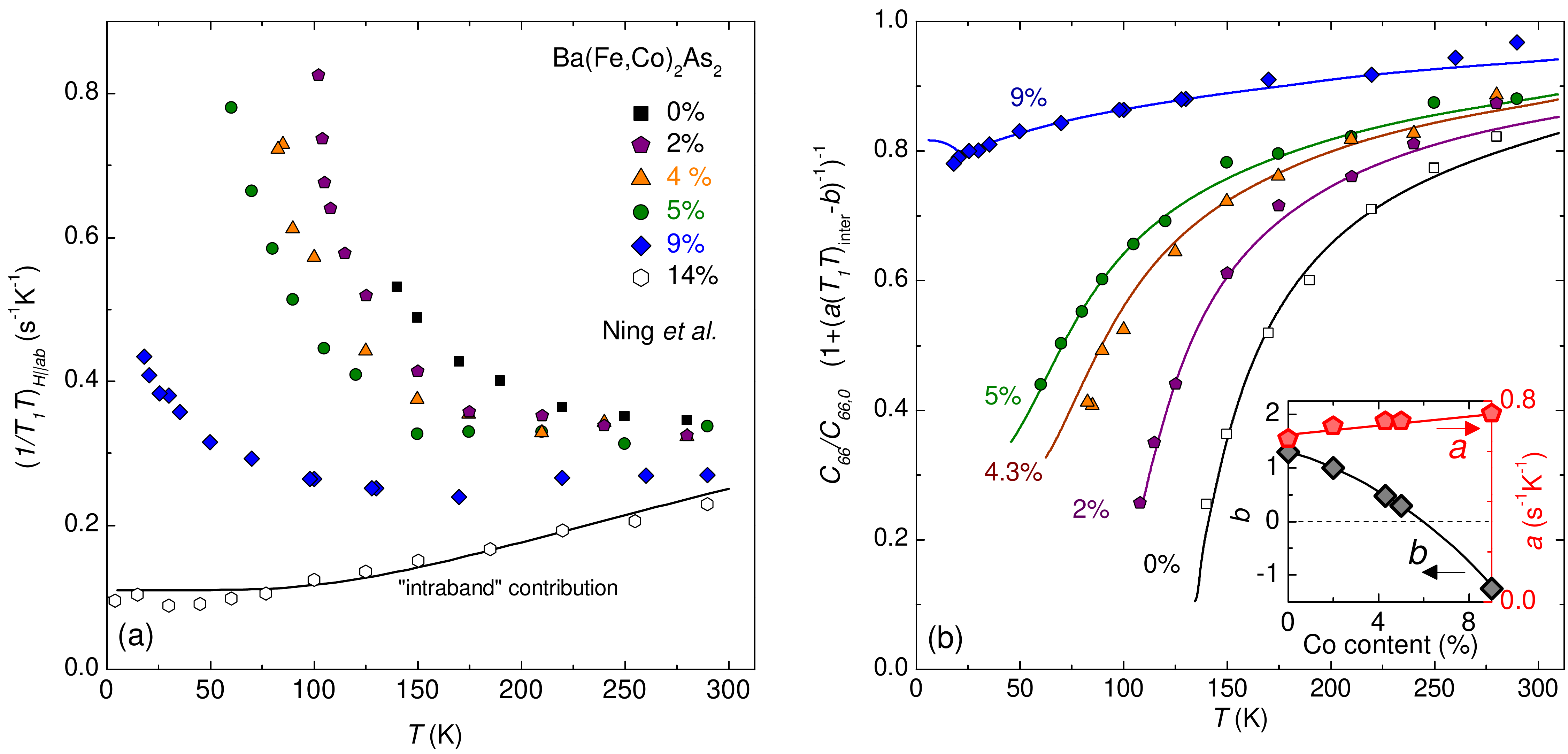}
\caption{(a) Spin-lattice relaxation rate divided by temperature $1/T_1T$ of Ba(Fe$_{1-x}$Co$_{x}$)$_2$As$_2$ with magnetic field $H||ab$, from Refs. \cite{Ning2010,Ning2014}. The data for the 14\% doped sample are taken as a background (black line). (b) Scaling analysis of the shear modulus $C_{66}/C_{66,0}$ and spin-lattice relaxation rate. The lines are obtained from three-point bending, while the symbols show the data of (a), scaled according to eq. \ref{eq:scaling} after subtraction of the background. The scaling parameters are reported in the inset.}
\label{fig:9}
\end{figure}

\begin{figure}
\includegraphics[width=\textwidth]{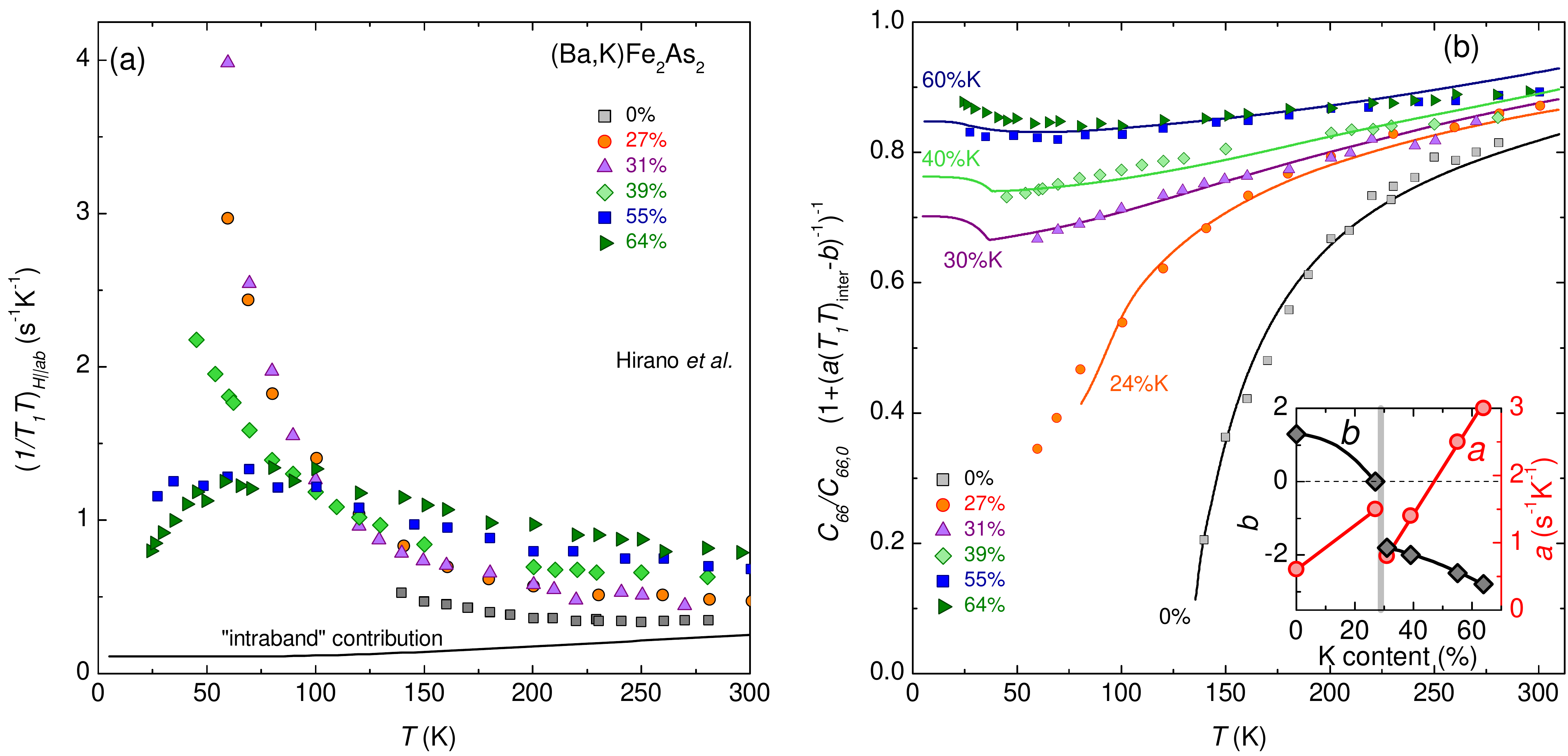}
\caption{Same as Fig. \ref{fig:9} but for Ba$_{1-x}$K$_{x}$Fe$_2$As$_2$. NMR data with field $H||ab$ are taken from Ref. \cite{Hirano2012}. The vertical line in the inset marks the K content at which the structural transition is suppressed in a first-order-like fashion and the temperature dependence of the nematic susceptibility changes abruptly (see Fig. \ref{fig:7})}
\label{fig:10}
\end{figure}

Figure \ref{fig:9}a shows $1/T_1T$ data of Ba(Fe$_{1-x}$Co$_{x}$)$_2$As$_2$ from Refs. \cite{Ning2010,Ning2014}, measured under an in-plane magnetic field and Fig. \ref{fig:10}a shows the equivalent data for Ba$_{1-x}$K$_{x}$Fe$_2$As$_2$ from Ref. \cite{Hirano2012}. We first discuss the Ba(Fe$_{1-x}$Co$_{x}$)$_2$As$_2$ system. Eq. \ref{eq:scaling} considers only the fluctuations around the AFM wave vector $Q$, which are referred to as ``interband'' contribution in Ref. \cite{Ning2010}. The non-critical ``intraband'' contribution has to be subtracted from the data as a background. Following Ref. \cite{Ning2010}, this is achieved by subtracting the data for the strongly overdoped Ba(Fe$_{0.86}$Co$_{0.14}$)$_2$As$_2$, modeled as $\left(1/T_1T\right)_{\mathrm{intra}}=0.11\textnormal{ K$^{-1}$s$^{-1}$}+0.63\textnormal{ K$^{-1}$s$^{-1}$}\exp(-450\textnormal{ K}/T)$ (black line in Fig. \ref{fig:9}a).
The obtained $\left(1/T_1T\right)_{\mathrm{inter}}$ can be scaled onto the elastic data according to equation \ref{eq:scaling} and the result is shown in Fig. \ref{fig:9}b, with the scaling parameters $a$ and $b$ are given in the inset. The scaling works very well for all Co substitution levels, supporting a magnetic origin of the shear-modulus softening\cite{Fernandes2013}. Alternatively, a phenomenological linear scaling between $C_{66}$ and $1/T_1T$ was proposed in Ref. \cite{Nakai2013}, which, however, does not work as well. Note that the parameter $b$ is proportional to the difference between the Weiss temperatures $T_0$ and $T_{0,T_1T}$ of the nematic susceptibility $\chi_\varphi$ and $1/T_1T$, respectively, which are shown in Fig. \ref{fig:13}. Interestingly, $T_0$ and $T_{0,T_1T}$ cross and $b$ changes sign around the critical composition where the structural transition is suppressed. However, the assumption of a finite temperature critical point used to derive the scaling relation is not strictly valid in this doping region and, hence, this sign change of $b$ might be an artefact. However, if $b$, or, equivalently, $g$ really becomes negative for overdoped Ba(Fe$_{1-x}$Co$_{x}$)$_2$As$_2$, it would indicate that a non-nematic, $C_4$-symmetric magnetic state is preferred over the stripe-type one. Such a $C_4$-symmetric magnetic phase is, indeed, found to be induced by Na or K substitution in BaFe$_2$As$_2$ very close to the critical point \cite{Avci2014,Boehmer2015II}. 

We note that a similar scaling attempt in Ref. \cite{Gallais2014}, between the nematic charge susceptibility $\chi_0^{x^2-y^2}$ measured using electronic Raman scattering, and the elastic modulus failed. Here, the authors equated the $\chi_0^{x^2-y^2}$ with the renormalized nematic susceptibility $\tilde{\chi}_\varphi$ so that the scaling takes the form ${C_{66}}/{C_{66,0}}=\left(1+a\chi_0^{x^2-y^2}\right)^{-1}$ (deduced using eq. \ref{eq:CeffFernandes}) with a single parameter $a$. However, when assuming that $\chi_0^{x^2-y^2}$ reflects the bare nematic susceptibility $\chi_\varphi$\cite{Kontani2014,Gallaisprivatecommunication}, the form of the scaling should rather be identical to eq. \ref{eq:scaling}, which does work well also for the Raman data of Ba(Fe$_{1-x}$Co$_{x}$)$_2$As$_2$. 

It is interesting to consider the scaling relation of $1/T_1T$ and $C_{66}$ also in the Ba$_{1-x}$K$_{x}$Fe$_2$As$_2$ system, for which both sets of data have been published\cite{Hirano2012,Boehmer2014}, but the scaling has not been attempted. The striking feature of this system is that both $1/T_1T$ and $C_{66}$ do not follow a Curie-Weiss temperature dependence over the whole doping region. In particular, the nematic susceptibility as obtained from the shear modulus changes its temperature dependence abruptly between 24\% and 30\% K content. The NMR data from Ref. \cite{Hirano2012} show a similar change of temperature dependence, however, occurring rather between 39\% and 55\%. In spite of the lack of a Curie-Weiss-like temperature dependence, it is remarkable that these data can be scaled quite well by Eq. \ref{eq:scaling} (Fig. \ref{fig:10}). Tentatively, the same ``intraband'' background as for the Ba(Fe$_{1-x}$Co$_{x}$)$_2$As$_2$ system has been subtracted from the raw $1/T_1T$ data. The parameters $a$ and $b$ used in the scaling are shown in the inset of Fig. \ref{fig:10}b. There are several points of interest here. First, there is an abrupt change of both $a$ and $b$ at $\sim25\%$ K content (marked by a vertical line), reflecting that the temperature dependence of $C_{66}$ but not of $T_1T$ changes abruptly. $b$ is found to be close to zero just before the structural transition disappears, which is similar to the Ba(Fe$_{1-x}$Co$_{x}$)$_2$As$_2$ system. Beyond this concentration, $b$ is negative indicating that the systems tends to a non-nematic magnetic order. Interestingly, the magnetic ground state of Ba$_{1-x}$K$_{x}$Fe$_2$As$_2$ close to this K concentration seems, indeed, to be tetragonal and it would be fascinating to study NMR and shear modulus in detail for the respective substitution range. The parameter $a$ also shows a strong dependence on K content, while it is roughly independent of Co content. Examining the scaling relation (Eq. \ref{eq:scaling}) shows that the parameter $a$ renormalizes the magnitude of $T_1T$. The strong increase of $a$ with doping reflects that the high-temperature values of $1/T_1T$ increase significantly with K content, which might reflect either a change in the hyperfine coupling (as would be correctly captured by $a$ \cite{Fernandes2013}), or a strongly doping dependent background contribution. Note that $1/T_1T$ of the pure KFe$_2$As$_2$ has the largest values of all samples\cite{Hirano2012}, though it is supposedly far away from a magnetic instability. 

\subsection{Shear-modulus softening and magnetic fluctuations in FeSe}

\begin{figure}
\includegraphics[width=\textwidth]{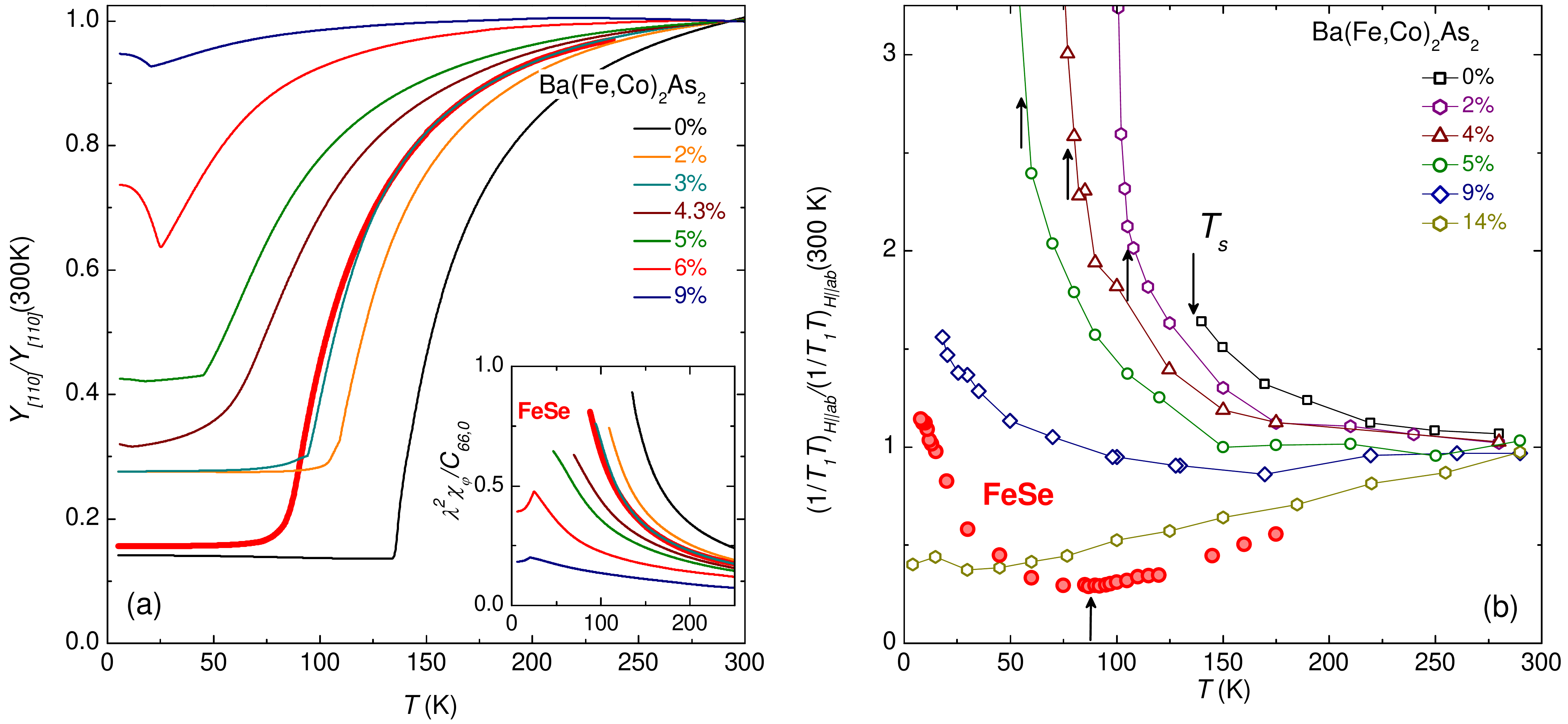}
\caption{(a) Young's modulus $Y_{[110]}$ of FeSe, superimposed on the data for Ba(Fe$_{1-x}$Co$_{x}$)$_2$As$_2$. FeSe fits very well in the series. The inset shows the nematic susceptibility (see eq. \ref{eq:Ceff}). (b) $1/T_1T$ normalized by its (extrapolated) value at room temperature of FeSe and Ba(Fe$_{1-x}$Co$_{x}$)$_2$As$_2$ (data taken from Refs. \cite{Ning2010,Ning2014}). Vertical arrows indicate $T_s$. While $1/T_1T$ of underdoped Ba(Fe$_{1-x}$Co$_{x}$)$_2$As$_2$ shows a strong increase on cooling towards $T_s$ (and a steeper increase below), the data for FeSe only increase below $T_s$.}
\label{fig:11}
\end{figure}

\begin{figure}
\includegraphics[width=\textwidth]{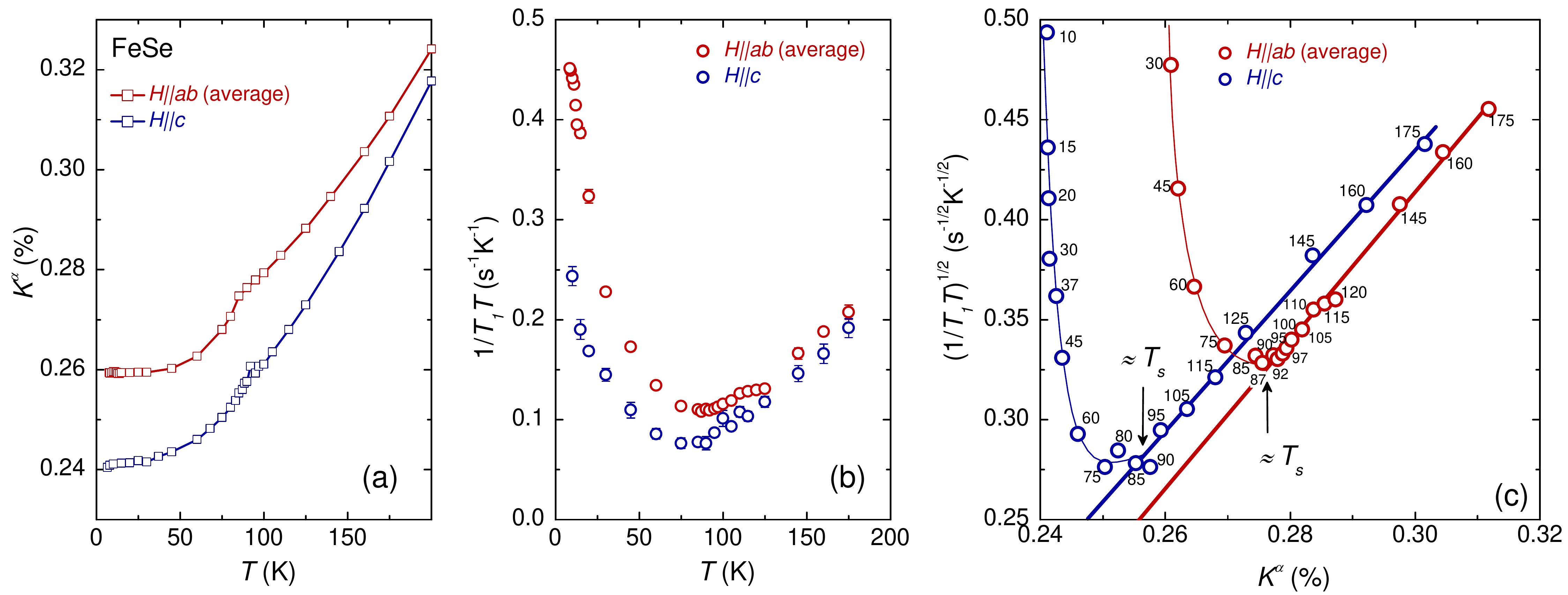}
\caption{(a) NMR spectral shift $K^\alpha$ and (b) spin-lattice relaxation rate divided by temperature $1/T_1T$ of single-crystalline FeSe with the field direction, $\alpha$, in the $ab$ plane (average of $a$ and $b$ axis in the orthorhombic state) and along the $c$ axis. (c) shows a plot of $\sqrt{1/T_1T}$ vs. $K^\alpha$, with the temperature as an implicit parameter indicated in units of K. The straight lines correspond to the Fermi-liquid type Korringa relation eq. \ref{eq:Korringa}, deviations from which show the emergence of a spin-fluctuation contribution. Such an additional contribution is evident only below $T_s$. The thin lines are a guide to the eye.   }
\label{fig:12}
\end{figure}

As described in the introduction, the iron-based superconductor FeSe is particularly interesting with respect to the relation of structure and magnetism. In particular, its large paramagnetic, orthorhombic (i.e., nematic) phase makes FeSe an interesting test case to study the origin of nematicity. As discernable from Figs. \ref{fig:4} and \ref{fig:6}, $Y_{[110]}$ of FeSe is very similar to underdoped Ba(Fe$_{1-x}$Co$_x$)$_2$As$_2$. In Fig. \ref{fig:11}, we compare the two systems in detail. Curiously, $Y_{[110]}$ and $\chi_\varphi$ of FeSe are nearly identical to Ba(Fe$_{0.97}$Co$_{0.03}$)$_2$As$_2$, which has a similar $T_s$\cite{Boehmer2015}. The nearly identical temperature dependences indicate that the coupling $\lambda^2/aC_{66,0}$ has the same value in FeSe and underdoped Ba(Fe$_{1-x}$Co$_{x}$)$_2$As$_2$, which is remarkable considering the differences between the two systems.

In light of this similarity of the Young's modulus, it is somewhat surprising to find that the evidence for spin fluctuations from the spin-lattice relaxation data of FeSe is much less pronounced than in the Ba(Fe$_{1-x}$Co$_{x}$)$_2$As$_2$ system \cite{Imai2009,Boehmer2015} (Fig. \ref{fig:11}b). In contrast to underdoped Ba(Fe$_{1-x}$Co$_{x}$)$_2$As$_2$, $1/T_1T$ of FeSe shows an increase upon cooling only below $T_s$ and, actually, decreases upon cooling from room temperature down to $T_s$. This drastic difference already suggests that spin fluctuations may not be the origin of the diverging nematic susceptibility in FeSe. In order to exclude the possibility that a strongly temperature dependent ``intraband'' background contribution hampers the determination of the relevant $\left(1/T_1T\right)_{\mathrm{inter}}$ and, hence, the scaling analysis in FeSe, the NMR data have been analyzed in detail (see Fig. \ref{fig:12}, \cite{Boehmer2015}).

Fig. \ref{fig:12}(a),(b) show the spectral shift $K^\alpha$ and $1/T_1T$ of a collection of $\sim 10$ single crystals of FeSe, measured in a field of 9 T with the field direction $\alpha$ both in-plane and along the $c$ axis. $K^\alpha$ has a pronounced temperature dependence, which can presumably be explained in a Fermi-liquid picture by the small Fermi energy \cite{Sales2010} found in FeSe \cite{Maletz2014,Shimojima2014,Terashima2014,Kasahara2014}. Since $K$ and $1/T_1T$ are related by the Korringa relation for a Fermi liquid
\begin{equation}
{\left(\frac{1}{T_1T}\right)_{\mathrm{FL}}}\propto K^2_{spin}.\label{eq:Korringa}
\end{equation}
the temperature dependent $K^\alpha$ will lead to a temperature dependent $1/T_1T$. In order to isolate a possible contribution from spin-fluctuations to $1/T_1T$, we test eq. \ref{eq:Korringa} by plotting $\sqrt{1/T_1T}$ vs. $K^\alpha$ with temperature as an implicit parameter in Fig. \ref{fig:12}c. For both field directions, we find a linear relation between the two quantities, which shows that the Fermi-liquid Korringa relation is indeed satisfied down to $T_s$. Hence, this analysis shows the absence of any measurable spin fluctuation contribution to $1/T_1T$ for $T>T_s$, in strong contrast to underdoped BaFe$_2$As$_2$. 
This means that such spin-fluctuations cannot be the origin of the shear-modulus softening and, in consequence, the structural transition and nematicity in FeSe, and it is likely that the alternative orbital order drives the structural transition, as also suggested by ARPES studies \cite{Shimojima2014,Nakayama2014}. A similar conclusion was reached by another NMR work by Baek et al. \cite{Baek2015}.
Curiously, the nematic susceptibility $\chi_\varphi$ of underdoped Ba(Fe$_{1-x}$Co$_{x}$)$_2$As$_2$ and FeSe is very similar, which may raise doubt on the magnetic origin of nematicity in Ba(Fe$_{1-x}$Co$_{x}$)$_2$As$_2$ as well.
Finally, the NMR data on FeSe (Figs. \ref{fig:11}, \ref{fig:12}) seem to suggest that the structural transition at $T_s$ triggers the emergence of spin fluctuations. However, there is no such correlation under hydrostatic pressure, which enhances spin fluctuations \cite{Imai2009} but suppresses $T_s$ \cite{Miyoshi2014,Terashima2015,Knoener2015}. This result suggests an unusual relation between magnetic order and the structural transition in FeSe.

As we pointed out previously\cite{Boehmer2015}, inelastic neutron scattering experiments are needed to determine the nature of the magnetic fluctuations, and very recently the first of such experiments have been reported on \cite{Rahn2015,Wang2015}. Surprisingly, these studies provide evidence of magnetic stripe-like $(\pi,0)$ fluctuations in FeSe, similar to the 122 compounds. As in the other Fe-based materials, these magnetic fluctuations are found to occur already above $T_s$, in apparent contradiction to the NMR results. There is, however, evidence for a sizeable spin-gap in the neutron data already at 110 K, which can explain the absence of spin fluctuations in the NMR experiment, a probe that is sensitive only to the low-energy fluctuations. However, the neutron experiments, which have been performed both on polycrystalline \cite{Rahn2015} and single-crystalline material \cite{Wang2015}, are also not fully consistent with each other. In Ref. \cite{Rahn2015} the strength of the magnetic signal is temperature independent between 8 K and 104 K, whereas a strong temperature dependence is observed in Ref. \cite{Wang2015}. These differences may also reflect differences in samples. The properties of FeSe samples depend strongly on the preparation technique. In particular, the samples prepared using a floating-zone technique used in Ref. \cite{Wang2015} have undergone a phase transformation to the tetragonal, superconducting $\beta$-FeSe phase on cooling to room temperature. Such samples are not single phase\cite{Ma2014} and do not show a clear tetragonal-to-orthorhombic structural transition in the resistivity\cite{Ma2014}. The samples prepared by solid-state synthesis used in Ref. \cite{Rahn2015} show an unusual difference of zero-field-cooled and field-cooled magnetic susceptibility around $T_s$\cite{Rahn2015}, different from the vapor-grown crystals\cite{Boehmer2013}. Hence, the issue seems not fully resolved yet and more careful neutron experiments on vapor-grown crystals are still desirable. Finally, we mention that these puzzling features of FeSe, i.e. the nematic phase and the absence of magnetic order, have attracted the attention of theorists, and there are now several different proposed theoretical scenarios, including strong magnetic frustration \cite{Glasbrenner2015}, spin-quadropolar order \cite{Yu2015}, the formation of a quantum paramagnet \cite{Wang2015II}, and charge-current density wave \cite{Chubukov2015}.

\section{Summary and outlook\label{sec:summary}}

In this review we have compared the electronic nematic susceptibility of various iron-based superconducting materials derived from measurements of the shear-modulus, the elastoresistivity, and the Raman response function. Particular emphasis has been put on our own studies of the Young modulus in Ba(Fe$_{1-x}$Co$_x$)$_2$As$_2$, Ba$_{1-x}$K$_x$Fe$_2$As$_2$ and FeSe obtained via a three-point bending technique in a capacitance dilatometer. In a Landau formalism, in which an electronic nematic order parameter drives the structural transition via bilinear coupling to the orthorhombic lattice distortion, the nematic susceptibility was obtained from the elastic data. The relation of this ``thermodynamic'' nematic susceptibility to spin- and orbital degrees of freedom is found to be intricate, and may be different for the different iron-based systems. Notably, the nematic susceptibility from the shear-modulus data seems to be closely related to the orbital nematic susceptibility from electronic Raman scattering in all the systems. On the other hand, the elastoresistivity behaves somewhat differently, in particular for optimally doped Ba$_{1-x}$K$_x$Fe$_2$As$_2$.  The excellent scaling of shear-modulus softening and spin-lattice relaxation rate $1/T_1$ in both Ba(Fe$_{1-x}$Co$_x$)$_2$As$_2$ and Ba$_{1-x}$K$_x$Fe$_2$As$_2$ supports the notion that the structural transition is driven by magnetic fluctuations. FeSe seems to be an unusual case, in that there is no spin-fluctuation contribution to $1/T_1$ above the structural transition, even though the nematic susceptibility of FeSe and underdoped Ba(Fe$_{1-x}$Co$_x$)$_2$As$_2$ have a very similar temperature dependence, suggesting the importance of orbital degrees of freedom. However, very recent inelastic neutron scattering experiments suggest that spin-fluctuations in FeSe are similar to the BaFe$_2$As$_2$-based systems. Further neutron studies on high-quality vapor grown FeSe crystals are thus highly desirable.

The interrelationship of the various types of coupled order in iron-based systems---e.g., structural, orbital, magnetic and superconducting---has turned out to be a very rich field of study and electronic nematicity has grown into one of the most intensively studied concepts in the field. The relationship between nematic fluctuations and superconductivity remains an interesting and open problem. Recently, evidence for a nematic resonance in the superconducting state was presented in Raman scattering experiments\cite{Gallais2015III} suggesting a close link between nematic fluctuations and superconductivity. Further, the enhancement of superconductivity near a nematic quantum critical point has been investigated\cite{Lederer2015}. We believe that in order to obtain further insight into the microscopic origin of nematicity and its possible relation to superconducitivity, the study of systems in which structural and magnetic transitions do not closely follow each other is particularly promising. In this context, FeSe and the magnetic $C_4$ phase in hole-doped BaFe$_2$As$_2$ have recently attracted a great deal of attention and may very well hold further surprises in the near future.

\section*{Acknowledgements}

First, we would like to acknowledge Thomas Wolf for the excellent iron-based single crystals, without which our studies would not have been possible. We would like to especially thank W. Schranz and M. Reinecker for the opportunity to perform the dynamic three-point bending experiments and for their hospitality in Vienna, and also Kenji Ishida and his group in Kyoto (T. Arai, T. Iye, T. Hattori) for the opportunity to make NMR measurements on FeSe and for their hospitality.  We thank R. M. Fernandes and J. Schmalian for the very fruitful collaboration concerning the scaling of bending and NMR data. Further, we acknowledge fruitful and rewarding collaborations and discussions with F. Hardy, P. Burger, L. Wang, P. Schweiss, H. v. Löhneysen, U. Karahasanovic, M. Hoyer, P. Hlobil,  A. Chubukov, S. Kasahara, T. Shibauchi, Y. Matsuda, I. R. Fisher, Y. Gallais, Y. Furukawa, A. Kreyssig, S. L. Bud'ko and P. C. Canfield.  Part of this work was supported by the Japan-Germany Research Cooperative Program, KAKENHI from JSPS and Project No. 56393598 from DAAD, and by the DFG through SPP 1458.


\section*{References}

\begin{thebibliography}{141}
\expandafter\ifx\csname natexlab\endcsname\relax\def\natexlab#1{#1}\fi
\providecommand{\url}[1]{\texttt{#1}}
\providecommand{\href}[2]{#2}
\providecommand{\path}[1]{#1}
\providecommand{\DOIprefix}{doi:}
\providecommand{\ArXivprefix}{arXiv:}
\providecommand{\URLprefix}{URL: }
\providecommand{\Pubmedprefix}{pmid:}
\providecommand{\doi}[1]{\href{http://dx.doi.org/#1}{\path{#1}}}
\providecommand{\Pubmed}[1]{\href{pmid:#1}{\path{#1}}}
\providecommand{\bibinfo}[2]{#2}
\ifx\xfnm\relax \def\xfnm[#1]{\unskip,\space#1}\fi
\bibitem[{Fradkin et~al.(2010)Fradkin, Kivelson, Lawler, Eisenstein, and
  Mackenzie}]{Fradkin2010}
\bibinfo{author}{E.~Fradkin}, \bibinfo{author}{S.~A. Kivelson},
  \bibinfo{author}{M.~J. Lawler}, \bibinfo{author}{J.~P. Eisenstein},
  \bibinfo{author}{A.~P. Mackenzie},
\newblock \bibinfo{title}{Nematic {Fermi} fluids in condensed matter physics},
\newblock \bibinfo{journal}{Annu. Rev. Condens. Matter Phys.}
  \bibinfo{volume}{1} (\bibinfo{year}{2010}) \bibinfo{pages}{153}.
\bibitem[{Fernandes et~al.(2014)Fernandes, Chubukov, and
  Schmalian}]{Fernandes2014}
\bibinfo{author}{R.~M. Fernandes}, \bibinfo{author}{A.~V. Chubukov},
  \bibinfo{author}{J.~Schmalian},
\newblock \bibinfo{title}{What drives nematic order in iron-based
  superconductors?},
\newblock \bibinfo{journal}{Nature Physics} \bibinfo{volume}{10}
  (\bibinfo{year}{2014}) \bibinfo{pages}{97--104}.
\bibitem[{Fernandes and Schmalian(2012)}]{Fernandes2012}
\bibinfo{author}{R.~M. Fernandes}, \bibinfo{author}{J.~Schmalian},
\newblock \bibinfo{title}{Manifestations of nematic degrees of freedom in the
  magnetic, elastic, and superconducting properties of the iron pnictides},
\newblock \bibinfo{journal}{Superconductor Science and Technology}
  \bibinfo{volume}{25} (\bibinfo{year}{2012}) \bibinfo{pages}{084005}.
\bibitem[{de~la Cruz et~al.(2008)de~la Cruz, Huang, Lynn, Li, II, Zarestky,
  Mook, Chen, Luo, Wang, and Dai}]{Cruz2008}
\bibinfo{author}{C.~de~la Cruz}, \bibinfo{author}{Q.~Huang},
  \bibinfo{author}{J.~W. Lynn}, \bibinfo{author}{J.~Li}, \bibinfo{author}{W.~R.
  II}, \bibinfo{author}{J.~L. Zarestky}, \bibinfo{author}{H.~A. Mook},
  \bibinfo{author}{G.~F. Chen}, \bibinfo{author}{J.~L. Luo},
  \bibinfo{author}{N.~L. Wang}, \bibinfo{author}{P.~Dai},
\newblock \bibinfo{title}{Magnetic order close to superconductivity in the
  iron-based layered {LaO$_{1-x}$F$_x$FeAs} systems},
\newblock \bibinfo{journal}{Nature} \bibinfo{volume}{453}
  (\bibinfo{year}{2008}) \bibinfo{pages}{899--902}.
\bibitem[{Rotter et~al.(2008)Rotter, Tegel, Johrendt, Schellenberg, Hermes, and
  P\"ottgen}]{Rotter2008III}
\bibinfo{author}{M.~Rotter}, \bibinfo{author}{M.~Tegel},
  \bibinfo{author}{D.~Johrendt}, \bibinfo{author}{I.~Schellenberg},
  \bibinfo{author}{W.~Hermes}, \bibinfo{author}{R.~P\"ottgen},
\newblock \bibinfo{title}{Spin-density-wave anomaly at {140 K} in the ternary
  iron arsenide {BaFe$_2$As$_2$}},
\newblock \bibinfo{journal}{Phys. Rev. B} \bibinfo{volume}{78}
  (\bibinfo{year}{2008}) \bibinfo{pages}{020503}.
\bibitem[{Nomura et~al.(2008)Nomura, Kim, Kamihara, Hirano, Sushko, Kato,
  Takata, Shluger, and Hosono}]{Nomura2008}
\bibinfo{author}{T.~Nomura}, \bibinfo{author}{S.~W. Kim},
  \bibinfo{author}{Y.~Kamihara}, \bibinfo{author}{M.~Hirano},
  \bibinfo{author}{P.~V. Sushko}, \bibinfo{author}{K.~Kato},
  \bibinfo{author}{M.~Takata}, \bibinfo{author}{A.~L. Shluger},
  \bibinfo{author}{H.~Hosono},
\newblock \bibinfo{title}{Crystallographic phase transition and high-{$T_c$}
  superconductivity in {LaFeAsO:F}},
\newblock \bibinfo{journal}{Superconductor Science and Technology}
  \bibinfo{volume}{21} (\bibinfo{year}{2008}) \bibinfo{pages}{125028}.
\bibitem[{Lumsden and Christianson(2010)}]{Lumsden2010}
\bibinfo{author}{M.~D. Lumsden}, \bibinfo{author}{A.~D. Christianson},
\newblock \bibinfo{title}{Magnetism in {Fe}-based superconductors},
\newblock \bibinfo{journal}{Journal of Physics: Condensed Matter}
  \bibinfo{volume}{22} (\bibinfo{year}{2010}) \bibinfo{pages}{203203}.
\bibitem[{Cano et~al.(2010)Cano, Civelli, Eremin, and Paul}]{Cano2010}
\bibinfo{author}{A.~Cano}, \bibinfo{author}{M.~Civelli},
  \bibinfo{author}{I.~Eremin}, \bibinfo{author}{I.~Paul},
\newblock \bibinfo{title}{Interplay of magnetic and structural transitions in
  iron-based pnictide superconductors},
\newblock \bibinfo{journal}{Phys. Rev. B} \bibinfo{volume}{82}
  (\bibinfo{year}{2010}) \bibinfo{pages}{020408}.
\bibitem[{Avci et~al.(2011)Avci, Chmaissem, Goremychkin, Rosenkranz, Castellan,
  Chung, Todorov, Schlueter, Claus, Kanatzidis, Daoud-Aladine, Khalyavin, and
  Osborn}]{Avci2011}
\bibinfo{author}{S.~Avci}, \bibinfo{author}{O.~Chmaissem},
  \bibinfo{author}{E.~A. Goremychkin}, \bibinfo{author}{S.~Rosenkranz},
  \bibinfo{author}{J.-P. Castellan}, \bibinfo{author}{D.~Y. Chung},
  \bibinfo{author}{I.~S. Todorov}, \bibinfo{author}{J.~A. Schlueter},
  \bibinfo{author}{H.~Claus}, \bibinfo{author}{M.~G. Kanatzidis},
  \bibinfo{author}{A.~Daoud-Aladine}, \bibinfo{author}{D.~Khalyavin},
  \bibinfo{author}{R.~Osborn},
\newblock \bibinfo{title}{Magnetoelastic coupling in the phase diagram of
  {Ba$_{1-x}$K$_x$Fe$_2$As$_2$} as seen via neutron diffraction},
\newblock \bibinfo{journal}{Phys. Rev. B} \bibinfo{volume}{83}
  (\bibinfo{year}{2011}) \bibinfo{pages}{172503}.
\bibitem[{Kim et~al.(2011)Kim, Fernandes, Kreyssig, Kim, Thaler, Bud'ko,
  Canfield, McQueeney, Schmalian, and Goldman}]{Kim2011}
\bibinfo{author}{M.~G. Kim}, \bibinfo{author}{R.~M. Fernandes},
  \bibinfo{author}{A.~Kreyssig}, \bibinfo{author}{J.~W. Kim},
  \bibinfo{author}{A.~Thaler}, \bibinfo{author}{S.~L. Bud'ko},
  \bibinfo{author}{P.~C. Canfield}, \bibinfo{author}{R.~J. McQueeney},
  \bibinfo{author}{J.~Schmalian}, \bibinfo{author}{A.~I. Goldman},
\newblock \bibinfo{title}{Character of the structural and magnetic phase
  transitions in the parent and electron-doped {BaFe$_2$As$_2$} compounds},
\newblock \bibinfo{journal}{Phys. Rev. B} \bibinfo{volume}{83}
  (\bibinfo{year}{2011}) \bibinfo{pages}{134522}.
\bibitem[{Ni et~al.(2008)Ni, Tillman, Yan, Kracher, Hannahs, Bud'ko, and
  Canfield}]{Ni2008}
\bibinfo{author}{N.~Ni}, \bibinfo{author}{M.~E. Tillman},
  \bibinfo{author}{J.-Q. Yan}, \bibinfo{author}{A.~Kracher},
  \bibinfo{author}{S.~T. Hannahs}, \bibinfo{author}{S.~L. Bud'ko},
  \bibinfo{author}{P.~C. Canfield},
\newblock \bibinfo{title}{Effects of {Co} substitution on thermodynamic and
  transport properties and anisotropic ${H}_{c2}$ in
  {$\text{Ba}{({\text{Fe}}_{1-x}{\text{Co}}_{x})}_{2}\text{As}_{2}$} single
  crystals},
\newblock \bibinfo{journal}{Phys. Rev. B} \bibinfo{volume}{78}
  (\bibinfo{year}{2008}) \bibinfo{pages}{214515}.
\bibitem[{Chu et~al.(2009)Chu, Analytis, Kucharczyk, and Fisher}]{Chu2009}
\bibinfo{author}{J.-H. Chu}, \bibinfo{author}{J.~G. Analytis},
  \bibinfo{author}{C.~Kucharczyk}, \bibinfo{author}{I.~R. Fisher},
\newblock \bibinfo{title}{Determination of the phase diagram of the
  electron-doped superconductor {B}a({F}e$_{1-x}${C}o$_x$)$_2${A}s$_2$},
\newblock \bibinfo{journal}{Phys. Rev. B} \bibinfo{volume}{79}
  (\bibinfo{year}{2009}) \bibinfo{pages}{014506}.
\bibitem[{Lester et~al.(2009)Lester, Chu, Analytis, Capelli, Erickson, Condron,
  Toney, Fisher, and Hayden}]{Lester2009}
\bibinfo{author}{C.~Lester}, \bibinfo{author}{J.-H. Chu},
  \bibinfo{author}{J.~G. Analytis}, \bibinfo{author}{S.~C. Capelli},
  \bibinfo{author}{A.~S. Erickson}, \bibinfo{author}{C.~L. Condron},
  \bibinfo{author}{M.~F. Toney}, \bibinfo{author}{I.~R. Fisher},
  \bibinfo{author}{S.~M. Hayden},
\newblock \bibinfo{title}{Neutron scattering study of the interplay between
  structure and magnetism in
  {$\text{Ba}{({\text{Fe}}_{1-x}{\text{Co}}_{x})}_{2}{\text{As}}_{2}$}},
\newblock \bibinfo{journal}{Phys. Rev. B} \bibinfo{volume}{79}
  (\bibinfo{year}{2009}) \bibinfo{pages}{144523}.
\bibitem[{Kreyssig et~al.(2010)Kreyssig, Kim, Nandi, Pratt, Tian, Zarestky, Ni,
  Thaler, Bud'ko, Canfield, McQueeney, and Goldman}]{Kreyssig2010}
\bibinfo{author}{A.~Kreyssig}, \bibinfo{author}{M.~G. Kim},
  \bibinfo{author}{S.~Nandi}, \bibinfo{author}{D.~K. Pratt},
  \bibinfo{author}{W.~Tian}, \bibinfo{author}{J.~L. Zarestky},
  \bibinfo{author}{N.~Ni}, \bibinfo{author}{A.~Thaler}, \bibinfo{author}{S.~L.
  Bud'ko}, \bibinfo{author}{P.~C. Canfield}, \bibinfo{author}{R.~J. McQueeney},
  \bibinfo{author}{A.~I. Goldman},
\newblock \bibinfo{title}{Suppression of antiferromagnetic order and
  orthorhombic distortion in superconducting
  {B}a({F}e$_{0.0.961}${R}h$_{0.039}$)$_{2}${A}s$_{2}$},
\newblock \bibinfo{journal}{Phys. Rev. B} \bibinfo{volume}{81}
  (\bibinfo{year}{2010}) \bibinfo{pages}{134512}.
\bibitem[{Ni et~al.(2010)Ni, Thaler, Yan, Kracher, Colombier, Bud'ko, Canfield,
  and Hannahs}]{Ni2010}
\bibinfo{author}{N.~Ni}, \bibinfo{author}{A.~Thaler}, \bibinfo{author}{J.~Q.
  Yan}, \bibinfo{author}{A.~Kracher}, \bibinfo{author}{E.~Colombier},
  \bibinfo{author}{S.~L. Bud'ko}, \bibinfo{author}{P.~C. Canfield},
  \bibinfo{author}{S.~T. Hannahs},
\newblock \bibinfo{title}{Temperature versus doping phase diagrams for
  {Ba(Fe$_{1-x}$TM$_x$)$_2$As$_2$} {(TM=Ni,Cu,Cu/Co)} single crystals},
\newblock \bibinfo{journal}{Phys. Rev. B} \bibinfo{volume}{82}
  (\bibinfo{year}{2010}) \bibinfo{pages}{024519}.
\bibitem[{Luetkens et~al.(2009)Luetkens, Klauss, Kraken, Litterst, Dellmann,
  Klingeler, Hess, Khasanov, Amato, Baines, Kosmala, Schumann, Braden,
  Hamann-Borrero, Leps, Kondrat, Behr, Werner, and B\"uchner}]{Luetkens2009}
\bibinfo{author}{H.~Luetkens}, \bibinfo{author}{H.-H. Klauss},
  \bibinfo{author}{M.~Kraken}, \bibinfo{author}{F.~J. Litterst},
  \bibinfo{author}{T.~Dellmann}, \bibinfo{author}{R.~Klingeler},
  \bibinfo{author}{C.~Hess}, \bibinfo{author}{R.~Khasanov},
  \bibinfo{author}{A.~Amato}, \bibinfo{author}{C.~Baines},
  \bibinfo{author}{M.~Kosmala}, \bibinfo{author}{O.~J. Schumann},
  \bibinfo{author}{M.~Braden}, \bibinfo{author}{J.~Hamann-Borrero},
  \bibinfo{author}{N.~Leps}, \bibinfo{author}{A.~Kondrat},
  \bibinfo{author}{G.~Behr}, \bibinfo{author}{J.~Werner},
  \bibinfo{author}{B.~B\"uchner},
\newblock \bibinfo{title}{The electronic phase diagram of the
  {L}a({O}$_{1-x}${F}$_x$){F}e{A}s superconductor},
\newblock \bibinfo{journal}{Nat. Mater.} \bibinfo{volume}{8}
  (\bibinfo{year}{2009}) \bibinfo{pages}{305--309}.
\bibitem[{Parker et~al.(2010)Parker, Smith, Lancaster, Steele, Franke, Baker,
  Pratt, Pitcher, Blundell, and Clarke}]{Parker2010}
\bibinfo{author}{D.~R. Parker}, \bibinfo{author}{M.~J.~P. Smith},
  \bibinfo{author}{T.~Lancaster}, \bibinfo{author}{A.~J. Steele},
  \bibinfo{author}{I.~Franke}, \bibinfo{author}{P.~J. Baker},
  \bibinfo{author}{F.~L. Pratt}, \bibinfo{author}{M.~J. Pitcher},
  \bibinfo{author}{S.~J. Blundell}, \bibinfo{author}{S.~J. Clarke},
\newblock \bibinfo{title}{Control of the competition between a magnetic phase
  and a superconducting phase in cobalt-doped and nickel-doped {NaFeAs} using
  electron count},
\newblock \bibinfo{journal}{Phys. Rev. Lett.} \bibinfo{volume}{104}
  (\bibinfo{year}{2010}) \bibinfo{pages}{057007}.
\bibitem[{Nandi et~al.(2010)Nandi, Kim, Kreyssig, Fernandes, Pratt, Thaler, Ni,
  Bud'ko, Canfield, Schmalian, McQueeney, and Goldman}]{Nandi2010}
\bibinfo{author}{S.~Nandi}, \bibinfo{author}{M.~G. Kim},
  \bibinfo{author}{A.~Kreyssig}, \bibinfo{author}{R.~M. Fernandes},
  \bibinfo{author}{D.~K. Pratt}, \bibinfo{author}{A.~Thaler},
  \bibinfo{author}{N.~Ni}, \bibinfo{author}{S.~L. Bud'ko},
  \bibinfo{author}{P.~C. Canfield}, \bibinfo{author}{J.~Schmalian},
  \bibinfo{author}{R.~J. McQueeney}, \bibinfo{author}{A.~I. Goldman},
\newblock \bibinfo{title}{Anomalous suppression of the orthorhombic lattice
  distortion in superconducting {Ba(Fe$_{1-x}$Co$_x$)$_2$As$_2$} single
  crystals},
\newblock \bibinfo{journal}{Phys. Rev. Lett.} \bibinfo{volume}{104}
  (\bibinfo{year}{2010}) \bibinfo{pages}{057006}.
\bibitem[{Fernandes et~al.(2010)Fernandes, VanBebber, Bhattacharya, Chandra,
  Keppens, Mandrus, McGuire, Sales, Sefat, and Schmalian}]{Fernandes2010}
\bibinfo{author}{R.~M. Fernandes}, \bibinfo{author}{L.~H. VanBebber},
  \bibinfo{author}{S.~Bhattacharya}, \bibinfo{author}{P.~Chandra},
  \bibinfo{author}{V.~Keppens}, \bibinfo{author}{D.~Mandrus},
  \bibinfo{author}{M.~A. McGuire}, \bibinfo{author}{B.~C. Sales},
  \bibinfo{author}{A.~S. Sefat}, \bibinfo{author}{J.~Schmalian},
\newblock \bibinfo{title}{Effects of nematic fluctuations on the elastic
  properties of iron arsenide superconductors},
\newblock \bibinfo{journal}{Phys. Rev. Lett.} \bibinfo{volume}{105}
  (\bibinfo{year}{2010}) \bibinfo{pages}{157003}.
\bibitem[{McQueen et~al.(2009)McQueen, Williams, Stephens, Tao, Zhu,
  Ksenofontov, Casper, Felser, and Cava}]{McQueen2009}
\bibinfo{author}{T.~M. McQueen}, \bibinfo{author}{A.~J. Williams},
  \bibinfo{author}{P.~W. Stephens}, \bibinfo{author}{J.~Tao},
  \bibinfo{author}{Y.~Zhu}, \bibinfo{author}{V.~Ksenofontov},
  \bibinfo{author}{F.~Casper}, \bibinfo{author}{C.~Felser},
  \bibinfo{author}{R.~J. Cava},
\newblock \bibinfo{title}{Tetragonal-to-orthorhombic structural phase
  transition at 90 {K} in the superconductor {Fe$_{1.01}$Se}},
\newblock \bibinfo{journal}{Phys. Rev. Lett.} \bibinfo{volume}{103}
  (\bibinfo{year}{2009}) \bibinfo{pages}{057002}.
\bibitem[{B{\"o}hmer et~al.(2014)B{\"o}hmer, Burger, Hardy, Wolf, Schweiss,
  Fromknecht, Reinecker, Schranz, and Meingast}]{Boehmer2014}
\bibinfo{author}{A.~E. B{\"o}hmer}, \bibinfo{author}{P.~Burger},
  \bibinfo{author}{F.~Hardy}, \bibinfo{author}{T.~Wolf},
  \bibinfo{author}{P.~Schweiss}, \bibinfo{author}{R.~Fromknecht},
  \bibinfo{author}{M.~Reinecker}, \bibinfo{author}{W.~Schranz},
  \bibinfo{author}{C.~Meingast},
\newblock \bibinfo{title}{Nematic susceptibility of hole-doped and
  electron-doped {BaFe$_2$As$_2$} iron-based superconductors from shear modulus
  measurements},
\newblock \bibinfo{journal}{Phys. Rev. Lett.} \bibinfo{volume}{112}
  (\bibinfo{year}{2014}) \bibinfo{pages}{047001}.
\bibitem[{B{\"{o}}hmer(2014)}]{Boehmerthesis}
\bibinfo{author}{A.~E. B{\"{o}}hmer}, \bibinfo{title}{Competing Phases in
  Iron-Based Superconductors Studied by High-Resolution Thermal-Expansion and
  Shear-Modulus Measurements}, Ph.D. thesis, Fakult{\"{a}}t f{\"{u}}r {P}hysik,
  {K}arlsruhe {I}nstitute of {T}echnology, \bibinfo{year}{2014}. \URLprefix
  \url{http://digbib.ubka.uni-karlsruhe.de/volltexte/1000042623},
  \bibinfo{note}{{K}arlsruhe, {KIT}, {D}iss., 2014}.
\bibitem[{B\"ohmer et~al.(2015)B\"ohmer, Arai, Hardy, Hattori, Iye, Wolf,
  L\"ohneysen, Ishida, and Meingast}]{Boehmer2015}
\bibinfo{author}{A.~E. B\"ohmer}, \bibinfo{author}{T.~Arai},
  \bibinfo{author}{F.~Hardy}, \bibinfo{author}{T.~Hattori},
  \bibinfo{author}{T.~Iye}, \bibinfo{author}{T.~Wolf}, \bibinfo{author}{H.~v.
  L\"ohneysen}, \bibinfo{author}{K.~Ishida}, \bibinfo{author}{C.~Meingast},
\newblock \bibinfo{title}{Origin of the tetragonal-to-orthorhombic phase
  transition in {FeSe}: {A} combined thermodynamic and {NMR} study of
  nematicity},
\newblock \bibinfo{journal}{Phys. Rev. Lett.} \bibinfo{volume}{114}
  (\bibinfo{year}{2015}) \bibinfo{pages}{027001}.
\bibitem[{Fernandes et~al.(2013)Fernandes, B{\"o}hmer, Meingast, and
  Schmalian}]{Fernandes2013}
\bibinfo{author}{R.~M. Fernandes}, \bibinfo{author}{A.~E. B{\"o}hmer},
  \bibinfo{author}{C.~Meingast}, \bibinfo{author}{J.~Schmalian},
\newblock \bibinfo{title}{Scaling between magnetic and lattice fluctuations in
  iron pnictide superconductors},
\newblock \bibinfo{journal}{Phys. Rev. Lett.} \bibinfo{volume}{111}
  (\bibinfo{year}{2013}) \bibinfo{pages}{137001}.
\bibitem[{Chu et~al.(2010)Chu, Analytis, De~Greve, McMahon, Islam, Yamamoto,
  and Fisher}]{Chu2010}
\bibinfo{author}{J.-H. Chu}, \bibinfo{author}{J.~G. Analytis},
  \bibinfo{author}{K.~De~Greve}, \bibinfo{author}{P.~L. McMahon},
  \bibinfo{author}{Z.~Islam}, \bibinfo{author}{Y.~Yamamoto},
  \bibinfo{author}{I.~R. Fisher},
\newblock \bibinfo{title}{In-plane resistivity anisotropy in an underdoped iron
  arsenide superconductor},
\newblock \bibinfo{journal}{Science} \bibinfo{volume}{329}
  (\bibinfo{year}{2010}) \bibinfo{pages}{824--826}.
\bibitem[{Ying et~al.(2011)Ying, Wang, Wu, Xiang, Liu, Yan, Wang, Zhang, Ye,
  Cheng, Hu, and Chen}]{Ying2011}
\bibinfo{author}{J.~J. Ying}, \bibinfo{author}{X.~F. Wang},
  \bibinfo{author}{T.~Wu}, \bibinfo{author}{Z.~J. Xiang},
  \bibinfo{author}{R.~H. Liu}, \bibinfo{author}{Y.~J. Yan},
  \bibinfo{author}{A.~F. Wang}, \bibinfo{author}{M.~Zhang},
  \bibinfo{author}{G.~J. Ye}, \bibinfo{author}{P.~Cheng},
  \bibinfo{author}{J.~P. Hu}, \bibinfo{author}{X.~H. Chen},
\newblock \bibinfo{title}{Measurements of the anisotropic in-plane resistivity
  of underdoped {FeAs}-based pnictide superconductors},
\newblock \bibinfo{journal}{Phys. Rev. Lett.} \bibinfo{volume}{107}
  (\bibinfo{year}{2011}) \bibinfo{pages}{067001}.
\bibitem[{Nakajima et~al.(2012)Nakajima, Ishida, Tomioka, Kihou, Lee, Iyo, Ito,
  Kakeshita, Eisaki, and Uchida}]{Nakajima2012}
\bibinfo{author}{M.~Nakajima}, \bibinfo{author}{S.~Ishida},
  \bibinfo{author}{Y.~Tomioka}, \bibinfo{author}{K.~Kihou},
  \bibinfo{author}{C.~H. Lee}, \bibinfo{author}{A.~Iyo},
  \bibinfo{author}{T.~Ito}, \bibinfo{author}{T.~Kakeshita},
  \bibinfo{author}{H.~Eisaki}, \bibinfo{author}{S.~Uchida},
\newblock \bibinfo{title}{Effect of {Co} doping on the in-plane anisotropy in
  the optical spectrum of underdoped
  {$\mathrm{Ba}({\mathrm{Fe}}_{1-x}{\mathrm{Co}}_{x})_{2}{\mathrm{As}}_{2}$}},
\newblock \bibinfo{journal}{Phys. Rev. Lett.} \bibinfo{volume}{109}
  (\bibinfo{year}{2012}) \bibinfo{pages}{217003}.
\bibitem[{Ishida et~al.(2013{\natexlab{a}})Ishida, Nakajima, Liang, Kihou, Lee,
  Iyo, Eisaki, Kakeshita, Tomioka, Ito, and Uchida}]{Ishida2013}
\bibinfo{author}{S.~Ishida}, \bibinfo{author}{M.~Nakajima},
  \bibinfo{author}{T.~Liang}, \bibinfo{author}{K.~Kihou},
  \bibinfo{author}{C.~H. Lee}, \bibinfo{author}{A.~Iyo},
  \bibinfo{author}{H.~Eisaki}, \bibinfo{author}{T.~Kakeshita},
  \bibinfo{author}{Y.~Tomioka}, \bibinfo{author}{T.~Ito},
  \bibinfo{author}{S.~Uchida},
\newblock \bibinfo{title}{Anisotropy of the in-plane resistivity of underdoped
  {Ba(Fe$_{1-x}$Co$_x$)$_2$As$_2$} superconductors induced by impurity
  scattering in the antiferromagnetic orthorhombic phase},
\newblock \bibinfo{journal}{Phys. Rev. Lett.} \bibinfo{volume}{110}
  (\bibinfo{year}{2013}{\natexlab{a}}) \bibinfo{pages}{207001}.
\bibitem[{Ishida et~al.(2013{\natexlab{b}})Ishida, Nakajima, Liang, Kihou, Lee,
  Iyo, Eisaki, Kakeshita, Tomioka, Ito, and Uchida}]{Ishida2013II}
\bibinfo{author}{S.~Ishida}, \bibinfo{author}{M.~Nakajima},
  \bibinfo{author}{T.~Liang}, \bibinfo{author}{K.~Kihou},
  \bibinfo{author}{C.-H. Lee}, \bibinfo{author}{A.~Iyo},
  \bibinfo{author}{H.~Eisaki}, \bibinfo{author}{T.~Kakeshita},
  \bibinfo{author}{Y.~Tomioka}, \bibinfo{author}{T.~Ito},
  \bibinfo{author}{S.-i. Uchida},
\newblock \bibinfo{title}{Effect of doping on the magnetostructural ordered
  phase of iron arsenides: A comparative study of the resistivity anisotropy in
  doped {BaFe$_2$As$_2$} with doping into three different sites},
\newblock \bibinfo{journal}{Journal of the American Chemical Society}
  \bibinfo{volume}{135} (\bibinfo{year}{2013}{\natexlab{b}})
  \bibinfo{pages}{3158--3163}.
\bibitem[{Blomberg et~al.(2013)Blomberg, Tanatar, Fernandes, Mazin, Shen, Wen,
  Johannes, Schmalian, and Prozorov}]{Blomberg2013}
\bibinfo{author}{E.~C. Blomberg}, \bibinfo{author}{M.~A. Tanatar},
  \bibinfo{author}{R.~M. Fernandes}, \bibinfo{author}{I.~I. Mazin},
  \bibinfo{author}{B.~Shen}, \bibinfo{author}{H.-H. Wen},
  \bibinfo{author}{M.~D. Johannes}, \bibinfo{author}{J.~Schmalian},
  \bibinfo{author}{R.~Prozorov},
\newblock \bibinfo{title}{Sign-reversal of the in-plane resistivity anisotropy
  in hole-doped iron pnictides},
\newblock \bibinfo{journal}{Nature Communications} \bibinfo{volume}{4}
  (\bibinfo{year}{2013}).
\bibitem[{{Liu} et~al.(2015){Liu}, {Mikami}, {Ishida}, {Koshiishi}, {Okazaki},
  {Yoshida}, {Suzuki}, {Horio}, {Ambolode}, {Xu}, {Kumigashira}, {Ono},
  {Nakajima}, {Kihou}, {Lee}, {Iyo}, {Eisaki}, {Kakeshita}, {Uchida}, and
  {Fujimori}}]{Liu2015}
\bibinfo{author}{L.~{Liu}}, \bibinfo{author}{T.~{Mikami}},
  \bibinfo{author}{S.~{Ishida}}, \bibinfo{author}{K.~{Koshiishi}},
  \bibinfo{author}{K.~{Okazaki}}, \bibinfo{author}{T.~{Yoshida}},
  \bibinfo{author}{H.~{Suzuki}}, \bibinfo{author}{M.~{Horio}},
  \bibinfo{author}{L.~C.~C. {Ambolode}, II}, \bibinfo{author}{J.~{Xu}},
  \bibinfo{author}{H.~{Kumigashira}}, \bibinfo{author}{K.~{Ono}},
  \bibinfo{author}{M.~{Nakajima}}, \bibinfo{author}{K.~{Kihou}},
  \bibinfo{author}{C.~H. {Lee}}, \bibinfo{author}{A.~{Iyo}},
  \bibinfo{author}{H.~{Eisaki}}, \bibinfo{author}{T.~{Kakeshita}},
  \bibinfo{author}{S.~{Uchida}}, \bibinfo{author}{A.~{Fujimori}},
\newblock \bibinfo{title}{In-plane electronic anisotropy in the
  antiferromagnetic-orthorhombic phase of isovalent-substituted
  {Ba(Fe$_{1-x}$Ru$_x$)$_2$As$_2$}},
\newblock \bibinfo{journal}{ArXiv e-prints}  (\bibinfo{year}{2015})
  \bibinfo{pages}{1503.02855}.
\bibitem[{Fernandes et~al.(2011)Fernandes, Abrahams, and
  Schmalian}]{Fernandes2011}
\bibinfo{author}{R.~M. Fernandes}, \bibinfo{author}{E.~Abrahams},
  \bibinfo{author}{J.~Schmalian},
\newblock \bibinfo{title}{Anisotropic in-plane resistivity in the nematic phase
  of the iron pnictides},
\newblock \bibinfo{journal}{Phys. Rev. Lett.} \bibinfo{volume}{107}
  (\bibinfo{year}{2011}) \bibinfo{pages}{217002}.
\bibitem[{Gastiasoro et~al.(2014)Gastiasoro, Paul, Wang, Hirschfeld, and
  Andersen}]{Gastiasoro2014}
\bibinfo{author}{M.~N. Gastiasoro}, \bibinfo{author}{I.~Paul},
  \bibinfo{author}{Y.~Wang}, \bibinfo{author}{P.~J. Hirschfeld},
  \bibinfo{author}{B.~M. Andersen},
\newblock \bibinfo{title}{Emergent defect states as a source of resistivity
  anisotropy in the nematic phase of iron pnictides},
\newblock \bibinfo{journal}{Phys. Rev. Lett.} \bibinfo{volume}{113}
  (\bibinfo{year}{2014}) \bibinfo{pages}{127001}.
\bibitem[{Song et~al.(2012)Song, Wang, Jiang, Wang, He, Chen, Hoffman, Ma, and
  Xue}]{Song2012}
\bibinfo{author}{C.-L. Song}, \bibinfo{author}{Y.-L. Wang},
  \bibinfo{author}{Y.-P. Jiang}, \bibinfo{author}{L.~Wang},
  \bibinfo{author}{K.~He}, \bibinfo{author}{X.~Chen}, \bibinfo{author}{J.~E.
  Hoffman}, \bibinfo{author}{X.-C. Ma}, \bibinfo{author}{Q.-K. Xue},
\newblock \bibinfo{title}{Suppression of superconductivity by twin boundaries
  in {FeSe}},
\newblock \bibinfo{journal}{Phys. Rev. Lett.} \bibinfo{volume}{109}
  (\bibinfo{year}{2012}) \bibinfo{pages}{137004}.
\bibitem[{Allan et~al.(2013)Allan, Chuang, Massee, Xie, Ni, Bud/'ko, Boebinger,
  Wang, Dessau, Canfield, Golden, and Davis}]{Allen2013}
\bibinfo{author}{M.~P. Allan}, \bibinfo{author}{T.-M. Chuang},
  \bibinfo{author}{F.~Massee}, \bibinfo{author}{Y.~Xie},
  \bibinfo{author}{N.~Ni}, \bibinfo{author}{S.~L. Bud/'ko},
  \bibinfo{author}{G.~S. Boebinger}, \bibinfo{author}{Q.~Wang},
  \bibinfo{author}{D.~S. Dessau}, \bibinfo{author}{P.~C. Canfield},
  \bibinfo{author}{M.~S. Golden}, \bibinfo{author}{J.~C. Davis},
\newblock \bibinfo{title}{Anisotropic impurity states, quasiparticle scattering
  and nematic transport in underdoped {Ca(Fe$_{1-x}$Co$_x$)$_2$As$_2$}},
\newblock \bibinfo{journal}{Nature Physics} \bibinfo{volume}{9}
  (\bibinfo{year}{2013}) \bibinfo{pages}{220--224}.
\bibitem[{Rosenthal et~al.(2014)Rosenthal, Andrade, Arguello, Fernandes, Xing,
  Wang, Jin, Millis, and Pasupathy}]{Rosenthal2014}
\bibinfo{author}{E.~P. Rosenthal}, \bibinfo{author}{E.~F. Andrade},
  \bibinfo{author}{C.~J. Arguello}, \bibinfo{author}{R.~M. Fernandes},
  \bibinfo{author}{L.~Y. Xing}, \bibinfo{author}{X.~C. Wang},
  \bibinfo{author}{C.~Q. Jin}, \bibinfo{author}{A.~J. Millis},
  \bibinfo{author}{A.~N. Pasupathy},
\newblock \bibinfo{title}{Visualization of electron nematicity and
  unidirectional antiferroic fluctuations at high temperatures in {NaFeAs}},
\newblock \bibinfo{journal}{Nature Physics} \bibinfo{volume}{10}
  (\bibinfo{year}{2014}) \bibinfo{pages}{225--232}.
\bibitem[{{Deng} et~al.(2015){Deng}, {Xing}, {Liu}, {Yang}, and
  {Wen}}]{Deng2015}
\bibinfo{author}{Q.~{Deng}}, \bibinfo{author}{J.~{Xing}},
  \bibinfo{author}{J.~{Liu}}, \bibinfo{author}{H.~{Yang}},
  \bibinfo{author}{H.-H. {Wen}},
\newblock \bibinfo{title}{Anisotropic electronic mobilities in the nematic
  state of the parent phase {NaFeAs}},
\newblock \bibinfo{journal}{ArXiv e-prints}  (\bibinfo{year}{2015})
  \bibinfo{pages}{1503.07090}.
\bibitem[{Dusza et~al.(2011)Dusza, Lucarelli, Pfuner, Chu, Fisher, and
  Degiorgi}]{Dusza2011}
\bibinfo{author}{A.~Dusza}, \bibinfo{author}{A.~Lucarelli},
  \bibinfo{author}{F.~Pfuner}, \bibinfo{author}{J.-H. Chu},
  \bibinfo{author}{I.~R. Fisher}, \bibinfo{author}{L.~Degiorgi},
\newblock \bibinfo{title}{Anisotropic charge dynamics in detwinned
  {Ba(Fe$_{1-x}$Co$_x$)$_2$As$_2$}},
\newblock \bibinfo{journal}{EPL (Europhysics Letters)} \bibinfo{volume}{93}
  (\bibinfo{year}{2011}) \bibinfo{pages}{37002}.
\bibitem[{Mirri et~al.(2014{\natexlab{a}})Mirri, Dusza, Bastelberger, Chu, Kuo,
  Fisher, and Degiorgi}]{Mirri2014}
\bibinfo{author}{C.~Mirri}, \bibinfo{author}{A.~Dusza},
  \bibinfo{author}{S.~Bastelberger}, \bibinfo{author}{J.-H. Chu},
  \bibinfo{author}{H.-H. Kuo}, \bibinfo{author}{I.~R. Fisher},
  \bibinfo{author}{L.~Degiorgi},
\newblock \bibinfo{title}{Hysteretic behavior in the optical response of the
  underdoped {Fe}-arsenide {Ba(Fe$_{1-x}$Co$_x$)$_2$As$_2$} in the electronic
  nematic phase},
\newblock \bibinfo{journal}{Phys. Rev. B} \bibinfo{volume}{89}
  (\bibinfo{year}{2014}{\natexlab{a}}) \bibinfo{pages}{060501}.
\bibitem[{Mirri et~al.(2014{\natexlab{b}})Mirri, Dusza, Bastelberger, Chu, Kuo,
  Fisher, and Degiorgi}]{Mirri2014II}
\bibinfo{author}{C.~Mirri}, \bibinfo{author}{A.~Dusza},
  \bibinfo{author}{S.~Bastelberger}, \bibinfo{author}{J.-H. Chu},
  \bibinfo{author}{H.-H. Kuo}, \bibinfo{author}{I.~R. Fisher},
  \bibinfo{author}{L.~Degiorgi},
\newblock \bibinfo{title}{Nematic-driven anisotropic electronic properties of
  underdoped detwinned {Ba(Fe$_{1-x}$Co$_x$)$_2$As$_2$} revealed by optical
  spectroscopy},
\newblock \bibinfo{journal}{Phys. Rev. B} \bibinfo{volume}{90}
  (\bibinfo{year}{2014}{\natexlab{b}}) \bibinfo{pages}{155125}.
\bibitem[{{Mirri} et~al.(2015){Mirri}, {Dusza}, {Bastelberger}, {Chinotti},
  {Chu}, {Kuo}, {Fisher}, and {Degiorgi}}]{Mirri2015}
\bibinfo{author}{C.~{Mirri}}, \bibinfo{author}{A.~{Dusza}},
  \bibinfo{author}{S.~{Bastelberger}}, \bibinfo{author}{M.~{Chinotti}},
  \bibinfo{author}{J.-H. {Chu}}, \bibinfo{author}{H.-H. {Kuo}},
  \bibinfo{author}{I.~R. {Fisher}}, \bibinfo{author}{L.~{Degiorgi}},
\newblock \bibinfo{title}{Origin of the resistive anisotropy in the electronic
  nematic phase of {BaFe$_2$As$_2$} revealed by optical spectroscopy},
\newblock \bibinfo{journal}{ArXiv e-prints}  (\bibinfo{year}{2015})
  \bibinfo{pages}{1504.06829}.
\bibitem[{Kuo and Fisher(2014)}]{Kuo2014}
\bibinfo{author}{H.~Kuo}, \bibinfo{author}{I.~R. Fisher},
\newblock \bibinfo{title}{Effect of disorder on the resistivity anisotropy near
  the electronic nematic phase transition in pure and electron-doped
  {BaFe$_{2}$As$_{2}$}},
\newblock \bibinfo{journal}{Phys. Rev. Lett.} \bibinfo{volume}{112}
  (\bibinfo{year}{2014}) \bibinfo{pages}{227001}.
\bibitem[{Yi et~al.(2011)Yi, Lu, Chu, Analytis, Sorini, Kemper, Moritz, Mo,
  Moore, Hashimoto, Lee, Hussain, Devereaux, Fisher, and Shen}]{Yi2011}
\bibinfo{author}{M.~Yi}, \bibinfo{author}{D.~Lu}, \bibinfo{author}{J.-H. Chu},
  \bibinfo{author}{J.~G. Analytis}, \bibinfo{author}{A.~P. Sorini},
  \bibinfo{author}{A.~F. Kemper}, \bibinfo{author}{B.~Moritz},
  \bibinfo{author}{S.-K. Mo}, \bibinfo{author}{R.~G. Moore},
  \bibinfo{author}{M.~Hashimoto}, \bibinfo{author}{W.-S. Lee},
  \bibinfo{author}{Z.~Hussain}, \bibinfo{author}{T.~P. Devereaux},
  \bibinfo{author}{I.~R. Fisher}, \bibinfo{author}{Z.-X. Shen},
\newblock \bibinfo{title}{Symmetry-breaking orbital anisotropy observed for
  detwinned {Ba(Fe$_{1-x}$Co$_x$)$_2$As$_2$} above the spin density wave
  transition},
\newblock \bibinfo{journal}{Proceedings of the National Academy of Sciences}
  \bibinfo{volume}{108} (\bibinfo{year}{2011}) \bibinfo{pages}{6878--6883}.
\bibitem[{Jiang et~al.(2013)Jiang, Jeevan, Dong, and Gegenwart}]{Jiang2013}
\bibinfo{author}{S.~Jiang}, \bibinfo{author}{H.~S. Jeevan},
  \bibinfo{author}{J.~Dong}, \bibinfo{author}{P.~Gegenwart},
\newblock \bibinfo{title}{Thermopower as a sensitive probe of electronic
  nematicity in iron pnictides},
\newblock \bibinfo{journal}{Phys. Rev. Lett.} \bibinfo{volume}{110}
  (\bibinfo{year}{2013}) \bibinfo{pages}{067001}.
\bibitem[{Fu et~al.(2012)Fu, Torchetti, Imai, Ning, Yan, and Sefat}]{Fu2012}
\bibinfo{author}{M.~Fu}, \bibinfo{author}{D.~A. Torchetti},
  \bibinfo{author}{T.~Imai}, \bibinfo{author}{F.~L. Ning},
  \bibinfo{author}{J.-Q. Yan}, \bibinfo{author}{A.~S. Sefat},
\newblock \bibinfo{title}{{NMR} search for the spin nematic state in a
  {LaFeAsO} single crystal},
\newblock \bibinfo{journal}{Phys. Rev. Lett.} \bibinfo{volume}{109}
  (\bibinfo{year}{2012}) \bibinfo{pages}{247001}.
\bibitem[{Harriger et~al.(2011)Harriger, Luo, Liu, Frost, Hu, Norman, and
  Dai}]{Harriger2011}
\bibinfo{author}{L.~W. Harriger}, \bibinfo{author}{H.~Q. Luo},
  \bibinfo{author}{M.~S. Liu}, \bibinfo{author}{C.~Frost},
  \bibinfo{author}{J.~P. Hu}, \bibinfo{author}{M.~R. Norman},
  \bibinfo{author}{P.~Dai},
\newblock \bibinfo{title}{Nematic spin fluid in the tetragonal phase of
  {BaFe$_{2}$As$_{2}$}},
\newblock \bibinfo{journal}{Phys. Rev. B} \bibinfo{volume}{84}
  (\bibinfo{year}{2011}) \bibinfo{pages}{054544}.
\bibitem[{Lu et~al.(2014)Lu, Park, Zhang, Luo, Nevidomskyy, Si, and
  Dai}]{Lu2014}
\bibinfo{author}{X.~Lu}, \bibinfo{author}{J.~T. Park},
  \bibinfo{author}{R.~Zhang}, \bibinfo{author}{H.~Luo}, \bibinfo{author}{A.~H.
  Nevidomskyy}, \bibinfo{author}{Q.~Si}, \bibinfo{author}{P.~Dai},
\newblock \bibinfo{title}{Nematic spin correlations in the tetragonal state of
  uniaxial-strained {BaFe$_{2-x}$Ni$_x$As$_2$}},
\newblock \bibinfo{journal}{Science} \bibinfo{volume}{345}
  (\bibinfo{year}{2014}) \bibinfo{pages}{657--660}.
\bibitem[{Inosov(2015)}]{Inosov2015}
\bibinfo{author}{D.~Inosov},
\newblock \bibinfo{title}{Spin fluctuations in iron pnictides and
  chalcogenides: {F}rom antiferromagnetism to superconductivity.},
\newblock \bibinfo{journal}{C. R. Physique}  (\bibinfo{year}{2015}).
  \bibinfo{note}{(in press)}.
\bibitem[{Chu et~al.(2012)Chu, Kuo, Analytis, and Fisher}]{Chu2012}
\bibinfo{author}{J.-H. Chu}, \bibinfo{author}{H.-H. Kuo},
  \bibinfo{author}{J.~G. Analytis}, \bibinfo{author}{I.~R. Fisher},
\newblock \bibinfo{title}{Divergent nematic susceptibility in an iron arsenide
  superconductor},
\newblock \bibinfo{journal}{Science} \bibinfo{volume}{337}
  (\bibinfo{year}{2012}) \bibinfo{pages}{710--712}.
\bibitem[{Patz et~al.(2014)Patz, Li, Ran, Fernandes, Schmalian, Bud{'}ko,
  Canfield, Perakis, and Wang}]{Patz2014}
\bibinfo{author}{A.~Patz}, \bibinfo{author}{T.~Li}, \bibinfo{author}{S.~Ran},
  \bibinfo{author}{R.~M. Fernandes}, \bibinfo{author}{J.~Schmalian},
  \bibinfo{author}{S.~L. Bud{'}ko}, \bibinfo{author}{P.~C. Canfield},
  \bibinfo{author}{I.~E. Perakis}, \bibinfo{author}{J.~Wang},
\newblock \bibinfo{title}{Ultrafast observation of critical nematic
  fluctuations and giant magnetoelastic coupling in iron pnictides},
\newblock \bibinfo{journal}{Nat Commun} \bibinfo{volume}{5}
  (\bibinfo{year}{2014}) \bibinfo{pages}{10.1038}.
\bibitem[{Fisher et~al.(2011)Fisher, Degiorgi, and Shen}]{Fisher2011}
\bibinfo{author}{I.~R. Fisher}, \bibinfo{author}{L.~Degiorgi},
  \bibinfo{author}{Z.~X. Shen},
\newblock \bibinfo{title}{In-plane electronic anisotropy of underdoped '122'
  {Fe}-arsenide superconductors revealed by measurements of detwinned single
  crystals},
\newblock \bibinfo{journal}{Reports on Progress in Physics}
  \bibinfo{volume}{74} (\bibinfo{year}{2011}) \bibinfo{pages}{124506}.
\bibitem[{Ruff et~al.(2012)Ruff, Chu, Kuo, Das, Nojiri, Fisher, and
  Islam}]{Ruff2012}
\bibinfo{author}{J.~P.~C. Ruff}, \bibinfo{author}{J.-H. Chu},
  \bibinfo{author}{H.-H. Kuo}, \bibinfo{author}{R.~K. Das},
  \bibinfo{author}{H.~Nojiri}, \bibinfo{author}{I.~R. Fisher},
  \bibinfo{author}{Z.~Islam},
\newblock \bibinfo{title}{Susceptibility anisotropy in an iron arsenide
  superconductor revealed by x-ray diffraction in pulsed magnetic fields},
\newblock \bibinfo{journal}{Phys. Rev. Lett.} \bibinfo{volume}{109}
  (\bibinfo{year}{2012}) \bibinfo{pages}{027004}.
\bibitem[{Zapf et~al.(2014)Zapf, Stingl, Post, Maiwald, Bach, Pietsch,
  Neubauer, L{\"o}hle, Clauss, Jiang, Jeevan, Basov, Gegenwart, and
  Dressel}]{Zapf2014}
\bibinfo{author}{S.~Zapf}, \bibinfo{author}{C.~Stingl}, \bibinfo{author}{K.~W.
  Post}, \bibinfo{author}{J.~Maiwald}, \bibinfo{author}{N.~Bach},
  \bibinfo{author}{I.~Pietsch}, \bibinfo{author}{D.~Neubauer},
  \bibinfo{author}{A.~L{\"o}hle}, \bibinfo{author}{C.~Clauss},
  \bibinfo{author}{S.~Jiang}, \bibinfo{author}{H.~S. Jeevan},
  \bibinfo{author}{D.~N. Basov}, \bibinfo{author}{P.~Gegenwart},
  \bibinfo{author}{M.~Dressel},
\newblock \bibinfo{title}{Persistent detwinning of iron-pnictide
  {EuFe$_{2}$As$_{2}$} crystals by small external magnetic fields},
\newblock \bibinfo{journal}{Phys. Rev. Lett.} \bibinfo{volume}{113}
  (\bibinfo{year}{2014}) \bibinfo{pages}{227001}.
\bibitem[{Blomberg et~al.(2012)Blomberg, Kreyssig, Tanatar, Fernandes, Kim,
  Thaler, Schmalian, Bud'ko, Canfield, Goldman, and Prozorov}]{Blomberg2012}
\bibinfo{author}{E.~C. Blomberg}, \bibinfo{author}{A.~Kreyssig},
  \bibinfo{author}{M.~A. Tanatar}, \bibinfo{author}{R.~M. Fernandes},
  \bibinfo{author}{M.~G. Kim}, \bibinfo{author}{A.~Thaler},
  \bibinfo{author}{J.~Schmalian}, \bibinfo{author}{S.~L. Bud'ko},
  \bibinfo{author}{P.~C. Canfield}, \bibinfo{author}{A.~I. Goldman},
  \bibinfo{author}{R.~Prozorov},
\newblock \bibinfo{title}{Effect of tensile stress on the in-plane resistivity
  anisotropy in {BaFe$_2$As$_2$}},
\newblock \bibinfo{journal}{Phys. Rev. B} \bibinfo{volume}{85}
  (\bibinfo{year}{2012}) \bibinfo{pages}{144509}.
\bibitem[{Dhital et~al.(2012)Dhital, Yamani, Tian, Zeretsky, Sefat, Wang,
  Birgeneau, and Wilson}]{Dhital2012}
\bibinfo{author}{C.~Dhital}, \bibinfo{author}{Z.~Yamani},
  \bibinfo{author}{W.~Tian}, \bibinfo{author}{J.~Zeretsky},
  \bibinfo{author}{A.~S. Sefat}, \bibinfo{author}{Z.~Wang},
  \bibinfo{author}{R.~J. Birgeneau}, \bibinfo{author}{S.~D. Wilson},
\newblock \bibinfo{title}{Effect of uniaxial strain on the structural and
  magnetic phase transitions in {BaFe$_2$As$_2$}},
\newblock \bibinfo{journal}{Phys. Rev. Lett.} \bibinfo{volume}{108}
  (\bibinfo{year}{2012}) \bibinfo{pages}{087001}.
\bibitem[{Hu et~al.(2012)Hu, Setty, and Kivelson}]{Hu2012}
\bibinfo{author}{J.~Hu}, \bibinfo{author}{C.~Setty},
  \bibinfo{author}{S.~Kivelson},
\newblock \bibinfo{title}{Pressure effects on magnetically driven electronic
  nematic states in iron pnictide superconductors},
\newblock \bibinfo{journal}{Phys. Rev. B} \bibinfo{volume}{85}
  (\bibinfo{year}{2012}) \bibinfo{pages}{100507}.
\bibitem[{Gallais et~al.(2013)Gallais, Fernandes, Paul, Chauvi\`ere, Yang,
  M\'easson, Cazayous, Sacuto, Colson, and Forget}]{Gallais2014}
\bibinfo{author}{Y.~Gallais}, \bibinfo{author}{R.~M. Fernandes},
  \bibinfo{author}{I.~Paul}, \bibinfo{author}{L.~Chauvi\`ere},
  \bibinfo{author}{Y.-X. Yang}, \bibinfo{author}{M.-A. M\'easson},
  \bibinfo{author}{M.~Cazayous}, \bibinfo{author}{A.~Sacuto},
  \bibinfo{author}{D.~Colson}, \bibinfo{author}{A.~Forget},
\newblock \bibinfo{title}{Observation of incipient charge nematicity in
  {Ba(Fe$_{1-x}$Co$_x$)$_2$As$_2$}},
\newblock \bibinfo{journal}{Phys. Rev. Lett.} \bibinfo{volume}{111}
  (\bibinfo{year}{2013}) \bibinfo{pages}{267001}.
\bibitem[{Kontani et~al.(2011)Kontani, Saito, and Onari}]{Kontani2011}
\bibinfo{author}{H.~Kontani}, \bibinfo{author}{T.~Saito},
  \bibinfo{author}{S.~Onari},
\newblock \bibinfo{title}{Origin of orthorhombic transition, magnetic
  transition, and shear-modulus softening in iron pnictide superconductors:
  {A}nalysis based on the orbital fluctuations theory},
\newblock \bibinfo{journal}{Phys. Rev. B} \bibinfo{volume}{84}
  (\bibinfo{year}{2011}) \bibinfo{pages}{024528}.
\bibitem[{Yamase and Zeyher(2013)}]{Yamase2013}
\bibinfo{author}{H.~Yamase}, \bibinfo{author}{R.~Zeyher},
\newblock \bibinfo{title}{Superconductivity from orbital nematic fluctuations},
\newblock \bibinfo{journal}{Phys. Rev. B} \bibinfo{volume}{88}
  (\bibinfo{year}{2013}) \bibinfo{pages}{180502}.
\bibitem[{Yoshizawa and Simayi(2012)}]{Yoshizawa2012II}
\bibinfo{author}{M.~Yoshizawa}, \bibinfo{author}{S.~Simayi},
\newblock \bibinfo{title}{Anomalous elastic behavior and its correlation with
  superconductivity in iron-based superconductor
  {Ba(Fe$_{1-x}$Co$_x$)$_2$As$_2$}},
\newblock \bibinfo{journal}{Modern Physics Letters B} \bibinfo{volume}{26}
  (\bibinfo{year}{2012}) \bibinfo{pages}{1230011}.
\bibitem[{Margadonna et~al.(2009)Margadonna, Takabayashi, Ohishi, Mizuguchi,
  Takano, Kagayama, Nakagawa, Takata, and Prassides}]{Margadonna2009}
\bibinfo{author}{S.~Margadonna}, \bibinfo{author}{Y.~Takabayashi},
  \bibinfo{author}{Y.~Ohishi}, \bibinfo{author}{Y.~Mizuguchi},
  \bibinfo{author}{Y.~Takano}, \bibinfo{author}{T.~Kagayama},
  \bibinfo{author}{T.~Nakagawa}, \bibinfo{author}{M.~Takata},
  \bibinfo{author}{K.~Prassides},
\newblock \bibinfo{title}{Pressure evolution of the low-temperature crystal
  structure and bonding of the superconductor {FeSe} ({$T_c=37$ K})},
\newblock \bibinfo{journal}{Phys. Rev. B} \bibinfo{volume}{80}
  (\bibinfo{year}{2009}) \bibinfo{pages}{064506}.
\bibitem[{Bendele et~al.(2010)Bendele, Amato, Conder, Elender, Keller, Klauss,
  Luetkens, Pomjakushina, Raselli, and Khasanov}]{Bendele2010}
\bibinfo{author}{M.~Bendele}, \bibinfo{author}{A.~Amato},
  \bibinfo{author}{K.~Conder}, \bibinfo{author}{M.~Elender},
  \bibinfo{author}{H.~Keller}, \bibinfo{author}{H.-H. Klauss},
  \bibinfo{author}{H.~Luetkens}, \bibinfo{author}{E.~Pomjakushina},
  \bibinfo{author}{A.~Raselli}, \bibinfo{author}{R.~Khasanov},
\newblock \bibinfo{title}{Pressure induced static magnetic order in
  superconducting {FeSe$_{1-x}$}},
\newblock \bibinfo{journal}{Phys. Rev. Lett.} \bibinfo{volume}{104}
  (\bibinfo{year}{2010}) \bibinfo{pages}{087003}.
\bibitem[{Imai et~al.(2009)Imai, Ahilan, Ning, McQueen, and Cava}]{Imai2009}
\bibinfo{author}{T.~Imai}, \bibinfo{author}{K.~Ahilan}, \bibinfo{author}{F.~L.
  Ning}, \bibinfo{author}{T.~M. McQueen}, \bibinfo{author}{R.~J. Cava},
\newblock \bibinfo{title}{Why does undoped {FeSe} become a high-{$T_c$}
  superconductor under pressure?},
\newblock \bibinfo{journal}{Phys. Rev. Lett.} \bibinfo{volume}{102}
  (\bibinfo{year}{2009}) \bibinfo{pages}{177005}.
\bibitem[{B\"ohmer et~al.(2013)B\"ohmer, Hardy, Eilers, Ernst, Adelmann,
  Schweiss, Wolf, and Meingast}]{Boehmer2013}
\bibinfo{author}{A.~E. B\"ohmer}, \bibinfo{author}{F.~Hardy},
  \bibinfo{author}{F.~Eilers}, \bibinfo{author}{D.~Ernst},
  \bibinfo{author}{P.~Adelmann}, \bibinfo{author}{P.~Schweiss},
  \bibinfo{author}{T.~Wolf}, \bibinfo{author}{C.~Meingast},
\newblock \bibinfo{title}{Lack of coupling between superconductivity and
  orthorhombic distortion in stoichiometric single-crystalline {FeSe}},
\newblock \bibinfo{journal}{Phys. Rev. B} \bibinfo{volume}{87}
  (\bibinfo{year}{2013}) \bibinfo{pages}{180505}.
\bibitem[{Lin et~al.(2011)Lin, Hsieh, Chareev, Vasiliev, Parsons, and
  Yang}]{Lin2011}
\bibinfo{author}{J.-Y. Lin}, \bibinfo{author}{Y.~S. Hsieh},
  \bibinfo{author}{D.~A. Chareev}, \bibinfo{author}{A.~N. Vasiliev},
  \bibinfo{author}{Y.~Parsons}, \bibinfo{author}{H.~D. Yang},
\newblock \bibinfo{title}{Coexistence of isotropic and extended $s$-wave order
  parameter in {FeSe} as revealed by low-temperature specific heat},
\newblock \bibinfo{journal}{Phys. Rev. B} \bibinfo{volume}{84}
  (\bibinfo{year}{2011}) \bibinfo{pages}{220507(R)}.
\bibitem[{Chareev et~al.(2013)Chareev, Osadchii, Kuzmicheva, Lin, Kuzmichev,
  Volkova, and Vasiliev}]{Chareev2013}
\bibinfo{author}{D.~Chareev}, \bibinfo{author}{E.~Osadchii},
  \bibinfo{author}{T.~Kuzmicheva}, \bibinfo{author}{J.-Y. Lin},
  \bibinfo{author}{S.~Kuzmichev}, \bibinfo{author}{O.~Volkova},
  \bibinfo{author}{A.~Vasiliev},
\newblock \bibinfo{title}{Single crystal growth and characterization of
  tetragonal {FeSe$_{1-x}$} superconductors},
\newblock \bibinfo{journal}{CrystEngComm} \bibinfo{volume}{15}
  (\bibinfo{year}{2013}) \bibinfo{pages}{1989--1993}.
\bibitem[{Baek et~al.(2015)Baek, Efremov, Ok, Kim, van~den Brink, and
  B{\"u}chner}]{Baek2015}
\bibinfo{author}{S.-H. Baek}, \bibinfo{author}{D.~V. Efremov},
  \bibinfo{author}{J.~M. Ok}, \bibinfo{author}{J.~S. Kim},
  \bibinfo{author}{J.~van~den Brink}, \bibinfo{author}{B.~B{\"u}chner},
\newblock \bibinfo{title}{Orbital-driven nematicity in {FeSe}},
\newblock \bibinfo{journal}{Nature Materials} \bibinfo{volume}{14}
  (\bibinfo{year}{2015}) \bibinfo{pages}{210--214}.
\bibitem[{Watson et~al.(2015)Watson, Kim, Haghighirad, Davies, McCollam,
  Narayanan, Blake, Chen, Ghannadzadeh, Schofield, Hoesch, Meingast, Wolf, and
  Coldea}]{Watson2015}
\bibinfo{author}{M.~D. Watson}, \bibinfo{author}{T.~K. Kim},
  \bibinfo{author}{A.~A. Haghighirad}, \bibinfo{author}{N.~R. Davies},
  \bibinfo{author}{A.~McCollam}, \bibinfo{author}{A.~Narayanan},
  \bibinfo{author}{S.~F. Blake}, \bibinfo{author}{Y.~L. Chen},
  \bibinfo{author}{S.~Ghannadzadeh}, \bibinfo{author}{A.~J. Schofield},
  \bibinfo{author}{M.~Hoesch}, \bibinfo{author}{C.~Meingast},
  \bibinfo{author}{T.~Wolf}, \bibinfo{author}{A.~I. Coldea},
\newblock \bibinfo{title}{Emergence of the nematic electronic state in {FeSe}},
\newblock \bibinfo{journal}{Phys. Rev. B} \bibinfo{volume}{91}
  (\bibinfo{year}{2015}) \bibinfo{pages}{155106}.
\bibitem[{Shimojima et~al.(2014)Shimojima, Suzuki, Sonobe, Nakamura, Sakano,
  Omachi, Yoshioka, Kuwata-Gonokami, Ono, Kumigashira, B\"ohmer, Hardy, Wolf,
  Meingast, L\"ohneysen, Ikeda, and Ishizaka}]{Shimojima2014}
\bibinfo{author}{T.~Shimojima}, \bibinfo{author}{Y.~Suzuki},
  \bibinfo{author}{T.~Sonobe}, \bibinfo{author}{A.~Nakamura},
  \bibinfo{author}{M.~Sakano}, \bibinfo{author}{J.~Omachi},
  \bibinfo{author}{K.~Yoshioka}, \bibinfo{author}{M.~Kuwata-Gonokami},
  \bibinfo{author}{K.~Ono}, \bibinfo{author}{H.~Kumigashira},
  \bibinfo{author}{A.~E. B\"ohmer}, \bibinfo{author}{F.~Hardy},
  \bibinfo{author}{T.~Wolf}, \bibinfo{author}{C.~Meingast},
  \bibinfo{author}{H.~v. L\"ohneysen}, \bibinfo{author}{H.~Ikeda},
  \bibinfo{author}{K.~Ishizaka},
\newblock \bibinfo{title}{Lifting of $xz/yz$ orbital degeneracy at the
  structural transition in detwinned {FeSe}},
\newblock \bibinfo{journal}{Phys. Rev. B} \bibinfo{volume}{90}
  (\bibinfo{year}{2014}) \bibinfo{pages}{121111}.
\bibitem[{Maletz et~al.(2014)Maletz, Zabolotnyy, Evtushinsky, Thirupathaiah,
  Wolter, Harnagea, Yaresko, Vasiliev, Chareev, B\"ohmer, Hardy, Wolf,
  Meingast, Rienks, B\"uchner, and Borisenko}]{Maletz2014}
\bibinfo{author}{J.~Maletz}, \bibinfo{author}{V.~B. Zabolotnyy},
  \bibinfo{author}{D.~V. Evtushinsky}, \bibinfo{author}{S.~Thirupathaiah},
  \bibinfo{author}{A.~U.~B. Wolter}, \bibinfo{author}{L.~Harnagea},
  \bibinfo{author}{A.~N. Yaresko}, \bibinfo{author}{A.~N. Vasiliev},
  \bibinfo{author}{D.~A. Chareev}, \bibinfo{author}{A.~E. B\"ohmer},
  \bibinfo{author}{F.~Hardy}, \bibinfo{author}{T.~Wolf},
  \bibinfo{author}{C.~Meingast}, \bibinfo{author}{E.~D.~L. Rienks},
  \bibinfo{author}{B.~B\"uchner}, \bibinfo{author}{S.~V. Borisenko},
\newblock \bibinfo{title}{Unusual band renormalization in the simplest
  iron-based superconductor {FeSe$_{1-x}$}},
\newblock \bibinfo{journal}{Phys. Rev. B} \bibinfo{volume}{89}
  (\bibinfo{year}{2014}) \bibinfo{pages}{220506}.
\bibitem[{Nakayama et~al.(2014)Nakayama, Miyata, Phan, Sato, Tanabe, Urata,
  Tanigaki, and Takahashi}]{Nakayama2014}
\bibinfo{author}{K.~Nakayama}, \bibinfo{author}{Y.~Miyata},
  \bibinfo{author}{G.~Phan}, \bibinfo{author}{T.~Sato},
  \bibinfo{author}{Y.~Tanabe}, \bibinfo{author}{T.~Urata},
  \bibinfo{author}{K.~Tanigaki}, \bibinfo{author}{T.~Takahashi},
\newblock \bibinfo{title}{Reconstruction of band structure induced by
  electronic nematicity in an {FeSe} superconductor},
\newblock \bibinfo{journal}{Phys. Rev. Lett.} \bibinfo{volume}{113}
  (\bibinfo{year}{2014}) \bibinfo{pages}{237001}.
\bibitem[{Rahn et~al.(2015)Rahn, Ewings, Sedlmaier, Clarke, and
  Boothroyd}]{Rahn2015}
\bibinfo{author}{M.~C. Rahn}, \bibinfo{author}{R.~A. Ewings},
  \bibinfo{author}{S.~J. Sedlmaier}, \bibinfo{author}{S.~J. Clarke},
  \bibinfo{author}{A.~T. Boothroyd},
\newblock \bibinfo{title}{Strong {$(\pi,0)$} spin fluctuations in
  $\beta-${FeSe} observed by neutron spectroscopy},
\newblock \bibinfo{journal}{Phys. Rev. B} \bibinfo{volume}{91}
  (\bibinfo{year}{2015}) \bibinfo{pages}{180501}.
\bibitem[{{Wang} et~al.(2015){Wang}, {Shen}, {Pan}, {Hao}, {Ma}, {Zhou},
  {Steffens}, {Schmalzl}, {Forrest}, {Abdel-Hafiez}, {Chareev}, {Vasiliev},
  {Bourges}, {Sidis}, {Cao}, and {Zhao}}]{Wang2015}
\bibinfo{author}{Q.~{Wang}}, \bibinfo{author}{Y.~{Shen}},
  \bibinfo{author}{B.~{Pan}}, \bibinfo{author}{Y.~{Hao}},
  \bibinfo{author}{M.~{Ma}}, \bibinfo{author}{F.~{Zhou}},
  \bibinfo{author}{P.~{Steffens}}, \bibinfo{author}{K.~{Schmalzl}},
  \bibinfo{author}{T.~R. {Forrest}}, \bibinfo{author}{M.~{Abdel-Hafiez}},
  \bibinfo{author}{D.~A. {Chareev}}, \bibinfo{author}{A.~N. {Vasiliev}},
  \bibinfo{author}{P.~{Bourges}}, \bibinfo{author}{Y.~{Sidis}},
  \bibinfo{author}{H.~{Cao}}, \bibinfo{author}{J.~{Zhao}},
\newblock \bibinfo{title}{Strong interplay between stripe spin fluctuations,
  nematicity and superconductivity in {FeSe}},
\newblock \bibinfo{journal}{ArXiv e-prints}  (\bibinfo{year}{2015})
  \bibinfo{pages}{1502.07544}.
\bibitem[{Garbarino et~al.(2009)Garbarino, Sow, Lejay, Sulpice, Toulemonde,
  Mezouar, and Núñez-Regueiro}]{Garbarino2009}
\bibinfo{author}{G.~Garbarino}, \bibinfo{author}{A.~Sow},
  \bibinfo{author}{P.~Lejay}, \bibinfo{author}{A.~Sulpice},
  \bibinfo{author}{P.~Toulemonde}, \bibinfo{author}{M.~Mezouar},
  \bibinfo{author}{M.~Núñez-Regueiro},
\newblock \bibinfo{title}{High-temperature superconductivity ({$T_c$} onset at
  {34~K}) in the high-pressure orthorhombic phase of {FeSe}},
\newblock \bibinfo{journal}{EPL (Europhysics Letters)} \bibinfo{volume}{86}
  (\bibinfo{year}{2009}) \bibinfo{pages}{27001}.
\bibitem[{Bendele et~al.(2012)Bendele, Ichsanow, Pashkevich, Keller,
  Str\"assle, Gusev, Pomjakushina, Conder, Khasanov, and Keller}]{Bendele2012}
\bibinfo{author}{M.~Bendele}, \bibinfo{author}{A.~Ichsanow},
  \bibinfo{author}{Y.~Pashkevich}, \bibinfo{author}{L.~Keller},
  \bibinfo{author}{T.~Str\"assle}, \bibinfo{author}{A.~Gusev},
  \bibinfo{author}{E.~Pomjakushina}, \bibinfo{author}{K.~Conder},
  \bibinfo{author}{R.~Khasanov}, \bibinfo{author}{H.~Keller},
\newblock \bibinfo{title}{Coexistence of superconductivity and magnetism in
  {FeSe$_{1-x}$} under pressure},
\newblock \bibinfo{journal}{Phys. Rev. B} \bibinfo{volume}{85}
  (\bibinfo{year}{2012}) \bibinfo{pages}{064517}.
\bibitem[{Miyoshi et~al.(2014)Miyoshi, Morishita, Mutou, Kondo, Seida,
  Fujiwara, Takeuchi, and Nishigori}]{Miyoshi2014}
\bibinfo{author}{K.~Miyoshi}, \bibinfo{author}{K.~Morishita},
  \bibinfo{author}{E.~Mutou}, \bibinfo{author}{M.~Kondo},
  \bibinfo{author}{O.~Seida}, \bibinfo{author}{K.~Fujiwara},
  \bibinfo{author}{J.~Takeuchi}, \bibinfo{author}{S.~Nishigori},
\newblock \bibinfo{title}{Enhanced superconductivity on the tetragonal lattice
  in {FeSe} under hydrostatic pressure},
\newblock \bibinfo{journal}{Journal of the Physical Society of Japan}
  \bibinfo{volume}{83} (\bibinfo{year}{2014}) \bibinfo{pages}{013702}.
\bibitem[{Terashima et~al.(2015)Terashima, Kikugawa, Kasahara, Watashige,
  Shibauchi, Matsuda, Wolf, B\"ohmer, Hardy, Meingast, v.~L\"ohneysen, and
  Uji}]{Terashima2015}
\bibinfo{author}{T.~Terashima}, \bibinfo{author}{N.~Kikugawa},
  \bibinfo{author}{S.~Kasahara}, \bibinfo{author}{T.~Watashige},
  \bibinfo{author}{T.~Shibauchi}, \bibinfo{author}{Y.~Matsuda},
  \bibinfo{author}{T.~Wolf}, \bibinfo{author}{A.~E. B\"ohmer},
  \bibinfo{author}{F.~Hardy}, \bibinfo{author}{C.~Meingast},
  \bibinfo{author}{H.~v.~L\"ohneysen}, \bibinfo{author}{S.~Uji},
\newblock \bibinfo{title}{Pressure-induced antiferromagnetic transition and
  phase diagram in {FeSe}},
\newblock \bibinfo{journal}{Journal of the Physical Society of Japan}
  \bibinfo{volume}{84} (\bibinfo{year}{2015}) \bibinfo{pages}{063701}.
\bibitem[{Kn{\"o}ner et~al.(2015)Kn{\"o}ner, Zielke, K{\"o}hler, Wolf, Wolf,
  Wang, B{\"o}hmer, Meingast, and Lang}]{Knoener2015}
\bibinfo{author}{S.~Kn{\"o}ner}, \bibinfo{author}{D.~Zielke},
  \bibinfo{author}{S.~K{\"o}hler}, \bibinfo{author}{B.~Wolf},
  \bibinfo{author}{T.~Wolf}, \bibinfo{author}{L.~Wang},
  \bibinfo{author}{A.~B{\"o}hmer}, \bibinfo{author}{C.~Meingast},
  \bibinfo{author}{M.~Lang},
\newblock \bibinfo{title}{Resistivity and magnetoresistance of {FeSe} single
  crystals under helium-gas pressure},
\newblock \bibinfo{journal}{Phys. Rev. B} \bibinfo{volume}{91}
  (\bibinfo{year}{2015}) \bibinfo{pages}{174510}.
\bibitem[{Mizuguchi et~al.(2008)Mizuguchi, Tomioka, Tsuda, Yamaguchi, and
  Takano}]{Mizuguchi2008}
\bibinfo{author}{Y.~Mizuguchi}, \bibinfo{author}{F.~Tomioka},
  \bibinfo{author}{S.~Tsuda}, \bibinfo{author}{T.~Yamaguchi},
  \bibinfo{author}{Y.~Takano},
\newblock \bibinfo{title}{Superconductivity at {27 K} in tetragonal {FeSe}
  under high pressure},
\newblock \bibinfo{journal}{Applied Physics Letters} \bibinfo{volume}{93}
  (\bibinfo{year}{2008}) \bibinfo{pages}{152505}.
\bibitem[{Medvedev et~al.(2009)Medvedev, McQueen, Troyan, Palasyuk, Eremets,
  Cava, Naghavi, Casper, Ksenofontov, Wortmann, and Felser}]{Medvedev2009}
\bibinfo{author}{S.~Medvedev}, \bibinfo{author}{T.~M. McQueen},
  \bibinfo{author}{I.~A. Troyan}, \bibinfo{author}{T.~Palasyuk},
  \bibinfo{author}{M.~I. Eremets}, \bibinfo{author}{R.~J. Cava},
  \bibinfo{author}{S.~Naghavi}, \bibinfo{author}{F.~Casper},
  \bibinfo{author}{V.~Ksenofontov}, \bibinfo{author}{G.~Wortmann},
  \bibinfo{author}{C.~Felser},
\newblock \bibinfo{title}{Electronic and magnetic phase diagram of
  {$\beta$-Fe$_{1.01}$} with superconductivity at {36.7 K} under pressure},
\newblock \bibinfo{journal}{Nature Mat.} \bibinfo{volume}{8}
  (\bibinfo{year}{2009}) \bibinfo{pages}{630--633}.
\bibitem[{{Yu} and {Si}(2015)}]{Yu2015}
\bibinfo{author}{R.~{Yu}}, \bibinfo{author}{Q.~{Si}},
\newblock \bibinfo{title}{Antiferroquadrupolar and {I}sing-nematic orders of a
  frustrated bilinear-biquadratic {H}eisenberg model and implications for the
  magnetism of {FeSe}},
\newblock \bibinfo{journal}{ArXiv e-prints}  (\bibinfo{year}{2015})
  \bibinfo{pages}{1501.05926}.
\bibitem[{{Glasbrenner} et~al.(2015){Glasbrenner}, {Mazin}, {Jeschke},
  {Hirschfeld}, and {Valent{\'{\i}}}}]{Glasbrenner2015}
\bibinfo{author}{J.~K. {Glasbrenner}}, \bibinfo{author}{I.~I. {Mazin}},
  \bibinfo{author}{H.~O. {Jeschke}}, \bibinfo{author}{P.~J. {Hirschfeld}},
  \bibinfo{author}{R.~{Valent{\'{\i}}}},
\newblock \bibinfo{title}{Effect of magnetic frustration on nematicity and
  superconductivity in {Fe} chalcogenides},
\newblock \bibinfo{journal}{ArXiv e-prints}  (\bibinfo{year}{2015})
  \bibinfo{pages}{1501.04946}.
\bibitem[{{Wang} et~al.(2015){Wang}, {Kivelson}, and {Lee}}]{Wang2015II}
\bibinfo{author}{F.~{Wang}}, \bibinfo{author}{S.~{Kivelson}},
  \bibinfo{author}{D.-H. {Lee}},
\newblock \bibinfo{title}{Is {FeSe} a nematic quantum paramagnet?},
\newblock \bibinfo{journal}{ArXiv e-prints}  (\bibinfo{year}{2015})
  \bibinfo{pages}{1501.00844}.
\bibitem[{{Mukherjee} et~al.(2015){Mukherjee}, {Kreisel}, {Hirschfeld}, and
  {Andersen}}]{Mukherjee2015}
\bibinfo{author}{S.~{Mukherjee}}, \bibinfo{author}{A.~{Kreisel}},
  \bibinfo{author}{P.~J. {Hirschfeld}}, \bibinfo{author}{B.~{Andersen}},
\newblock \bibinfo{title}{Model of electronic structure and superconductivity
  in orbitally ordered {FeSe}},
\newblock \bibinfo{journal}{ArXiv e-prints}  (\bibinfo{year}{2015})
  \bibinfo{pages}{1502.03354}.
\bibitem[{{Chubukov} et~al.(2015){Chubukov}, {Fernandes}, and
  {Schmalian}}]{Chubukov2015}
\bibinfo{author}{A.~V. {Chubukov}}, \bibinfo{author}{R.~M. {Fernandes}},
  \bibinfo{author}{J.~{Schmalian}},
\newblock \bibinfo{title}{The origin of nematic order in {FeSe}},
\newblock \bibinfo{journal}{ArXiv e-prints}  (\bibinfo{year}{2015})
  \bibinfo{pages}{1504.02315}.
\bibitem[{Avci et~al.(2014)Avci, , Chmaissem, Allred, Rosenkranz, Eremin,
  Chubukov, Bugaris, Chung, Kanatzidis, Castellan, Schlueter, Claus, Khalyavin,
  Manuel, Daoud-Aladine, and Osborn}]{Avci2014}
\bibinfo{author}{S.~Avci}, , \bibinfo{author}{O.~Chmaissem},
  \bibinfo{author}{J.~Allred}, \bibinfo{author}{S.~Rosenkranz},
  \bibinfo{author}{I.~Eremin}, \bibinfo{author}{A.~Chubukov},
  \bibinfo{author}{D.~Bugaris}, \bibinfo{author}{D.~Chung},
  \bibinfo{author}{M.~Kanatzidis}, \bibinfo{author}{J.-P. Castellan},
  \bibinfo{author}{J.~Schlueter}, \bibinfo{author}{H.~Claus},
  \bibinfo{author}{D.~Khalyavin}, \bibinfo{author}{P.~Manuel},
  \bibinfo{author}{A.~Daoud-Aladine}, \bibinfo{author}{R.~Osborn},
\newblock \bibinfo{title}{Magnetically driven suppression of nematic order in
  an iron-based superconductor},
\newblock \bibinfo{journal}{Nature Communications} \bibinfo{volume}{5}
  (\bibinfo{year}{2014}) \bibinfo{pages}{3845}.
\bibitem[{Wa\ss{}er et~al.(2015)Wa\ss{}er, Schneidewind, Sidis, Wurmehl,
  Aswartham, B\"uchner, and Braden}]{Wasser2015}
\bibinfo{author}{F.~Wa\ss{}er}, \bibinfo{author}{A.~Schneidewind},
  \bibinfo{author}{Y.~Sidis}, \bibinfo{author}{S.~Wurmehl},
  \bibinfo{author}{S.~Aswartham}, \bibinfo{author}{B.~B\"uchner},
  \bibinfo{author}{M.~Braden},
\newblock \bibinfo{title}{Spin reorientation in
  {Ba$_{0.65}$Na$_{0.35}$Fe$_{2}$As$_{2}$} studied by single-crystal neutron
  diffraction},
\newblock \bibinfo{journal}{Phys. Rev. B} \bibinfo{volume}{91}
  (\bibinfo{year}{2015}) \bibinfo{pages}{060505}.
\bibitem[{Khalyavin et~al.(2014)Khalyavin, Lovesey, Manuel, Kr\"uger,
  Rosenkranz, Allred, Chmaissem, and Osborn}]{Khalyavin2014}
\bibinfo{author}{D.~D. Khalyavin}, \bibinfo{author}{S.~W. Lovesey},
  \bibinfo{author}{P.~Manuel}, \bibinfo{author}{F.~Kr\"uger},
  \bibinfo{author}{S.~Rosenkranz}, \bibinfo{author}{J.~M. Allred},
  \bibinfo{author}{O.~Chmaissem}, \bibinfo{author}{R.~Osborn},
\newblock \bibinfo{title}{Symmetry of reentrant tetragonal phase in
  {Ba$_{1-x}$Na$_{x}$Fe$_{2}$As$_{2}$}: {M}agnetic versus orbital ordering
  mechanism},
\newblock \bibinfo{journal}{Phys. Rev. B} \bibinfo{volume}{90}
  (\bibinfo{year}{2014}) \bibinfo{pages}{174511}.
\bibitem[{Kang et~al.(2015)Kang, Wang, Chubukov, and Fernandes}]{Kang2014}
\bibinfo{author}{J.~Kang}, \bibinfo{author}{X.~Wang}, \bibinfo{author}{A.~V.
  Chubukov}, \bibinfo{author}{R.~M. Fernandes},
\newblock \bibinfo{title}{Interplay between tetragonal magnetic order, stripe
  magnetism, and superconductivity in iron-based materials},
\newblock \bibinfo{journal}{Phys. Rev. B} \bibinfo{volume}{91}
  (\bibinfo{year}{2015}) \bibinfo{pages}{121104}.
\bibitem[{{Gastiasoro} and {Andersen}(2015)}]{Gastiasoro2015}
\bibinfo{author}{M.~N. {Gastiasoro}}, \bibinfo{author}{B.~M. {Andersen}},
\newblock \bibinfo{title}{Competing magnetic double-{$Q$} phases and
  superconductivity-induced re-entrance of {$C_2$} magnetic stripe order in
  iron pnictides},
\newblock \bibinfo{journal}{ArXiv e-prints}  (\bibinfo{year}{2015})
  \bibinfo{pages}{1502.05859}.
\bibitem[{Taddei and {\textit{et al.}}(2015)}]{Taddei2015}
\bibinfo{author}{Taddei}, \bibinfo{author}{{\textit{et al.}}},
  \bibinfo{year}{2015}. \bibinfo{note}{APS March Meeting 2015, L5.00006}.
\bibitem[{B{\"o}hmer et~al.(2014)B{\"o}hmer, Hardy, Wang, Wolf, Schweiss, and
  Meingast}]{Boehmer2015II}
\bibinfo{author}{A.~E. B{\"o}hmer}, \bibinfo{author}{F.~Hardy},
  \bibinfo{author}{L.~Wang}, \bibinfo{author}{T.~Wolf},
  \bibinfo{author}{P.~Schweiss}, \bibinfo{author}{C.~Meingast},
\newblock \bibinfo{title}{Superconductivity-induced reentrance of orthorhombic
  distortion in {Ba$_{1-x}$K$_x$Fe$_2$As$_2$}},
\newblock \bibinfo{journal}{ArXiv e-prints}  (\bibinfo{year}{2014})
  \bibinfo{pages}{1412.7038}.
\bibitem[{Salje(1993)}]{Salje1990}
\bibinfo{author}{E.~Salje}, \bibinfo{title}{Phase transitions in ferroelastic
  and co-elastic crystals}, \bibinfo{publisher}{Cambridge university press},
  \bibinfo{year}{1993}. \URLprefix
  \url{http://www.cambridge.org/ve/academic/subjects/earth-and-environmental-science/mineralogy-petrology-and-volcanology/phase-transitions-ferroelastic-and-co-elastic-crystals}.
\bibitem[{Goto et~al.(2011)Goto, Kurihara, Araki, Mitsumoto, Akatsu, Nemoto,
  Tatematsu, and Sato}]{Goto2011}
\bibinfo{author}{T.~Goto}, \bibinfo{author}{R.~Kurihara},
  \bibinfo{author}{K.~Araki}, \bibinfo{author}{K.~Mitsumoto},
  \bibinfo{author}{M.~Akatsu}, \bibinfo{author}{Y.~Nemoto},
  \bibinfo{author}{S.~Tatematsu}, \bibinfo{author}{M.~Sato},
\newblock \bibinfo{title}{Quadrupole effects in layered iron pnictide
  superconductor {Ba(Fe$_{0.9}$Co$_{0.1}$)$_2$As$_2$}},
\newblock \bibinfo{journal}{Journal of the Physical Society of Japan}
  \bibinfo{volume}{80} (\bibinfo{year}{2011}) \bibinfo{pages}{073702}.
\bibitem[{Yoshizawa et~al.(2012)Yoshizawa, Kimura, Chiba, Simayi, Nakanishi,
  Kihou, Lee, Iyo, Eisaki, Nakajima, and Uchida}]{Yoshizawa2012}
\bibinfo{author}{M.~Yoshizawa}, \bibinfo{author}{D.~Kimura},
  \bibinfo{author}{T.~Chiba}, \bibinfo{author}{S.~Simayi},
  \bibinfo{author}{Y.~Nakanishi}, \bibinfo{author}{K.~Kihou},
  \bibinfo{author}{C.-H. Lee}, \bibinfo{author}{A.~Iyo},
  \bibinfo{author}{H.~Eisaki}, \bibinfo{author}{M.~Nakajima},
  \bibinfo{author}{S.-i. Uchida},
\newblock \bibinfo{title}{Structural quantum criticality and superconductivity
  in iron-based superconductor {Ba(Fe$_{1-x}$Co$_x$)$_2$As$_2$}},
\newblock \bibinfo{journal}{Journal of the Physical Society of Japan}
  \bibinfo{volume}{81} (\bibinfo{year}{2012}) \bibinfo{pages}{024604}.
\bibitem[{Simayi et~al.(2013)Simayi, Sakano, Takezawa, Nakamura, Nakanishi,
  Kihou, Nakajima, Lee, Iyo, Eisaki, ichi Uchida, and Yoshizawa}]{Simayi2013}
\bibinfo{author}{S.~Simayi}, \bibinfo{author}{K.~Sakano},
  \bibinfo{author}{H.~Takezawa}, \bibinfo{author}{M.~Nakamura},
  \bibinfo{author}{Y.~Nakanishi}, \bibinfo{author}{K.~Kihou},
  \bibinfo{author}{M.~Nakajima}, \bibinfo{author}{C.-H. Lee},
  \bibinfo{author}{A.~Iyo}, \bibinfo{author}{H.~Eisaki},
  \bibinfo{author}{S.~ichi Uchida}, \bibinfo{author}{M.~Yoshizawa},
\newblock \bibinfo{title}{Strange inter-layer properties of
  {Ba(Fe$_{1-x}$Co$_x$)$_2$As$_2$} appearing in ultrasonic measurements},
\newblock \bibinfo{journal}{Journal of the Physical Society of Japan}
  \bibinfo{volume}{82} (\bibinfo{year}{2013}) \bibinfo{pages}{114604}.
\bibitem[{Zvyagina et~al.(2013)Zvyagina, Gaydamak, Zhekov, Bilich, Fil,
  Chareev, and Vasiliev}]{Zvyagina2013}
\bibinfo{author}{G.~A. Zvyagina}, \bibinfo{author}{T.~N. Gaydamak},
  \bibinfo{author}{K.~R. Zhekov}, \bibinfo{author}{I.~V. Bilich},
  \bibinfo{author}{V.~D. Fil}, \bibinfo{author}{D.~A. Chareev},
  \bibinfo{author}{A.~N. Vasiliev},
\newblock \bibinfo{title}{Acoustic characteristics of {FeSe} single crystals},
\newblock \bibinfo{journal}{EPL (Europhysics Letters)} \bibinfo{volume}{101}
  (\bibinfo{year}{2013}) \bibinfo{pages}{56005}.
\bibitem[{Liang et~al.(2013)Liang, Moreo, and Dagotto}]{Liang2013}
\bibinfo{author}{S.~Liang}, \bibinfo{author}{A.~Moreo},
  \bibinfo{author}{E.~Dagotto},
\newblock \bibinfo{title}{Nematic state of pnictides stabilized by interplay
  between spin, orbital, and lattice degrees of freedom},
\newblock \bibinfo{journal}{Phys. Rev. Lett.} \bibinfo{volume}{111}
  (\bibinfo{year}{2013}) \bibinfo{pages}{047004}.
\bibitem[{Liang et~al.(2014)Liang, Mukherjee, Patel, Bishop, Dagotto, and
  Moreo}]{Liang2014}
\bibinfo{author}{S.~Liang}, \bibinfo{author}{A.~Mukherjee},
  \bibinfo{author}{N.~D. Patel}, \bibinfo{author}{C.~B. Bishop},
  \bibinfo{author}{E.~Dagotto}, \bibinfo{author}{A.~Moreo},
\newblock \bibinfo{title}{Diverging nematic susceptibility, physical meaning of
  {${T}^{*}$} scale, and pseudogap in the spin fermion model for the
  pnictides},
\newblock \bibinfo{journal}{Phys. Rev. B} \bibinfo{volume}{90}
  (\bibinfo{year}{2014}) \bibinfo{pages}{184507}.
\bibitem[{Rehwald(1973)}]{Rehwald1973}
\bibinfo{author}{W.~Rehwald},
\newblock \bibinfo{title}{The study of structural phase transitions by means of
  ultrasonic experiments},
\newblock \bibinfo{journal}{Advances in Physics} \bibinfo{volume}{22}
  (\bibinfo{year}{1973}) \bibinfo{pages}{721--755}.
\bibitem[{Benoît et~al.(198)Benoît, Berger, Krauzman, and Godet}]{Benoit1986}
\bibinfo{author}{J.~P. Benoît}, \bibinfo{author}{J.~Berger},
  \bibinfo{author}{M.~Krauzman}, \bibinfo{author}{J.~L. Godet},
\newblock \bibinfo{title}{Experimental observation of a soft mode in ammonium
  hydrogen oxalate hemihydrate by {B}rillouin scattering},
\newblock \bibinfo{journal}{J. Phys. France} \bibinfo{volume}{47}
  (\bibinfo{year}{198}) \bibinfo{pages}{815--819}.
\bibitem[{Cowley(1976)}]{Cowley1976}
\bibinfo{author}{R.~A. Cowley},
\newblock \bibinfo{title}{Acoustic phonon instabilities and structural phase
  transitions},
\newblock \bibinfo{journal}{Phys. Rev. B} \bibinfo{volume}{13}
  (\bibinfo{year}{1976}) \bibinfo{pages}{4877--4885}.
\bibitem[{Folk et~al.(1976)Folk, Iro, and Schwabl}]{Folk1976}
\bibinfo{author}{R.~Folk}, \bibinfo{author}{H.~Iro},
  \bibinfo{author}{F.~Schwabl},
\newblock \bibinfo{title}{Critical statics of elastic phase transitions},
\newblock \bibinfo{journal}{Zeitschrift für Physik B Condensed Matter}
  \bibinfo{volume}{25} (\bibinfo{year}{1976}) \bibinfo{pages}{69--81}.
\bibitem[{Als?Nielsen and Birgeneau(1977)}]{AlsNielsen1977}
\bibinfo{author}{J.~Als?Nielsen}, \bibinfo{author}{R.~J. Birgeneau},
\newblock \bibinfo{title}{Mean field theory, the {G}inzburg criterion, and
  marginal dimensionality of phase transitions},
\newblock \bibinfo{journal}{American Journal of Physics} \bibinfo{volume}{45}
  (\bibinfo{year}{1977}) \bibinfo{pages}{554--560}.
\bibitem[{Kityk et~al.(1996)Kityk, Soprunyuk, Fuith, Schranz, and
  Warhanek}]{Kityk1996}
\bibinfo{author}{A.~V. Kityk}, \bibinfo{author}{V.~P. Soprunyuk},
  \bibinfo{author}{A.~Fuith}, \bibinfo{author}{W.~Schranz},
  \bibinfo{author}{H.~Warhanek},
\newblock \bibinfo{title}{Low-frequency elastic properties of the
  incommensurate ferroelastic {[N(CH$_3$)$_4$]$_2$CuCl$_4$}},
\newblock \bibinfo{journal}{Phys. Rev. B} \bibinfo{volume}{53}
  (\bibinfo{year}{1996}) \bibinfo{pages}{6337--6344}.
\bibitem[{Rossiter and Baetzold(1991)}]{Rossiter1991}
\bibinfo{editor}{B.~W. Rossiter}, \bibinfo{editor}{R.~C. Baetzold} (Eds.),
  volume~\bibinfo{volume}{7} of \textit{\bibinfo{series}{Physical methods of
  chemistry}}, \bibinfo{publisher}{Wiley}, \bibinfo{address}{New York, N.Y.},
  \bibinfo{year}{1991}. \URLprefix
  \url{http://eu.wiley.com/WileyCDA/WileyTitle/productCd-0471534382.html}.
\bibitem[{Meingast et~al.(1990)Meingast, Blank, B\"urkle, Obst, Wolf, W{\"u}hl,
  Selvamanickam, and Salama}]{Meingast1990}
\bibinfo{author}{C.~Meingast}, \bibinfo{author}{B.~Blank},
  \bibinfo{author}{H.~B\"urkle}, \bibinfo{author}{B.~Obst},
  \bibinfo{author}{T.~Wolf}, \bibinfo{author}{H.~W{\"u}hl},
  \bibinfo{author}{V.~Selvamanickam}, \bibinfo{author}{K.~Salama},
\newblock \bibinfo{title}{Anisotropic pressure dependence of {${T}_c$} in
  single crystal {YBa$_2$Cu$_3$O$_7$} via thermal expansion},
\newblock \bibinfo{journal}{Phys. Rev. B} \bibinfo{volume}{41}
  (\bibinfo{year}{1990}) \bibinfo{pages}{11299--11304}.
\bibitem[{Meingast et~al.(2012)Meingast, Hardy, Heid, Adelmann, B\"ohmer,
  Burger, Ernst, Fromknecht, Schweiss, and Wolf}]{Meingast2012}
\bibinfo{author}{C.~Meingast}, \bibinfo{author}{F.~Hardy},
  \bibinfo{author}{R.~Heid}, \bibinfo{author}{P.~Adelmann},
  \bibinfo{author}{A.~B\"ohmer}, \bibinfo{author}{P.~Burger},
  \bibinfo{author}{D.~Ernst}, \bibinfo{author}{R.~Fromknecht},
  \bibinfo{author}{P.~Schweiss}, \bibinfo{author}{T.~Wolf},
\newblock \bibinfo{title}{Thermal expansion and {G}rüneisen parameters of
  {Ba(Fe$_{1-x}$Co$_x$)$_2$As$_2$} - a thermodynamic quest for quantum
  criticality},
\newblock \bibinfo{journal}{Phys. Rev. Lett.} \bibinfo{volume}{108}
  (\bibinfo{year}{2012}) \bibinfo{pages}{177004}.
\bibitem[{Salje and Schranz(2010)}]{Salje2010}
\bibinfo{author}{E.~K.~H. Salje}, \bibinfo{author}{W.~Schranz},
\newblock \bibinfo{title}{Low amplitude, low frequency elastic measurements
  using dynamic mechanical analyzer {(DMA)} spectroscopy},
\newblock \bibinfo{journal}{Zeitschrift für Kristallographie Crystalline
  Materials} \bibinfo{volume}{226} (\bibinfo{year}{2010})
  \bibinfo{pages}{1--17}.
\bibitem[{currently unpublished(2015)}]{noteSchranz}
currently unpublished, \bibinfo{year}{2015}. \bibinfo{note}{Measurements were
  made in Prof. W. Schranz's group in Vienna.}
\bibitem[{Parshall et~al.(2015)Parshall, Pintschovius, Niedziela, Castellan,
  Lamago, Mittal, Wolf, and Reznik}]{Parshall2014}
\bibinfo{author}{D.~Parshall}, \bibinfo{author}{L.~Pintschovius},
  \bibinfo{author}{J.~L. Niedziela}, \bibinfo{author}{J.-P. Castellan},
  \bibinfo{author}{D.~Lamago}, \bibinfo{author}{R.~Mittal},
  \bibinfo{author}{T.~Wolf}, \bibinfo{author}{D.~Reznik},
\newblock \bibinfo{title}{Close correlation between magnetic properties and the
  soft phonon mode of the structural transition in {BaFe$_2$As$_2$} and
  {SrFe$_2$As$_2$}},
\newblock \bibinfo{journal}{Phys. Rev. B} \bibinfo{volume}{91}
  (\bibinfo{year}{2015}) \bibinfo{pages}{134426}.
\bibitem[{Schranz et~al.(2012{\natexlab{a}})Schranz, Kabelka, Sarras, and
  Burock}]{Schranz2012}
\bibinfo{author}{W.~Schranz}, \bibinfo{author}{H.~Kabelka},
  \bibinfo{author}{A.~Sarras}, \bibinfo{author}{M.~Burock},
\newblock \bibinfo{title}{Giant domain wall response of highly twinned
  ferroelastic materials},
\newblock \bibinfo{journal}{Applied Physics Letters} \bibinfo{volume}{101}
  (\bibinfo{year}{2012}{\natexlab{a}}) \bibinfo{pages}{141913}.
\bibitem[{Schranz et~al.(2012{\natexlab{b}})Schranz, Kabelka, and
  Tröster}]{Schranz2012II}
\bibinfo{author}{W.~Schranz}, \bibinfo{author}{H.~Kabelka},
  \bibinfo{author}{A.~Tröster},
\newblock \bibinfo{title}{Superelastic softening of ferroelastic multidomain
  crystals},
\newblock \bibinfo{journal}{Ferroelectrics} \bibinfo{volume}{426}
  (\bibinfo{year}{2012}{\natexlab{b}}) \bibinfo{pages}{242--250}.
\bibitem[{Varshni(1970)}]{Varshni1970}
\bibinfo{author}{Y.~P. Varshni},
\newblock \bibinfo{title}{Temperature dependence of the elastic constants},
\newblock \bibinfo{journal}{Phys. Rev. B} \bibinfo{volume}{2}
  (\bibinfo{year}{1970}) \bibinfo{pages}{3952--3958}.
\bibitem[{Kasahara et~al.(2012)Kasahara, Shi, Hashimoto, Tonegawa, Mizukami,
  Shibauchi, Sugimoto, Fukuda, Terashima, Nevimdomskyy, and
  Matsuda}]{Kasahara2012}
\bibinfo{author}{S.~Kasahara}, \bibinfo{author}{H.~J. Shi},
  \bibinfo{author}{K.~Hashimoto}, \bibinfo{author}{S.~Tonegawa},
  \bibinfo{author}{Y.~Mizukami}, \bibinfo{author}{T.~Shibauchi},
  \bibinfo{author}{K.~Sugimoto}, \bibinfo{author}{T.~Fukuda},
  \bibinfo{author}{T.~Terashima}, \bibinfo{author}{A.~H. Nevimdomskyy},
  \bibinfo{author}{Y.~Matsuda},
\newblock \bibinfo{title}{Electronic nematicity above the structural and
  superconducting transition in {BaFe$_2$(As$_{1-x}$P$_2$)$_2$}},
\newblock \bibinfo{journal}{Nature} \bibinfo{volume}{486}
  (\bibinfo{year}{2012}) \bibinfo{pages}{382}.
\bibitem[{B\"ohmer et~al.(2012)B\"ohmer, Burger, Hardy, Wolf, Schweiss,
  Fromknecht, v.~L\"ohneysen, Meingast, Mak, Lortz, Kasahara, Terashima,
  Shibauchi, and Matsuda}]{Boehmer2012}
\bibinfo{author}{A.~E. B\"ohmer}, \bibinfo{author}{P.~Burger},
  \bibinfo{author}{F.~Hardy}, \bibinfo{author}{T.~Wolf},
  \bibinfo{author}{P.~Schweiss}, \bibinfo{author}{R.~Fromknecht},
  \bibinfo{author}{H.~v.~L\"ohneysen}, \bibinfo{author}{C.~Meingast},
  \bibinfo{author}{H.~K. Mak}, \bibinfo{author}{R.~Lortz},
  \bibinfo{author}{S.~Kasahara}, \bibinfo{author}{T.~Terashima},
  \bibinfo{author}{T.~Shibauchi}, \bibinfo{author}{Y.~Matsuda},
\newblock \bibinfo{title}{Thermodynamic phase diagram, phase competition, and
  uniaxial pressure effects in {BaFe$_2$(As$_{1-x}$P$_x$)$_2$} studied by
  thermal expansion},
\newblock \bibinfo{journal}{Phys. Rev. B} \bibinfo{volume}{86}
  (\bibinfo{year}{2012}) \bibinfo{pages}{094521}.
\bibitem[{Luo et~al.(2015)Luo, Stanev, Shen, Fang, Ling, Osborn, Rosenkranz,
  Benseman, Divan, Kwok, and Welp}]{Luo2015}
\bibinfo{author}{X.~Luo}, \bibinfo{author}{V.~Stanev},
  \bibinfo{author}{B.~Shen}, \bibinfo{author}{L.~Fang}, \bibinfo{author}{X.~S.
  Ling}, \bibinfo{author}{R.~Osborn}, \bibinfo{author}{S.~Rosenkranz},
  \bibinfo{author}{T.~M. Benseman}, \bibinfo{author}{R.~Divan},
  \bibinfo{author}{W.-K. Kwok}, \bibinfo{author}{U.~Welp},
\newblock \bibinfo{title}{Antiferromagnetic and nematic phase transitions in
  {BaFe$_{2}$(As$_{1-x}$P$_{x}$)$_{2}$} studied by ac microcalorimetry and
  {SQUID} magnetometry},
\newblock \bibinfo{journal}{Phys. Rev. B} \bibinfo{volume}{91}
  (\bibinfo{year}{2015}) \bibinfo{pages}{094512}.
\bibitem[{Kontani and Yamakawa(2014)}]{Kontani2014}
\bibinfo{author}{H.~Kontani}, \bibinfo{author}{Y.~Yamakawa},
\newblock \bibinfo{title}{Linear response theory for shear modulus {$C_{66}$}
  and {R}aman quadrupole susceptibility: Evidence for nematic orbital
  fluctuations in {F}e-based superconductors},
\newblock \bibinfo{journal}{Phys. Rev. Lett.} \bibinfo{volume}{113}
  (\bibinfo{year}{2014}) \bibinfo{pages}{047001}.
\bibitem[{Pippard(1955)}]{Pippard1955}
\bibinfo{author}{A.~Pippard},
\newblock \bibinfo{title}{{CXXIII.} thermodynamics of a sheared
  superconductor},
\newblock \bibinfo{journal}{The London, Edinburgh, and Dublin Philosophical
  Magazine and Journal of Science} \bibinfo{volume}{46} (\bibinfo{year}{1955})
  \bibinfo{pages}{1115--1118}.
\bibitem[{Fernandes and Millis(2013)}]{Fernandes2013III}
\bibinfo{author}{R.~M. Fernandes}, \bibinfo{author}{A.~J. Millis},
\newblock \bibinfo{title}{Nematicity as a probe of superconducting pairing in
  iron-based superconductors},
\newblock \bibinfo{journal}{Phys. Rev. Lett.} \bibinfo{volume}{111}
  (\bibinfo{year}{2013}) \bibinfo{pages}{127001}.
\bibitem[{Terashima et~al.(2014)Terashima, Kikugawa, Kiswandhi, Choi, Brooks,
  Kasahara, Watashige, Ikeda, Shibauchi, Matsuda, Wolf, B\"ohmer, Hardy,
  Meingast, L\"ohneysen, Suzuki, Arita, and Uji}]{Terashima2014}
\bibinfo{author}{T.~Terashima}, \bibinfo{author}{N.~Kikugawa},
  \bibinfo{author}{A.~Kiswandhi}, \bibinfo{author}{E.-S. Choi},
  \bibinfo{author}{J.~S. Brooks}, \bibinfo{author}{S.~Kasahara},
  \bibinfo{author}{T.~Watashige}, \bibinfo{author}{H.~Ikeda},
  \bibinfo{author}{T.~Shibauchi}, \bibinfo{author}{Y.~Matsuda},
  \bibinfo{author}{T.~Wolf}, \bibinfo{author}{A.~E. B\"ohmer},
  \bibinfo{author}{F.~Hardy}, \bibinfo{author}{C.~Meingast},
  \bibinfo{author}{H.~v. L\"ohneysen}, \bibinfo{author}{M.-T. Suzuki},
  \bibinfo{author}{R.~Arita}, \bibinfo{author}{S.~Uji},
\newblock \bibinfo{title}{Anomalous {F}ermi surface in {FeSe} seen by
  {S}hubnikov-de {H}aas oscillation measurements},
\newblock \bibinfo{journal}{Phys. Rev. B} \bibinfo{volume}{90}
  (\bibinfo{year}{2014}) \bibinfo{pages}{144517}.
\bibitem[{Kasahara et~al.(2014)Kasahara, Watashige, Hanaguri, Kohsaka,
  Yamashita, Shimoyama, Mizukami, Endo, Ikeda, Aoyama, Terashima, Uji, Wolf,
  von Löhneysen, Shibauchi, and Matsuda}]{Kasahara2014}
\bibinfo{author}{S.~Kasahara}, \bibinfo{author}{T.~Watashige},
  \bibinfo{author}{T.~Hanaguri}, \bibinfo{author}{Y.~Kohsaka},
  \bibinfo{author}{T.~Yamashita}, \bibinfo{author}{Y.~Shimoyama},
  \bibinfo{author}{Y.~Mizukami}, \bibinfo{author}{R.~Endo},
  \bibinfo{author}{H.~Ikeda}, \bibinfo{author}{K.~Aoyama},
  \bibinfo{author}{T.~Terashima}, \bibinfo{author}{S.~Uji},
  \bibinfo{author}{T.~Wolf}, \bibinfo{author}{H.~von Löhneysen},
  \bibinfo{author}{T.~Shibauchi}, \bibinfo{author}{Y.~Matsuda},
\newblock \bibinfo{title}{Field-induced superconducting phase of {FeSe} in the
  {BCS-BEC} cross-over},
\newblock \bibinfo{journal}{Proceedings of the National Academy of Sciences}
  \bibinfo{volume}{111} (\bibinfo{year}{2014}) \bibinfo{pages}{16309--16313}.
\bibitem[{Hardy and \textit{et al.}(2015)}]{Hardyunpublished}
\bibinfo{author}{F.~Hardy}, \bibinfo{author}{\textit{et al.}},
  \bibinfo{year}{2015}. \bibinfo{note}{(in preparation)}.
\bibitem[{Hardy et~al.(2010)Hardy, Wolf, Fisher, Eder, Schweiss, Adelmann,
  v.~L\"ohneysen, and Meingast}]{Hardy2010}
\bibinfo{author}{F.~Hardy}, \bibinfo{author}{T.~Wolf}, \bibinfo{author}{R.~A.
  Fisher}, \bibinfo{author}{R.~Eder}, \bibinfo{author}{P.~Schweiss},
  \bibinfo{author}{P.~Adelmann}, \bibinfo{author}{H.~v.~L\"ohneysen},
  \bibinfo{author}{C.~Meingast},
\newblock \bibinfo{title}{Calorimetric evidence of multiband superconductivity
  in {Ba(Fe$_{0.925}$Co$_{0.075}$)$_2$As$_2$} single crystals},
\newblock \bibinfo{journal}{Phys. Rev. B} \bibinfo{volume}{81}
  (\bibinfo{year}{2010}) \bibinfo{pages}{060501}.
\bibitem[{Meingast(2015)}]{Meingastunpublished}
\bibinfo{author}{C.~Meingast}, \bibinfo{year}{2015}. \bibinfo{note}{(currently
  unpublished)}.
\bibitem[{Hardy et~al.(2010)Hardy, Burger, Wolf, Fisher, Schweiss, Adelmann,
  Heid, Fromknecht, Eder, Ernst, v.~Löhneysen, and Meingast}]{Hardy2010II}
\bibinfo{author}{F.~Hardy}, \bibinfo{author}{P.~Burger},
  \bibinfo{author}{T.~Wolf}, \bibinfo{author}{R.~A. Fisher},
  \bibinfo{author}{P.~Schweiss}, \bibinfo{author}{P.~Adelmann},
  \bibinfo{author}{R.~Heid}, \bibinfo{author}{R.~Fromknecht},
  \bibinfo{author}{R.~Eder}, \bibinfo{author}{D.~Ernst},
  \bibinfo{author}{H.~v.~Löhneysen}, \bibinfo{author}{C.~Meingast},
\newblock \bibinfo{title}{Doping evolution of superconducting gaps and
  electronic densities of states in {Ba(Fe$_{1-x}$Co$_x$)$_2$As$_2$} iron
  pnictides},
\newblock \bibinfo{journal}{EPL (Europhysics Letters)} \bibinfo{volume}{91}
  (\bibinfo{year}{2010}) \bibinfo{pages}{47008}.
\bibitem[{Kuo et~al.(2013)Kuo, Shapiro, Riggs, and Fisher}]{Kuo2013}
\bibinfo{author}{H.-H. Kuo}, \bibinfo{author}{M.~C. Shapiro},
  \bibinfo{author}{S.~C. Riggs}, \bibinfo{author}{I.~R. Fisher},
\newblock \bibinfo{title}{Measurement of the elastoresistivity coefficients of
  the underdoped iron arsenide {Ba(Fe$_{0.975}$Co$_{0.025}$)$_{2}$As$_{2}$}},
\newblock \bibinfo{journal}{Phys. Rev. B} \bibinfo{volume}{88}
  (\bibinfo{year}{2013}) \bibinfo{pages}{085113}.
\bibitem[{{Kuo} et~al.(2015){Kuo}, {Chu}, {Kivelson}, and {Fisher}}]{Kuo2015}
\bibinfo{author}{H.-H. {Kuo}}, \bibinfo{author}{J.-H. {Chu}},
  \bibinfo{author}{S.~A. {Kivelson}}, \bibinfo{author}{I.~R. {Fisher}},
\newblock \bibinfo{title}{Ubiquitous signatures of nematic quantum criticality
  in optimally doped {Fe}-based superconductors},
\newblock \bibinfo{journal}{ArXiv e-prints}  (\bibinfo{year}{2015})
  \bibinfo{pages}{1503.00402}.
\bibitem[{Yang et~al.(2013)Yang, Gallais, Fernandes, Paul, Chauvière,
  M{\'e}asson, Cazayous, Sacuto, Colson, and Forget}]{Yang2013}
\bibinfo{author}{Y.-X. Yang}, \bibinfo{author}{Y.~Gallais},
  \bibinfo{author}{R.~M. Fernandes}, \bibinfo{author}{I.~Paul},
  \bibinfo{author}{L.~Chauvière}, \bibinfo{author}{M.-A. M{\'e}asson},
  \bibinfo{author}{M.~Cazayous}, \bibinfo{author}{A.~Sacuto},
  \bibinfo{author}{D.~Colson}, \bibinfo{author}{A.~Forget},
\newblock \bibinfo{title}{Raman scattering as a probe of charge nematic
  fluctuations in iron based superconductors},
\newblock \bibinfo{journal}{JPS Conference Proceedings} \bibinfo{volume}{3}
  (\bibinfo{year}{2013}).
\bibitem[{Gallais(2015{\natexlab{a}})}]{Gallais2015II}
\bibinfo{author}{Y.~Gallais}, \bibinfo{year}{2015}{\natexlab{a}}.
  \bibinfo{note}{APS March Meeting, Y51.00003}.
\bibitem[{Gallais(2015{\natexlab{b}})}]{Gallaisprivatecommunication}
\bibinfo{author}{Y.~Gallais}, \bibinfo{year}{2015}{\natexlab{b}}.
  \bibinfo{note}{Private communication}.
\bibitem[{Gallais and Paul(2015)}]{Gallais2015}
\bibinfo{author}{Y.~Gallais}, \bibinfo{author}{I.~Paul},
\newblock \bibinfo{title}{Charge nematic fluctuations and electronic {R}aman
  response in {F}e superconductors},
\newblock \bibinfo{journal}{Comptes Rendus de Physique} \bibinfo{volume}{xx}
  (\bibinfo{year}{2015}) \bibinfo{pages}{xxx}.
\bibitem[{Ning et~al.(2010)Ning, Ahilan, Imai, Sefat, McGuire, Sales, Mandrus,
  Cheng, Shen, and Wen}]{Ning2010}
\bibinfo{author}{F.~L. Ning}, \bibinfo{author}{K.~Ahilan},
  \bibinfo{author}{T.~Imai}, \bibinfo{author}{A.~S. Sefat},
  \bibinfo{author}{M.~A. McGuire}, \bibinfo{author}{B.~C. Sales},
  \bibinfo{author}{D.~Mandrus}, \bibinfo{author}{P.~Cheng},
  \bibinfo{author}{B.~Shen}, \bibinfo{author}{H.-H. Wen},
\newblock \bibinfo{title}{Contrasting spin dynamics between underdoped and
  overdoped {B}a({Fe}$_{1-x}${Co}$_x$)$_2${A}s$_2$},
\newblock \bibinfo{journal}{Phys. Rev. Lett.} \bibinfo{volume}{104}
  (\bibinfo{year}{2010}) \bibinfo{pages}{037001}.
\bibitem[{Ning et~al.(2014)Ning, Fu, Torchetti, Imai, Sefat, Cheng, Shen, and
  Wen}]{Ning2014}
\bibinfo{author}{F.~L. Ning}, \bibinfo{author}{M.~Fu}, \bibinfo{author}{D.~A.
  Torchetti}, \bibinfo{author}{T.~Imai}, \bibinfo{author}{A.~S. Sefat},
  \bibinfo{author}{P.~Cheng}, \bibinfo{author}{B.~Shen}, \bibinfo{author}{H.-H.
  Wen},
\newblock \bibinfo{title}{Critical behavior of the spin density wave transition
  in underdoped {Ba(Fe$_{1-x}$Co$_{x}$)$_{2}$As$_{2}$} {($x\leq0.05$)}:
  {$^{75}$As} {NMR} investigation},
\newblock \bibinfo{journal}{Phys. Rev. B} \bibinfo{volume}{89}
  (\bibinfo{year}{2014}) \bibinfo{pages}{214511}.
\bibitem[{Smerald and Shannon(2011)}]{Smerald2011}
\bibinfo{author}{A.~Smerald}, \bibinfo{author}{N.~Shannon},
\newblock \bibinfo{title}{Angle-resolved {NMR}: {Q}uantitative theory of
  {$^{75}$As} {$T_1$} relaxation rate in {BaFe$_2$As$_2$}},
\newblock \bibinfo{journal}{Phys. Rev. B} \bibinfo{volume}{84}
  (\bibinfo{year}{2011}) \bibinfo{pages}{184437}.
\bibitem[{Hirano et~al.(2012)Hirano, Yamada, Saito, Nagashima, Konishi,
  Toriyama, Ohta, Fukazawa, Kohori, Furukawa, Kihou, Lee, Iyo, and
  Eisaki}]{Hirano2012}
\bibinfo{author}{M.~Hirano}, \bibinfo{author}{Y.~Yamada},
  \bibinfo{author}{T.~Saito}, \bibinfo{author}{R.~Nagashima},
  \bibinfo{author}{T.~Konishi}, \bibinfo{author}{T.~Toriyama},
  \bibinfo{author}{Y.~Ohta}, \bibinfo{author}{H.~Fukazawa},
  \bibinfo{author}{Y.~Kohori}, \bibinfo{author}{Y.~Furukawa},
  \bibinfo{author}{K.~Kihou}, \bibinfo{author}{C.-H. Lee},
  \bibinfo{author}{A.~Iyo}, \bibinfo{author}{H.~Eisaki},
\newblock \bibinfo{title}{Potential antiferromagnetic fluctuations in
  hole-doped iron-pnictide superconductor {Ba$_{1-x}$K$_x$Fe$_2$As$_2$} studied
  by {$^{75}$As} nuclear magnetic resonance measurement},
\newblock \bibinfo{journal}{Journal of the Physical Society of Japan}
  \bibinfo{volume}{81} (\bibinfo{year}{2012}) \bibinfo{pages}{054704}.
\bibitem[{Nakai et~al.(2013)Nakai, Iye, Kitagawa, Ishida, Kasahara, Shibauchi,
  Matsuda, Ikeda, and Terashima}]{Nakai2013}
\bibinfo{author}{Y.~Nakai}, \bibinfo{author}{T.~Iye},
  \bibinfo{author}{S.~Kitagawa}, \bibinfo{author}{K.~Ishida},
  \bibinfo{author}{S.~Kasahara}, \bibinfo{author}{T.~Shibauchi},
  \bibinfo{author}{Y.~Matsuda}, \bibinfo{author}{H.~Ikeda},
  \bibinfo{author}{T.~Terashima},
\newblock \bibinfo{title}{Normal-state spin dynamics in the iron-pnictide
  superconductors {BaFe$_{2}$(As$_{1-x}$P$_{x}$)$_{2}$} and
  {Ba(Fe$_{1-x}$Co$_{x}$)$_{2}$As$_{2}$} probed with {NMR} measurements},
\newblock \bibinfo{journal}{Phys. Rev. B} \bibinfo{volume}{87}
  (\bibinfo{year}{2013}) \bibinfo{pages}{174507}.
\bibitem[{Sales et~al.(2010)Sales, McGuire, Sefat, and Mandrus}]{Sales2010}
\bibinfo{author}{B.~Sales}, \bibinfo{author}{M.~McGuire},
  \bibinfo{author}{A.~Sefat}, \bibinfo{author}{D.~Mandrus},
\newblock \bibinfo{title}{A semimetal model of the normal state magnetic
  susceptibility and transport properties of
  {Ba(Fe$_{1?x}$Co$_{x}$)$_2$As$_2$}},
\newblock \bibinfo{journal}{Physica C: Superconductivity} \bibinfo{volume}{470}
  (\bibinfo{year}{2010}) \bibinfo{pages}{304 -- 308}.
\bibitem[{Ma et~al.(2014)Ma, Yuan, Wu, Zhou, Dong, and Zhou}]{Ma2014}
\bibinfo{author}{M.~Ma}, \bibinfo{author}{D.~Yuan}, \bibinfo{author}{Y.~Wu},
  \bibinfo{author}{H.~Zhou}, \bibinfo{author}{X.~Dong},
  \bibinfo{author}{F.~Zhou},
\newblock \bibinfo{title}{Flux-free growth of large superconducting crystal of
  {FeSe} by traveling-solvent floating-zone technique},
\newblock \bibinfo{journal}{Superconductor Science and Technology}
  \bibinfo{volume}{27} (\bibinfo{year}{2014}) \bibinfo{pages}{122001}.
\bibitem[{{Gallais} et~al.(2015){Gallais}, {Paul}, {Chauviere}, and
  {Schmalian}}]{Gallais2015III}
\bibinfo{author}{Y.~{Gallais}}, \bibinfo{author}{I.~{Paul}},
  \bibinfo{author}{L.~{Chauviere}}, \bibinfo{author}{J.~{Schmalian}},
\newblock \bibinfo{title}{Nematic resonance in the {R}aman response of
  iron-based superconductors},
\newblock \bibinfo{journal}{ArXiv e-prints}  (\bibinfo{year}{2015})
  \bibinfo{pages}{1504.04570}.
\bibitem[{Lederer et~al.(2015)Lederer, Schattner, Berg, and
  Kivelson}]{Lederer2015}
\bibinfo{author}{S.~Lederer}, \bibinfo{author}{Y.~Schattner},
  \bibinfo{author}{E.~Berg}, \bibinfo{author}{S.~A. Kivelson},
\newblock \bibinfo{title}{Enhancement of superconductivity near a nematic
  quantum critical point},
\newblock \bibinfo{journal}{Phys. Rev. Lett.} \bibinfo{volume}{114}
  (\bibinfo{year}{2015}) \bibinfo{pages}{097001}.

\end{thebibliography}
\end{document}